\documentclass[11pt, letterpaper]{amsart}
\usepackage[left=1in,right=1in,bottom=0.5in,top=0.8in]{geometry}
\usepackage{amsfonts}
\usepackage{amsfonts,amsbsy,amssymb,amsmath,graphicx,amsthm, bm} 
\usepackage{graphicx}
\usepackage[font=small,labelfont=bf]{caption}
\usepackage{epstopdf}
\usepackage{xcolor}
\usepackage{here}
\usepackage{float}
\usepackage{pgfplots}
\usepackage{listings}
\usepackage{longtable}
\usepackage{mathrsfs}
\usepackage{mathabx}
\usepackage{tikz}
\usepackage{calc}

\usetikzlibrary{calc,intersections}
\newsavebox{\mysquare}
\savebox{\mysquare}{\textcolor{red}{\rule{.5in}{.5in}}}

\newsavebox{\mysquareb}
\savebox{\mysquareb}{\textcolor{black}{\rule{.5in}{.5in}}}

\usepackage{subcaption}
\usepackage{lipsum}
\definecolor{CiteRed}{HTML}{B22222}  
\definecolor{LinkBlue}{HTML}{1A4FFF} 

\usepackage[colorlinks=true,
            linkcolor=LinkBlue,   
            citecolor=CiteRed,    
            urlcolor=LinkBlue     
           ]{hyperref}

\usepackage{epsfig}
\usepackage{dsfont}
\usepackage{natbib}
\usepackage{mathrsfs}
\usepackage{geometry}

\usepackage{booktabs}   
\usepackage{multirow}   
\usepackage{array}      

\usepackage{graphicx}

\usepackage{float} %

\newtheorem{thm}{Theorem}[section]
\newtheorem{cor}{Corollary}[section]

\newtheorem{lem}{Lemma}[section]

\theoremstyle{definition}

\newtheorem{assumption}{Assumption}

\theoremstyle{remark}
\newtheorem{rem}{Remark}[section]

\definecolor{energy}{RGB}{114,0,172}
\definecolor{freq}{RGB}{45,177,93}
\definecolor{spin}{RGB}{251,0,29}
\definecolor{signal}{RGB}{203,23,206}
\definecolor{circle}{RGB}{217,86,16}
\definecolor{average}{RGB}{203,23,206}

\colorlet{shadecolor}{gray!20}
\pgfplotsset{compat=1.9}

\usepgflibrary{fpu}

\newcommand{\R}{\mathbb{R}}

\newcommand{\F}{\mathcal{F}}

\def\1{1\kern-.20em {\rm l}}

\numberwithin{equation}{section}

\author[Mohamed Chaouch]{Mohamed Chaouch}
\address{(M. Chaouch)[corresponding author] Statistics Program, Department of Mathematics and Statistics, College of Arts and Sciences, Qatar University, Doha, Qatar}
\email{mchaouch@qu.edu.qa}

\author[Amina Angelika Bouchentouf ]{Amina Angelika Bouchentouf}
\address{(A. A. Bouchentouf) Laboratory of Mathematics, Djillali Liabes University of Sidi Bel Abbes, Sidi Bel Abbes, Algeria }
\email{bouchentouf\_amina@yahoo.fr}

\title[Conditional Extreme Market Risk Forecasting from Intraday Return Curves]{Conditional Extreme Market Risk Forecasting from Intraday Return Curves }
\begin{document}
\maketitle

\begin{abstract}
We develop a semiparametric framework for forecasting extreme conditional
Value-at-Risk (VaR) and Expected Shortfall (ES) when the conditioning
information is an intraday return curve rather than a finite set of daily
summary statistics. Under a functional location--scale specification, the
conditional mean and volatility of next-day returns are estimated by local
linear regression, and extreme tail risk is obtained by applying the Hill
estimator and Weissman extrapolation to the resulting standardized
residuals. We establish the consistency and asymptotic normality of the
proposed estimators for stationary, ergodic functional processes, and show
that the estimation of the location and scale operators is asymptotically
negligible relative to the extreme-value step. Monte Carlo experiments
confirm this ordering in finite samples: the accuracy of the feasible
estimator approaches that of its oracle counterpart as the probability
level becomes more extreme and as the tail becomes heavier, while at
moderate levels the estimation of the operators remains a material source
of error. In a decade-long out-of-sample application to the iShares
S\&P 500 Value ETF using one-minute bid--ask data, the functional model
attains the most accurate unconditional coverage among five competing
specifications at the $99.5\%$ level, and within periods of market stress
it is the only specification that passes both the conditional coverage and
the dynamic quantile tests. The advantage therefore lies in the timing
rather than in the average magnitude of extreme forecasts: a GARCH-EVT
filter attains a tighter average joint VaR--ES loss, and the functional
model is markedly conservative in calm markets. These findings indicate
that intraday price curves correct the dynamics of tail-risk forecasts
specifically during stress episodes, and motivate hybrid specifications
that combine parametric volatility filters with functional high-frequency
information.
\end{abstract}

\section{Introduction}\label{intro}

Accurate measurement of tail risk remains a central objective in financial
econometrics. Regulatory frameworks such as Basel~III and the Fundamental
Review of the Trading Book (FRTB) explicitly link capital requirements to
reliable estimation of Value-at-Risk (VaR) and Expected Shortfall (ES),
while portfolio allocation, risk budgeting, and stress testing all depend
critically on the stability and predictive accuracy of conditional risk
measures \citep{BCBS2019}. In financial markets, extreme losses are often
associated with heavy-tailed return distributions and rapidly changing
volatility conditions, which makes the accurate modeling of tail behavior
particularly challenging and has motivated a substantial body of
statistical and econometric work aimed at improving the estimation and
forecasting of extreme risk measures.

This paper develops a semiparametric estimator of extreme conditional VaR
and ES in which the conditioning information is not a finite vector of
lagged returns or volatility proxies, but the entire intraday return
curve observed on the preceding trading day. Combining functional
nonparametric regression with extreme-value tail extrapolation allows the
shape of the trading-day path, rather than a small set of daily summary
statistics, to inform the estimation of next-day tail risk.

Two main strands of the literature have addressed the problem of tail
risk estimation. The first relies on parametric or semiparametric
volatility models designed to capture time-varying second moments. Early
contributions include the GARCH framework of \cite{Bollerslev1986} and
its multivariate and dynamic extensions \citep{Engle2002}, as well as
stochastic volatility models \citep{Shephard2005}. These models provide
tractable representations of conditional heteroscedasticity and remain
widely used in practice. More recent approaches model conditional
quantiles or tail risk measures directly through regression-based
frameworks, including quantile regression \citep{Koenker2005} and dynamic
quantile specifications such as the CAViaR model of
\cite{EngleManganelli2004}. While these methods allow for flexible
modeling of conditional quantiles, they typically rely on
low-dimensional predictors constructed from lagged returns or volatility
measures.

A second line of research relies on extreme value theory (EVT), which
provides a rigorous probabilistic framework for analyzing tail behavior
under heavy-tailed distributions. Classical contributions establish the
asymptotic properties of tail index estimators and extreme quantile
estimators \citep{Embrechts1997,Beirlant2004,deHaanFerreira2006}.
Building on these foundations, a growing literature estimates
tail-related risk measures in financial time series by combining
volatility filtering with EVT-based tail extrapolation
\citep{McNeilFrey2000,ChavezDemoulin2005,CaiWang2008,WangTsai2009},
typically relying on the Weissman estimator \citep{Weissman1978} to
extrapolate extreme quantiles beyond the range of observed data. Recent
advances have further refined these methods by addressing bias reduction
and second-order tail behavior in extreme quantile estimation
\citep{gar10,Girardannals2021,DaouiaStupflerUsseglioCarleve2023}, thereby
improving the statistical accuracy of extreme risk forecasts. For instance, \citet{Girardannals2021} develop a general framework for residual-based extreme value estimation in heteroscedastic regression models, showing that under a high-level condition on the discrepancy between residuals and unobserved errors, residual-based estimators inherit the asymptotics of their unfeasible counterparts. Our Section~\ref{sec:mainresults} establishes that condition for the functional local linear estimator under ergodicity, thereby extending their framework to an infinite-dimensional covariate and more general dependence structure that includes processes satisfying the strong mixing condition.

Despite these advances, most empirical implementations continue to rely
on models that summarize market information through a limited set of
daily variables. Modern financial markets, however, generate vast
amounts of high-frequency data describing the intraday evolution of
prices, volatility, liquidity, and order flow. Intraday price paths often
exhibit volatility bursts, directional pressure, and liquidity shocks
that may contain valuable information about subsequent extreme returns;
aggregating these dynamics into daily summary statistics can obscure
nonlinear relationships between current market conditions and future
tail risk. This observation motivates the use of richer data
representations that capture the entire intraday trajectory of market
activity when modeling extreme financial risk.

Functional data analysis (FDA) provides a natural framework for
incorporating the entire intraday trajectory as a predictor
\citep{Kokoszka2017}. Rather than representing the trading day through a
small set of summary statistics, the intraday return process can be
viewed as a functional object whose geometry encodes the temporal
evolution of volatility, liquidity conditions, and information arrival
throughout the trading session. Several recent studies show that
modeling intraday returns as functional curves improves forecasting and
risk assessment relative to approaches based on aggregated measures. For
instance, \citet{RiceWirjantoZhao2020} show that intraday return curves
contain significant predictive information for Value-at-Risk forecasts,
\citet{KokoszkaMiaoStoevZheng2018} analyze the risk properties of
cumulative intraday return curves and highlight their relevance for tail
risk analysis, and \citet{ZhangZhangCucuringuQian2024} document strong
intraday commonality structures that can be exploited with machine
learning methods for volatility forecasting.

Within this perspective, the intraday return trajectory $X_t(\cdot)$ can
be treated as a functional covariate, allowing next-day risk measures to
be modeled conditionally on the shape of the trading-day path. This
framework lets the data determine which aspects of the intraday
evolution---early-day volatility bursts, persistent directional
movements, or late-day liquidity imbalances---carry predictive content
for subsequent downside risk.

We propose a functional location--scale model for conditional tail risk,
specified as
$$
Y_t = m(X_t) + \sigma(X_t)\,\varepsilon_t,
\qquad t=1,\dots,n,
$$
where $Y_t$ denotes the return realized over the trading day immediately
following the observation of the intraday curve $X_t(\cdot)$, $m(\cdot)$
captures the conditional mean component, and $\sigma(\cdot)$ represents the
conditional volatility. This specification induces a semiparametric
estimator of extreme conditional VaR and ES, obtained by combining
functional local linear estimation of the location and scale components
with EVT-based tail extrapolation applied to the standardized residuals.

A central feature of our approach is the use of local linear, rather than
local constant, estimation for both $m(\cdot)$ and $\sigma(\cdot)$. Local
linear smoothing offers two main advantages in this setting. First, it
reduces boundary bias, which is particularly relevant when neighborhoods
are defined through semi-metrics on an infinite-dimensional space, where
boundary effects are pervasive rather than confined to the edge of the
support. Second, it provides first-order bias correction without an
explicit parametric specification, preserving flexibility while
improving finite-sample performance relative to local constant smoothing \citep{BailloGrane2009,FerratyNagy2022}.


Modeling the volatility component through a flexible local specification
allows the conditional scale to vary nonparametrically with the geometry
of the intraday curve \citep{muller11,Hormann13,Chaouch2019}. This
flexibility matters because the relationship between market conditions
and future volatility is rarely linear or globally smooth: parametric
volatility models impose a fixed dynamic structure that can miss
nonlinear amplification effects during periods of market stress, whereas
a local modeling strategy lets the data determine how specific
features of the intraday trajectory---volatility clustering, directional
pressure, or end-of-day liquidity withdrawal---translate into next-day
dispersion.

To our knowledge, \citet{Ferraty2016} provided the first study of
conditional VaR and ES within a functional-data framework. Their approach
delivers nonparametric estimates of risk measures at conventional
probability levels but did not address the extrapolation problem that
arises when the target confidence level lies in the extreme tail. Our
methodology instead combines a functional location--scale representation
with extreme-value techniques applied to standardized residuals,
enabling inference on \emph{extreme} conditional VaR and ES through
tail-index estimation and Weissman-type extrapolation. This construction
decomposes tail risk into three economically interpretable components: a
conditional location effect, a conditional scale component driven by the
intraday curve, and a tail component governed by the heavy-tailed
innovation distribution.

The paper contributes to the econometrics of extreme risk measurement
along three dimensions. First, we extend the asymptotic theory of
functional nonparametric regression to the extreme-value setting under
stationarity and ergodicity rather than independence, establishing the
consistency and asymptotic normality of the residual-based Hill
estimator and of the resulting extreme conditional VaR and ES for a
functional covariate process. We also show that, under conditions on the relative rates, the estimation of the functional operators is asymptotically negligible relative to the extreme-value step. Second, we embed local linear functional
regression and Weissman-type tail extrapolation within a single
estimation pipeline that nests the corresponding scalar-covariate
procedure as a special case, so that the marginal contribution of
high-frequency intraday information can be isolated empirically rather
than confounded with differences in estimation method. Third, in a
decade-long out-of-sample application to a U.S. equity ETF, we show that
this contribution is regime-dependent: the functional covariate corrects
the \emph{timing}, rather than uniformly the depth, of extreme-risk
forecasts, yielding the only specification among five competing models that passes both the conditional coverage and the dynamic quantile tests during periods of market stress, where every competitor is rejected, while a parametric GARCH-EVT filter retains a tighter
average joint VaR--ES loss and the functional model remains markedly
conservative in calm markets. Taken together, these results indicate
that the value of functional high-frequency information for tail risk is
conditional on the market regime, and motivate the hybrid specification
we propose in Section~\ref{sec:conclusion}, combining parametric
volatility filtering with functional intraday information.

\subsection*{Layout of the paper}
The remainder of the paper is organized as follows. Sections~\ref{ESTCOMP}
and \ref{sec:estimation} introduce the functional location--scale
framework and present the proposed estimators of conditional VaR and ES
based on extreme value theory. Section~\ref{sec:mainresults} establishes
the asymptotic properties of the estimators under heavy-tailed
conditions. Section~\ref{sec:tuning} discusses the selection of the
tuning parameters involved in the estimation procedure.
Section~\ref{sec:simulation} investigates finite-sample performance
through Monte Carlo experiments. Section~\ref{sec:application} provides
an empirical application to the IVE ETF, using intraday high-frequency
data to construct functional covariates and evaluate the predictive
content of intraday price dynamics for extreme next-day losses.
Section~\ref{sec:conclusion} concludes. The proofs of the main results
are presented in the Appendix.

\subsection*{Notation}
We denote by $o_{a.s.}(v)$ a real random function $z$ such that $z(v)/v$
converges to zero almost surely as $v\rightarrow 0$. Similarly, we define
$\mathcal{O}_{a.s.}(v)$ as a real random function $z$ such that $z(v)/v$ is
almost surely bounded. We denote by $C$ a positive generic constant that may
change value. We use the notation $\overset{d}{\longrightarrow}$ to denote
convergence in distribution. We denote by $\1_A$ the indicator function that
evaluates to 1 if $A$ is satisfied and zero otherwise.

\section{Functional location-scale regression model with heavy-tailed errors}\label{ESTCOMP}
\subsection{The model}Let $(X_t,Y_t)_{t=1,\dots,n}$ be a pair of discrete-time stochastic processes taking values in $\mathcal{E}\times\mathbb{R}$ and distributed as $(X,Y)$. The variable $Y$ denotes the response of interest, while $X$ is a functional covariate taking values in an infinite-dimensional space $\mathcal{E}$. The space $\mathcal{E}$ is equipped with a semi-metric $d(\cdot,\cdot)$\footnote{A semi-metric (sometimes called pseudo-metric) $d(\cdot,\cdot)$ is a metric that allows $d(x_1,x_2)=0$ for some $x_1\neq x_2$.}, which defines a topology measuring the proximity between two elements of $\mathcal{E}$. Importantly, this semi-metric is chosen independently of the probabilistic definition of $X$ in order to avoid measurability issues.

We observe a sample $\mathcal{D}_n=\{(X_t,Y_t),\,t=1,\dots,n\}$ and assume throughout the paper that $(X_t,Y_t)_{t=1,\dots,n}$ forms a stationary and ergodic process. The data-generating mechanism is described by the following heteroscedastic functional regression model
\begin{equation}\label{model}
Y_t = m(X_t) + \sigma(X_t)\,\varepsilon_t , \qquad t=1,\dots,n,
\end{equation}
where $m(\cdot)=\mathbb{E}(Y\,|\,X=\cdot)$ denotes the regression operator and $\sigma^2(\cdot)\equiv U(\cdot)=\operatorname{Var}(Y\,|\,X=\cdot)$ the conditional variance operator. Both operators are assumed to be unknown.
\color{black}

\begin{rem}
It is worth mentioning that the strict stationarity and ergodicity of the joint process $(X_{t}, Y_{t})_{t \in \mathbb{Z}}$ are assumed as standard conditions in the functional data analysis setting (see, e.g., \citep{LL10}). From a theoretical point of view, the existence of such a stationary solution can be ensured by imposing Lipschitz-type contraction condition on the functional operators $m(\cdot)$ and $\sigma(\cdot)$. Since our asymptotic results are based on the stationarity and ergodicity of the observed process, we adopt these assumptions without imposing further structural restrictions on the unknown operators.
\end{rem}
\color{black}
The innovations $\{\varepsilon_t\}_{t=1}^n$ are assumed to be independent and identically distributed with
$
\mathbb{E}(\varepsilon_t)=0, 
\quad 
\mathbb{E}(\varepsilon_t^2)=1,
$
and independent of the functional covariate $X_t$ for all $t\in\{1,\dots,n\}$. Under the location--scale representation \eqref{model}, the standardized \color{black} errors (or innovations) \color{black} satisfy
$
\varepsilon_t = (Y_t - m(X_t))/\sigma(X_t).
$

Finally, we assume that the innovation distribution $F_\varepsilon$ is
heavy-tailed of Pareto type with tail index $\gamma \in (0,1/2)$, as
detailed in the following assumption. The upper bound is not an
additional restriction but a consequence of the normalisation
$\mathbb{E}(\varepsilon_t^2)=1$ adopted above: a distribution whose
survival function is regularly varying with index $-1/\gamma$ has a
finite second moment only if $\gamma<1/2$. We note that the risk
measures themselves are well defined under the weaker condition
$\gamma<1$, which is the condition retained in
Theorems~\ref{thm:extremeIQeps} and~\ref{thm:extremeQ} for the
finiteness of the Expected Shortfall.

\begin{assumption}\label{cgamma}
The innovation distribution $F_\varepsilon$ is heavy--tailed with tail index $\gamma \in (0, 1/2)$ such that its survival function satisfies
$
\overline F_\varepsilon(x):=1-F_\varepsilon(x)
\sim x^{-1/\gamma}L(x), 
\quad x\to\infty,
$
where $L$ is a slowly varying function. 

Let $\mathcal Q_\varepsilon$ denote the tail quantile function of $\varepsilon$ defined by
\[
\mathcal Q_\varepsilon(z)
=
\inf\left\{v\in\mathbb{R}:\frac{1}{1-F_\varepsilon(v)}\ge z\right\},
\qquad z>1.
\]
We further assume that $\mathcal Q_\varepsilon$ satisfies the following second--order regular variation condition: for all $v>0$,
\begin{equation}\label{eq:secondorder}
\lim_{z\to\infty}
\frac{1}{A(z)}
\left\{
\frac{\mathcal Q_\varepsilon(zv)}{\mathcal Q_\varepsilon(z)}-v^\gamma
\right\}
=
v^\gamma\frac{v^\rho-1}{\rho},
\end{equation}
for some $\rho<0$ and some function $A(z)\to0$ with constant sign as $z\to\infty$. When $\rho=0$, the expression $(v^\rho-1)/\rho$ is interpreted as $\log v$.
\end{assumption}

Assumption \ref{cgamma} corresponds to the standard second--order regular variation condition in extreme value theory and plays a key role in establishing the asymptotic normality of the estimator of the tail index $\gamma$ (see, e.g., \cite{deHaanFerreira2006}).

\subsection{Extreme conditional risk measures}

For  $\alpha \in (0,1)$, the unconditional quantile-based VaR of $\varepsilon$ is defined as the $\alpha$-th quantile of $F_\varepsilon$:
\[
\mu_\alpha(\varepsilon) \equiv \text{VaR}_\alpha(\varepsilon) = \inf\left\{ y \in \mathbb{R} : F_\varepsilon(y) \geq \alpha \right\}.
\]
The corresponding quantile-based Expected Shortfall (ES) is given by:
\[
\text{ES}_\alpha(\varepsilon) \equiv\mathbb{E}\left(\epsilon | \epsilon > \mu_\alpha(\epsilon) \right) = \frac{1}{1-\alpha} \int_\alpha^1 \mu_u(\varepsilon) du.
\]

Conditional quantile-based value-at-risk (CVaR) and conditional quantile-based expected shortfall (CES) are two of the most used synthetic measures of market risk in empirical finance (see \cite{Danielsson2011}).  Moreover, \cite{Beirlant2004} have shown that, as $\alpha \rightarrow 1$ and $0 < \gamma < 1$, the following relationship between the unconditional VaR and ES holds:
\begin{equation}\label{eq:VARESrelation}
\dfrac{\text{ES}_\alpha(\varepsilon)}{\mu_\alpha(\varepsilon)} \sim \dfrac{1}{1-\gamma},
\end{equation}
where the notation ``$\sim$" (or $\approx$) means asymptotic equivalence.

Let $Y_t$ be a stochastic process representing, e.g. the returns of a given stock, portfolio or market index. Then, for $\alpha \in (0,1)$ and $x\in \mathcal{E}$, we define the conditional value-at-risk of $Y_t$ given $X_t = x$, say $\text{CVaR}_\alpha(x)$, as the $\alpha$th conditional quantile associated with the conditional distribution of $Y_t$ given $X_t=x$, say $F_{Y_t|X_t=x}$.  Under location-scale model \eqref{model}, one can see the $\text{CVaR}_\alpha(x)$ as
\begin{eqnarray}\label{VaR}
\text{CVaR}_\alpha(x) &\equiv& \inf_y\{y: F_{Y_t|X_t}(y|x) \geq \alpha \} = \inf_y\left\{y: F_\epsilon\left(\dfrac{y - m(x)}{\sigma(x)}\right) \geq \alpha \right\}\nonumber\\
&=& m(x) + \sigma(x) F^{-1}_\epsilon(\alpha),
\end{eqnarray}
where $\sigma(x) := \sqrt{U(x)}$. Then it follows from equation \eqref{VaR} that $\mu_\alpha(x) := \text{CVaR}_\alpha(x)$ can be written, under model \eqref{model}, as follows:
\begin{equation}\label{defwithmu}
    \mu_\alpha(x) = m(x) + \sigma(x)\mu_\alpha(\epsilon).
\end{equation}

The conditional quantile-based expected shortfall $\text{CES}_\alpha(x)$ of $Y_t$ given $X_t = x$ is defined as the conditional expectation of $Y_t$ given that $Y_t$ exceeds its conditional value-at-risk $\mu_\alpha(x)$. Under model \eqref{model}, one defines 
\begin{eqnarray}\label{QCES}
 \text{CES}_\alpha(x) &\equiv& \mathbb{E}\left(Y_t | Y_t > \mu_\alpha(x) \right)\nonumber\\
 &=& m(x) + \sigma(x)\cdot \text{ES}_\alpha(\epsilon).
\end{eqnarray}

Extreme risk measures play a central role in modern financial econometrics because the most severe losses faced by market participants occur in the tails of return distributions. For illiquid securities and assets exposed to abrupt volatility shifts, these tail events are often driven by rare but economically meaningful market conditions, making conventional second-moment measures inadequate for risk assessment. High–confidence Value-at-Risk (VaR) and Expected Shortfall (ES) provide synthetic and interpretable summaries of tail risk that are directly relevant for capital allocation, portfolio stress testing, and regulatory requirements. Their conditional counterparts, which evaluate the distribution of returns given prevailing market states, are particularly informative in environments where volatility, liquidity, or trading activity vary systematically with macro–financial conditions. 

Within the functional location–scale framework \eqref{model}, extreme conditional quantiles and conditional expected shortfalls emerge naturally from the scale component $U(x)$ and reflect the heavy-tailed features of the error distribution $\varepsilon$. This provides a coherent link between extreme losses and the functional covariate $X_t$, motivating the formal definition and estimation of conditional VaR and ES at extreme probability levels. We now introduce these risk measures more precisely.

In many financial applications, interest lies not only in the estimation of 
$\text{CVaR}_\alpha(x)$ and $\text{CES}_\alpha(x)$ for moderate probability levels 
$\alpha \in (0,1)$, but in their behavior as $\alpha$ approaches one. 
The main issue then is that the straightforward empirical estimator of the quantile function of $\varepsilon$ (and thus the one of $Y|X=x$) is not consistent at extreme levels; in other words, direct estimation of a quantile exceeding the range covered by the available data is impossible, and this is of course an obstacle to meaningful estimation results in practice. This problem can be solved by considering the well-known Weissman's extrapolation procedure (see \cite{Weissman1978}) which consists in estimating the quantile of $\varepsilon$ at a smaller level $\alpha_n$, then, the targeted extreme quantile of higher order $\tau_n$, such that $\alpha_n< \tau_n \rightarrow 1,$ can be deduced as follows:
\begin{equation}\label{weissmanExtra}
\dfrac{\mu_{\tau_n}(\varepsilon)}{ \mu_{\alpha_n}(\varepsilon)} \approx \left(\dfrac{1- \alpha_n}{1 - \tau_n} \right)^{\gamma}.
\end{equation}

Combining the tail representation in \eqref{weissmanExtra} with the functional location–scale structure in \eqref{model}, one defines the extreme conditional value-at-risk of $Y|X=x$ at an extreme level $\tau_n \rightarrow 1$ as follows:
\begin{eqnarray}\label{eq:extreme_CVaR}
    \mu_{\tau_n}(x)
    &=& m(x) + \sigma(x)\, \mu_{\tau_n}(\varepsilon)\\
    &=& m(x) + \sigma(x)\, \left(\dfrac{1- \alpha_n}{1 - \tau_n} \right)^{\gamma} \,\mu_{\alpha_n}(\varepsilon)\nonumber
\end{eqnarray}
In the same spirit, the extreme conditional expected shortfall of $Y|X=x$ is defined, for $\tau_n \rightarrow 1,$ as follows
\begin{equation*}\label{eq:extreme_CES}
    \text{CES}_{\tau_n}(x)
    = m(x) + \sigma(x)\, \text{ES}_{\tau_n}(\varepsilon).
\end{equation*}

Note that thanks to the asymptotic equivalence \eqref{eq:VARESrelation} and \eqref{weissmanExtra}, one can show that, for $\alpha_n < \tau_n \rightarrow 1$, we have
\begin{equation}
\dfrac{\text{ES}_{\tau_n}(\varepsilon)}{\text{ES}_{\alpha_n}(\varepsilon)} \sim \dfrac{\mu_{\tau_n}(\varepsilon)}{\mu_{\alpha_n}(\varepsilon)} \approx \left(\dfrac{1- \alpha_n}{1 - \tau_n} \right)^{\gamma}.
\end{equation}
Thus, we finally obtain 
\begin{eqnarray}\label{extremeCES}
\text{CES}_{\tau_n}(x)  = m(x) + \sigma(x) \text{ES}_{\tau_n}(\varepsilon) &=& m(x) + \sigma(x)\, \left(\dfrac{1- \alpha_n}{1 - \tau_n} \right)^{\gamma} \, \text{ES}_{\alpha_n}(\varepsilon)\nonumber\\
&=& m(x) + \sigma(x)\, \left(\dfrac{1- \alpha_n}{1 - \tau_n} \right)^{\gamma} \, \dfrac{1}{1-\gamma}\, \mu_{\alpha_n}(\varepsilon).
\end{eqnarray}
Equations \eqref{eq:extreme_CVaR} and \eqref{extremeCES} 
show that the behavior of extreme conditional risk measures is governed entirely 
by the scale function $\sigma(x)$ and the tail quantile function of the innovations.  
The functional covariate therefore influences extreme conditional VaR and ES 
through its effect on the local volatility level $\sigma(x)$, 
while the heaviness of the error distribution determines the magnitude and rate of divergence 
of these tail-based risk measures.



\section{Estimation}\label{sec:estimation}

  \subsection{Local linear estimation of \color{black} the conditional mean $m(\cdot)$ and the conditional variance function $U(\cdot)$ \color{black} }
 Based on the sample $(X_t, Y_t)_{t=1}^n$, a local linear estimator of the 
regression operator $m(\cdot)$ is obtained by solving the weighted least squares problem

$$
\displaystyle\min_{(a, b) \in \mathbb{R}^2} \displaystyle\sum_{t=1}^n  \left(Y_t-a-b \rho\left(X_t, x\right)\right)^2  K\left(h^{-1} \varrho\left(x, X_t\right)\right).
$$
where $\rho(\cdot,\cdot)$ denotes a continuous bilinear form on $\mathcal{E}^2$ satisfying 
$\rho(o,o)=0$ for all $o \in \mathcal{E}$, and where the semi-metric is given by 
$d(\cdot,\cdot) = |\varrho(\cdot,\cdot)|$. 
     
For any fixed $x \in \mathcal{E}$, the resulting estimator admits the explicit representation
\begin{equation}\label{NWV1}
m_{n}(x)=\frac{\displaystyle\sum_{s=1}^n \Gamma_{s,l}(x)K_s(x) Y_s}{\displaystyle\sum_{s=1}^n \Gamma_{s,l}(x)K_s(x)}, 
\end{equation}
where
\begin{equation}\label{VS}
\Gamma_{s,l}(x) =
\begin{cases}
1, &  l=0 \\
\left(\displaystyle\sum_{t=1}^n \rho^2_t(x)K_t(x)\right) - \left(\displaystyle\sum_{t=1}^n \rho_t(x)K_t(x)\right)\rho_s(x), &  l=1 \\
\end{cases}
\end{equation}
 
 with the convention 0/0 = 0, and where
 \begin{equation*} \label{rK}\rho_t(x)=\rho\left(X_t, x\right) \quad  \text{and} \quad K_t(x)= K\left(h^{-1} \varrho\left(x, X_t\right)\right)
 \end{equation*}
 
Similarly, the conditional variance operator $U(\cdot)$ can be estimated by applying 
local linear smoothing to the squared residuals
\[
r_t = \left( Y_t - m_n(X_{t}) \right)^2.
\]
Specifically, $U(x)$ is characterized as the minimizer of
\[
\min_{(c,d) \in \mathbb{R}^2} 
\sum_{t=1}^n 
 \left( r_t - c - d\, \rho(X_t, x) \right)^2 
K\!\left( h^{-1}\, \varrho(x, X_t) \right),
\]
which yields the estimator
\begin{equation}\label{Un}
U_n(x) 
=
\frac{
\displaystyle\sum_{s=1}^n \Gamma_{s,l}(x)\, K_s(x)\, r_s
}{
\displaystyle\sum_{s=1}^n \Gamma_{s,l}(x)\, K_s(x)
},
\end{equation}
where $\Gamma_{s,l}(x)$ is defined in \eqref{VS}.\\

\subsection{Indirect extreme conditional risk measures estimation}

First, note that the estimation of the theoretical extreme conditional VaR and ES 
defined in \eqref{eq:extreme_CVaR} and \eqref{extremeCES} requires consistent estimates of 
the regression and volatility operators, as well as estimates of the tail index $\gamma$ 
and the intermediate quantile $\mu_{\alpha_n}(\varepsilon)$ of the innovation distribution. 
The local linear estimators of $m(\cdot)$ and $U(\cdot)$ were introduced in 
\eqref{NWV1} and \eqref{Un}, respectively. We now turn to the estimation of 
$\gamma$ and $\mu_{\alpha_n}(\varepsilon).$

\subsubsection{Estimation of the tail index}

\color{black} The model errors \color{black}
\[
\varepsilon_t = \sigma^{-1}(X_t)\left(Y_t - m(X_t)\right), \qquad t=1,\ldots,n,
\]
constitute a randomly i.i.d.\
sample from $F_\varepsilon$ and write
$\varepsilon_{1:n} \le \cdots \le \varepsilon_{n:n}$ their order
statistics and let $k_n$ be an intermediate sequence such that
$k_n \to \infty$ and $k_n/n \to 0$. The oracle Hill estimator based on the unobserved errors is then
\begin{equation}\label{HillMAR}
\widecheck{\gamma}
=
\frac{1}{k_n}
\sum_{j=1}^{k_n}
\left\{
\log \varepsilon_{n-j+1:n}
-
\log \varepsilon_{n-k_n:n}
\right\}.
\end{equation}

In practice, the true errors $\varepsilon_t$ are not observed, and the Hill 
estimator must therefore be constructed from residuals. Let
\begin{equation}\label{epsilonhat}
\widehat{\varepsilon}_t
=
\sigma_n^{-1}(X_t)\left(Y_t - m_n(X_t)\right), \qquad t = 1,\ldots,n,
\end{equation}
where $m_n(\cdot)$ and $\sigma_n(\cdot)$ are the local linear estimators defined 
in \eqref{NWV1} and \eqref{Un}. Let $\widehat{\varepsilon}_1,\ldots,\widehat{\varepsilon}_{n}$ denote the 
corresponding observed residuals, and write 
$\widehat{\varepsilon}_{1:n} \le \cdots \le 
 \widehat{\varepsilon}_{n:n}$ for their order statistics. The Hill estimator of the tail index $\gamma$ is then given by
\begin{equation}\label{HillResidual}
\widehat{\gamma}
=
\frac{1}{k_n}
\sum_{j=1}^{k_n}
\left\{
\log \widehat{\varepsilon}_{n-j+1:n}
-
\log \widehat{\varepsilon}_{n-k_n:n}
\right\},
\end{equation}
with $k_n \to \infty$ and $k_n/n \to 0$, as $n\to \infty.$ 


\subsubsection{Estimation of extreme conditional VaR and ES}
We now turn to the estimation of intermediate conditional quantiles of the
innovation distribution. Let $\alpha_n < \tau_n \to 1$ denote an intermediate
probability level. If the true errors $\varepsilon_t$ were observable, a direct estimation of the
intermediate quantile $\mu_{\alpha_n}(\varepsilon)$ could be obtained by
minimizing the weighted quantile regression criterion
\[
\sum_{t=1}^n \rho_{\alpha_n}(\varepsilon_t - u)
\quad\text{with respect to } u,
\]
where $\rho_{\alpha_n}(y) = |y|\,|\alpha_n - \1_{\{y \le 0\}}|$
denotes the check function. Since the true errors are not available, we replace
$\varepsilon_t$ by the residuals $\widehat{\varepsilon}_t$ defined in
\eqref{epsilonhat}, which yields the feasible estimator
\begin{equation}\label{mueps}
\widehat{\mu}_{\alpha_n}(\varepsilon)
=
\arg\min_{u \in \mathbb{R}}
\sum_{t=1}^n \rho_{\alpha_n}(\widehat{\varepsilon}_t - u).
\end{equation}
\begin{rem}
For the choice $\alpha_n = 1 - k_n/n$, the estimator 
$\widehat{\mu}_{\alpha_n}(\varepsilon)$ defined in \eqref{mueps} coincides with 
the $(n - k_n)$-th order statistic $\widehat{\varepsilon}_{n-k_n:n}$ of the observed residuals 
$\widehat{\varepsilon}_{1},\ldots,\widehat{\varepsilon}_{n}$, in direct 
analogy with the intermediate-quantile construction used in Weissman-type 
extrapolation (see, e.g., \cite{Cher05}).
\end{rem}

Combining the local linear estimators $m_n(\cdot)$ and $\sigma_n(\cdot)$ given
in \eqref{NWV1} and \eqref{Un} with the residual-based intermediate quantile
estimator \eqref{mueps}, the Hill estimator \eqref{HillResidual}, and the
extreme-value extrapolation formula \eqref{eq:extreme_CVaR}, we obtain the
estimator of the extreme conditional value-at-risk of $Y_t$ given $X_t = x$:
\begin{eqnarray}\label{estimatorcvar}
\widehat{\mu}_{\tau_n}(x)
&=& m_n(x)
+
\sigma_n(x) \widehat\mu_{\tau_n}(\varepsilon)\\
&=&
m_n(x)
+
\sigma_n(x)
\left(
\frac{1-\alpha_n}{1-\tau_n}
\right)^{\widehat{\gamma}}
\widehat{\mu}_{\alpha_n}(\varepsilon),
\qquad 
\alpha_n < \tau_n \to 1.\nonumber
\end{eqnarray}

Similarly, combining \eqref{NWV1}, \eqref{Un}, and \eqref{mueps} with the Hill
estimator \eqref{HillResidual} and the extreme conditional expected shortfall
representation \eqref{extremeCES}, we obtain the estimator of the extreme
conditional expected shortfall:
\begin{eqnarray}\label{estimatorcES}
\widehat{\text{CES}}_{\tau_n}(x)
&=& 
m_n(x)
+
\sigma_n(x)
\left(
\frac{1-\alpha_n}{1-\tau_n}
\right)^{\widehat{\gamma}} \, \widehat{\text{ES}}_{\alpha_n}(\varepsilon)\\
&=&
m_n(x)
+
\sigma_n(x)
\left(
\frac{1-\alpha_n}{1-\tau_n}
\right)^{\widehat{\gamma}}
\frac{1}{1-\widehat{\gamma}}\,
\widehat{\mu}_{\alpha_n}(\varepsilon),
\qquad 
\alpha_n < \tau_n \to 1.\nonumber
\end{eqnarray}

If we consider the intermediate level $\alpha_n = 1-k_n/n$, the intermediate $\alpha_n$-th quantile is replaced by the empirical order statistic $\widehat{\varepsilon}_{n-k_n:n}$ which now can be plugged-in equations \eqref{estimatorcvar} and \eqref{estimatorcES} to obtain the indirect extreme conditional VaR and ES of $Y|X=x.$

\section{Main results}\label{sec:mainresults}
Let $\mathcal C \subset \mathcal E$ be a class of functions such that the
covariate process satisfies $\mathbb{P}(X_t \in \mathcal C)=1$ for all $t\ge 1$.
For $\eta>0$, denote by $N(\eta,\mathcal C,\varrho)$ the $\eta$--covering number
of $\mathcal C$ with respect to the semi-metric $\varrho$, defined as
\[
N(\eta,\mathcal C,\varrho)
=
\min\Bigl\{n\ge 1 : \exists\, c_1,\dots,c_n\in\mathcal C
\ \text{such that}\
\forall x\in\mathcal C,\ \exists\, k\in\{1,\dots,n\} 
\ \text{with}\ \varrho(x,c_k)<\eta\Bigr\}.
\]
The quantity $N(\eta,\mathcal C,\varrho)$ describes the geometric complexity of
$\mathcal C$; in particular, finite covering numbers for all $\eta>0$ ensure that
$\mathcal C$ is sufficiently regular to allow uniform convergence rates for the estimators
$m_n(\cdot)$ and $\sigma_n(\cdot)$ defined in Section~\ref{sec:estimation}.

\subsection{Asymptotics for  $m_n(\cdot)$ and $U_n(\cdot)$} For $t=1, \dots, n$, let $\mathcal{F}_t$ denote  the $\sigma$-field  generated by $((X_1, Y_1), \dots, (X_t, Y_t)).$ 

\begin{assumption}[Kernel]\label{A1}
The kernel $K$ is nonnegative, bounded, Lipschitz continuous, and supported on $[-1,1]$.
\end{assumption}
\begin{assumption}[Small ball probability]\label{SBP}
For all $r > 0$,
\begin{itemize}
\item[(i)] $\forall x \in \mathcal{E}$, $\phi(r) = \mathbb{P}(X \in B(x, r)) > 0$, where $B(x, r) = \{x' \in \mathcal{E} : \mathbf{\varrho(x', x)} < r\}$.
\item [(ii)]
For all $t = 1,\ldots , n$, there exists a $\mathcal{F}_{t-1}$-measurable function $\phi_{t}(.)$  such that $0 < 
\mathbb{P}(X \in B(x, r)|\mathcal{F}_{t-1})\leq \phi_{t}(r).$ 
\color{black}
\item [(iii)] 
$\displaystyle \{n\phi(r)\}^{-1} \sum_{t=1}^{n} \phi_{t}(r)\overset{}{\rightarrow} 1 \quad \text{a.s.} $  as $r \rightarrow 0$.\color{black}
\end{itemize}
\end{assumption}

\begin{assumption}[Moments]\label{errors} We assume the following finite moments conditions.
\begin{itemize}
\item[(i)] There exists $\chi > 1$ such that  
$\mathbb{E}\{U(X_{t})^{\chi^{2}/2(\chi-1)}\}<\infty$ 
and $\mathbb{E}(|\epsilon_{t}|^{\chi})<\infty$ \quad \text{for all } $t=1,\dots,n.$

\item[(ii)] $\mathbb{E}\{(\epsilon^{2}_{t}-1)^{2}\} = \omega < \infty \quad \text{for all } \quad t=1,\dots,n.$

\item[(iii)] For some $\kappa > 0$, 
$\mathbb{E}\{(\epsilon^{2}_{t}-1)^{2+\kappa}\} < \infty \quad \text{for all } \quad t=1,\dots,n.$

\item[(iv)] $\mathbb{E}\{|U(X_{t})|^{\chi^{2}/(\chi-1)}\}<\infty$ 
and $\mathbb{E}(|\epsilon_{t}^{2}-1|^{\chi})<\infty$ \quad \text{for all } $t=1,\dots,n.$

\end{itemize}
\end{assumption}

\begin{assumption}[Semi-metric conditions]\label{SMC} We assume the following conditions on the semi-metrics:
\begin{itemize}
\item[(i)] $\forall z \in \mathcal{E},  C|\varrho(x,z)|\leq |\rho(x,z)|\leq C_0|\varrho(x,z)|$
\item[(ii)] $\forall (x_1,x_2) \in \mathcal{C}\times \mathcal{C}, |\rho(x_1,x)-\rho(x_2,x)|<C_0|\rho(x_1,x_2)|
$
\end{itemize}
\end{assumption}
\begin{assumption}(Regularity conditions on $m$ and $U$)\label{regularityconditions} We suppose the following smoothness conditions: 
\begin{itemize}
\item[(i)] For any $(x, x')\in \mathcal{E}^{2}$ and some  constant  $b_1 > 0$, $|m(x)-m(x')|<C\color{black}\varrho\color{black}^{b_1}(x,x')$.
\item [(ii)] For any $(x, x')\in \mathcal{E}^{2}$ and some  constant $b_2 > 0$, $|U(x)-U(x')|<C \color{black}\varrho\color{black}^{b_2}(x,x')$.
\item[(iii)] $U^{\kappa+2}(.)$ is continuous in a neighborhood of $x$ as $h \rightarrow 0$, that is
\begin{equation*}
\sup _{\{u: \varrho(x, u) \leq h\}}|U^{\kappa+2}(u)-U^{\kappa+2}(x)|=o(1).
\end{equation*}
\end{itemize}
\end{assumption}
\begin{assumption}[Bandwidth]\label{bdw}
Suppose that the bandwidth satisfies:
\begin{itemize}
 \item[(i)] $h$ is a positive sequence such that, $\displaystyle\lim _{n \rightarrow \infty}h=0$, $\displaystyle\lim _{n \rightarrow \infty}n\phi(h)=\infty$,  $\displaystyle\lim _{n \rightarrow \infty}\frac{\log n}{n\phi(h)}=0$,\\
and $\displaystyle\lim _{n \rightarrow \infty} \phi(h)^{m} n^{m}h^{2m}=0$ with $m > 0$ \color{black},
\item [(ii)]$\displaystyle-\frac{1}{\phi(h)}\int_{-1}^{1}\phi(zh,h)\frac{d}{dz}\left(z^2K(z)\right)dz>C$, 
\item [(iii)] $h\displaystyle \int_{B(x,h)}\rho(u,x)d\mathbb{P}(u)=o\left(\displaystyle \int_{B(x,h)}\rho^2(u,x)d\mathbb{P}(u)\right),$  where $d\mathbb{P}(x)$ is the cumulative distribution of $X$.
    \item[(iv)] $\displaystyle \int_{-1}^{1} (K^j)'(u) \phi(u) du < \infty$, for $j = 1, 2$.
\end{itemize}
\end{assumption}

\subsection*{Discussion of the assumptions}
The conditions above cover both the local constant ($l=0$) and the local
linear ($l=1$) estimator, where $l$ indexes the smoothing type through
$\Gamma_{s,l}$ in \eqref{VS}. We group them by the role they play.\\
\noindent\emph{Kernel and small ball probabilities.}
Assumption~\ref{A1} is standard. Assumption~\ref{SBP} is the usual
requirement when dealing with nonparametric estimation for dependent
functional data: parts (i)--(ii) ensure that the functional space
locally and conditionally charges the small balls, while (iii) controls
the asymptotic behavior of the conditional small ball probabilities
relative to their unconditional counterpart $\phi(r)$ as the radius
tends to zero. This last condition is what allows the asymptotic theory
to be developed under ergodicity, without imposing the mixing conditions
usually required in this literature \citep{LL10}. Such conditions relating the conditional and unconditional small ball probabilities are standard in this ergodic framework, and have been used in several subsequent works (see, e.g., \citep{Chaouch2019, Krebs2019}). \\
\noindent\emph{Moment conditions and the admissible range of $\gamma$.}
Assumption~\ref{errors} deserves particular comment, because its parts
place different and increasingly restrictive bounds on the tail index of
the innovations. Since $\overline F_\varepsilon$ is regularly varying
with index $-1/\gamma$, the variable $\varepsilon^2$ is regularly
varying with index $-1/(2\gamma)$, and a moment of order $p$ of
$\varepsilon$ is finite if and only if $\gamma<1/p$. Consequently:
\begin{itemize}
\item the normalisation $\mathbb E(\varepsilon_t^2)=1$ adopted in
Section~\ref{ESTCOMP}, which is what makes $U(\cdot)$ the conditional
variance and hence identifies the scale operator, already restricts
$\gamma<1/2$;
\item Assumption~\ref{errors}(i) requires $\mathbb E(|\varepsilon_t|^{\chi})
<\infty$ and therefore $\gamma<1/\chi$;
\item Assumption~\ref{errors}(iv) requires
$\mathbb E(|\varepsilon_t^{2}-1|^{\chi})<\infty$, that is a moment of
order $2\chi$ of $\varepsilon_t$, and therefore $\gamma<1/(2\chi)$;
\item Assumption~\ref{errors}(ii) requires a finite fourth moment of the
innovation and therefore $\gamma<1/4$;
\item Assumption~\ref{errors}(iii) requires a moment of order
$4+2\kappa$ and therefore $\gamma<1/(4+2\kappa)$; since $\kappa>0$ is
only required to exist, this again amounts to $\gamma<1/4$, the
admissible values of $\kappa$ shrinking to zero as $\gamma$ approaches
that boundary.
\end{itemize}
\noindent These conditions are not all used at the same place. The
uniform consistency results of Theorems~\ref{T1} and~\ref{T2} rely only
on Assumption~\ref{errors}(i) and (iv), and therefore hold whenever
$\gamma<1/(2\chi)$; the constant $\chi>1$ may be taken close to one when
the innovations are markedly heavy-tailed, at the cost of a slower
truncation rate through $\ell_n$ in Theorem~\ref{T1}. Assumption~(ii) and
(iii) are used exclusively for the asymptotic normality of $U_n(\cdot)$
in Theorem~\ref{T3}, whose limiting variance
$\frac{M_2}{M_1^2}\,\omega\,U^2(x)$ is finite only under a finite fourth
moment, and for the associated confidence interval \eqref{CIS}. They
therefore restrict the range of $\gamma$ for which inference on the
conditional variance is justified to $\gamma<1/4$, whereas the
extreme-value results of Section~\ref{sec:mainresults}, which build on
Theorems~\ref{T1}--\ref{T2} and require only $0<\gamma<1$ for the risk
measures themselves to be well defined, remain valid on the wider range.\\
\noindent Finally, since $\varepsilon_t$ is independent of $X_t$, the
conditional law of $Y_t$ given $X_t=x$ is the law of $\varepsilon$
shifted by $m(x)$ and scaled by $\sigma(x)$: the intraday curve may
raise or lower the level of next-day risk, but the shape of the tail,
and in particular the tail index $\gamma$, is the same at every $x$.
This is what allows the standardized residuals of
Section~\ref{sec:estimation} to be pooled and treated as a single sample
from $F_\varepsilon$ in the extreme-value step.\\
\noindent\emph{Conditions specific to the local linear construction.}
Assumptions~\ref{SMC}, \ref{regularityconditions}(i)--(ii) and
\ref{bdw}(ii)--(iii) are introduced to support the local linear
estimator: they relate the bilinear form $\rho$ to the semi-metric
$\varrho$, ensure the stability of the local design, and control the
bias terms generated by $\rho$. When $l=0$ one has
$\Gamma_{s,0}\equiv1$, the bilinear form disappears from the estimator
and these requirements become vacuous, so that the results reduce to the
classical Nadaraya--Watson theory for functional ergodic data
\citep{Chaouch2019}.\\ 
In the following we establish the main results for
$l\in\{0,1\}$ simultaneously, showing that the local linear
construction retains the convergence properties of its local constant
counterpart under this set of conditions.

\begin{thm}\label{T1}
Assume that Assumptions~\ref{A1}, \ref{SBP}, \ref{errors}(i), \ref{SMC}, \ref{regularityconditions}(i), and \ref{bdw}  hold true. In addition, for a sequence of positive real numbers $\lambda_{n}$ tending to zero, as $n \rightarrow \infty$ and  $\eta=\eta_{n}= o(h)$, assume that, we have
\begin{equation}\label{condunifconv}
\begin{aligned}
  \lim _{n \rightarrow \infty} \frac{\log N\left(\eta, \mathcal{C}, \varrho\right)}{\ell_{n}^{-2} \log n}=0 \quad & \text{and} \quad  \sum_{n \geq 1} \exp \left[-\log n\ell_{n}^{-2} \right]<\infty \,\  &
\end{aligned}
\end{equation}
where $\ell_{n}$ is a sequence of positive numbers that tends to infinity as $n \rightarrow \infty$ defined by\\ $\ell_{n}=\left(\frac{\log n}{\lambda_{n}\phi(h)^{\nu}n^{\nu}h^{2\nu}}\right)^{1/(\chi-1)},$ where $\nu=\frac{(\chi-1)}{\chi}, 
\chi$ is given in Assumption~\ref{errors}(i).Then, one finds
\begin{equation*}
\sup _{x \in \mathcal{C}}\left|m_{n}(x)-m(x)\right|=\mathcal{O}_{a . s .}\left(h^{b_1}\right)+\mathcal{O}_{a \cdot s .}\left( \sqrt{\frac{\log n}{n\phi(h)}}\right),
\end{equation*}
where \color{black} $m_n(\cdot)$ was defined in \eqref{NWV1}  \color{black} and $b_1$ is given in Assumption~\ref{regularityconditions}(i).
\end{thm}

\begin{thm}\label{T2}
Assume that Assumptions~\ref{A1}, \ref{SBP}, \ref{errors}(i), \ref{errors}(iv), \ref{SMC},  \ref{regularityconditions}(i)-(ii), \ref{bdw} and  condition  \eqref{condunifconv} hold true. Then,
\begin{equation*}
\sup _{x \in \mathcal{C}}\left|U_{n}(x)-U(x)\right|=\mathcal{O}_{a . s .}\left(h^{2b_1}+h^{b_2}\right)+\mathcal{O}_{a \cdot s .}\left( \sqrt{\frac{\log n}{n\phi(h)}} \right),
\end{equation*}
where $U_n(\cdot)$ was previously defined in  \eqref{Un} and $b_2$ is given in Assumption~\ref{regularityconditions}(ii).
\end{thm}


\begin{thm}\label{T3}
Assume that Assumptions~\ref{A1}-\ref{bdw} hold true. Additionally, suppose that the following conditions are satisfied:
\begin{equation}\label{BI}
\sqrt{n\phi(h)} h^{(l+1)b_1} \to 0 \quad \text{and} \quad \sqrt{n\phi(h)} \left( \frac{\log n}{n\phi(h)} \right) \to 0, \text{ as } n \to \infty, \quad \hbox{where} \quad l \in \{0, 1\}.
\end{equation}
Then, we have
\begin{itemize}
    \item[(i)] 
    $\displaystyle \sqrt{n\phi(h)} \left( \frac{{U}_n(x)}{U(x)} - 1 - B_n(x) \right) \xrightarrow{d} \mathcal{N}\left(0, \frac{M_2}{M_1^2}\omega\right)$, \quad \hbox{where} \quad $B_n(x) = \mathcal{O}(h^{b_2}).$ 

 \item[(ii)] If, in addition, we have
\begin{equation}\label{BI2}
\sqrt{n\phi(h)} h^{b_2} \to 0,
\end{equation}
then
    $\displaystyle \sqrt{n\phi(h)} \left( \frac{{U}_n(x)}{U(x)} - 1 \right) \xrightarrow{d} \mathcal{N}\left(0, \frac{M_2}{M_1^2}\omega\right)$,
where the constants $M_j$ (for $j=1,2$) are defined by
\begin{equation*}
M_j = K^j(1) - \int_{-1}^{1} (K^j)'(u) \phi(u) du.
\end{equation*}
\end{itemize}
\end{thm}

Our purpose now is to build asymptotic confidence intervals for $U(x)$ for any fixed curve $x \in \mathcal{E} $ based
on a normal approximation. Note that the asymptotic variance in Theorem \ref{T3} contains the unknown
quantities which are $\omega,$ and $U(\cdot).$  
Each of these parameters is replaced by its empirical version. We replace $U(x)$ by $U_{n}(x)$, and $\omega$ is estimated as follows:
\begin{equation*}
 \omega_{n}= n^{-1}\sum_{s=1}^n \left(\widehat{\varepsilon}_s^2 - 1 \right)^2,
\end{equation*}
where  $\widehat{\epsilon}_s$ is defined in \eqref{epsilonhat}.

\begin{cor}\label{Cor}
  Under the assumptions and conditions of Theorem \ref{T3}, we have
 \begin{equation}\label{Cor+1}
  \frac{\widehat{M}_1}{\sqrt{\widehat{M}_2}} \sqrt{\frac{n\phi(h)}{\omega_n}} \left( \frac{U_n(x)}{U(x)} - 1 \right) \xrightarrow{\mathcal{D}} \mathcal{N}(0, 1), \quad \hbox{as} \quad n\rightarrow\infty.
\end{equation}
\end{cor}


This corollary plays a key role in estimating confidence intervals of $U(x)$. More precisely, equation (\ref{Cor+1}) leads to obtain the following asymptotic $100(1-\alpha)\%$ confidence interval for the conditional variance $U(x)$
 \begin{equation}\label{CIS}
 \mathrm{CI}_{\alpha}=\left[ U_n(x) \pm q_{\alpha / 2} \frac{\sqrt{\widehat{M}_2}}{\widehat{M}_1} \sqrt{\frac{\omega_n}{n\phi(h)}} U_n(x) \right]\text {, }
\end{equation}
where $q_{\alpha/ 2}$ is the upper $\alpha / 2$ quantile of the standard normal distribution.

\subsection{Weak convergence of $\widehat{\gamma}$ and intermediate quantile $\widehat{\mu}_{\alpha_n}(\varepsilon)$}
We now derive the weak limit of the residual--based Hill estimator of the tail
index $\gamma$ in the functional location--scale model. Throughout this subsection we suppose there exist constants $0<\underline\sigma<\overline\sigma<\infty$ with
$
\underline\sigma\le \sigma(x)\le\overline\sigma
\quad\forall x\in\mathcal E.
$
\begin{thm}\label{thm:Hill-residual}
Suppose Assumption~\ref{cgamma} and Assumptions of Theorems \ref{T1} and \ref{T2} hold.  Let $(k_n)$ be an integer sequence such that
$
k = k_n = \left\lfloor n\left(1-\alpha_n\right)\right\rfloor \to\infty,
\,\text{where}\,\,
\alpha_n \uparrow 1.
$
Assume that the following conditions hold true as $n\to \infty$
\begin{equation}\label{cond1}
n(1-\alpha_n) \to \infty,\, \text{and}
\,
\sqrt{n(1-\alpha_n)}\,A((1-\alpha_n)^{-1})\to\lambda\in\R.
\end{equation}

\begin{equation}\label{condh}
n(1-\alpha_n)\log n/(n\phi(h)) \to 0\quad \text{and}\quad \sqrt{n(1-\alpha_n)}\,h^{min(b_1, b_2)} \to 0.
\end{equation}
Then, as $n\to\infty$,
\begin{equation}\label{eq:Hill-residual-CLT}
\sqrt{n(1-\alpha_n)}\,\left(\widehat\gamma-\gamma\right)
\overset{d}
{\longrightarrow} \mathcal N\left(\frac{\lambda}{1-\rho},\,\gamma^2\right).
\end{equation}

and 
\begin{equation}\label{CLTinter}
\sqrt{n(1-\alpha_n)}\,\left(\dfrac{\widehat{\mu}_{\alpha_n}(\varepsilon)}{\mu_{\alpha_n}(\varepsilon)} - 1\right) \overset{d}
{\longrightarrow} \mathcal N(0, \gamma^2).
\end{equation}

\end{thm}

\begin{cor}\label{corES}
    Under the assumptions and conditions of Theorem \ref{thm:Hill-residual}, one has, as $n\to\infty,$
    \begin{equation}
        \dfrac{\widehat{\text{ES}}_{\alpha_n}(\varepsilon)}{\text{ES}_{\alpha_n}(\varepsilon)} - 1 = \mathcal{O}_{\mathbb{P}}\left(\dfrac{1}{\sqrt{n(1-\alpha_n)}}\right).
    \end{equation}
\end{cor}

\subsection{Weak convergence of extreme conditional VaR and ES }
\begin{thm}\label{thm:extremeIQeps}
Assume that the assumptions of Theorem \ref{thm:Hill-residual} hold and that conditions \eqref{cond1} and \eqref{condh} are satisfied. Let $(\tau_n)$ be a sequence such that $\tau_n \uparrow 1$ such that, as $n\to\infty$,
\begin{equation}\label{condalpha}
 (1-\alpha_n)/(1-\tau_n) \to \infty \quad \text{and}\quad \sqrt{n(1-\alpha_n)}/\log((1-\alpha_n)/(1-\tau_n)) \to \infty.
\end{equation}
Then, as $n\to\infty$, 
\begin{equation}\label{CLTexteps}
\dfrac{\sqrt{n(1-\alpha_n)}}{\log\left((1-\alpha_n)/(1-\tau_n) \right)}\,\left(\dfrac{\widehat{\mu}_{\tau_n}(\varepsilon)}{\mu_{\tau_n}(\varepsilon)} - 1\right) \overset{d}
{\longrightarrow} \mathcal N\left(\dfrac{\lambda}{1-\rho}, \gamma^2\right).
\end{equation}
and
\begin{equation}\label{CLTESeps}
\dfrac{\sqrt{n(1-\alpha_n)}}{\log\left((1-\alpha_n)/(1-\tau_n) \right)}\,\left(\dfrac{\widehat{\text{ES}}_{\tau_n}(\varepsilon)}{\text{ES}_{\tau_n}(\varepsilon)} - 1\right) \overset{d}
{\longrightarrow} \mathcal N\left(\dfrac{\lambda}{1-\rho}, \gamma^2\right).
\end{equation}
\end{thm}

\begin{rem}
    The first condition in \eqref{condalpha} states that $\tau_n$ is a
properly extreme level relative to the intermediate level $\alpha_n$, so
that genuine extrapolation is required; the second limits the speed at
which the extrapolation range may grow relative to the
$\sqrt{n(1-\alpha_n)}$ rate at which $\widehat\gamma$ is estimated. Both
convergences in \eqref{CLTexteps}--\eqref{CLTESeps} are governed by the
limiting distribution of the tail-index estimator, which is why the rate
is slower than in Theorem~\ref{thm:Hill-residual} by the logarithmic
factor.
\end{rem}

\begin{thm}\label{thm:extremeQ}
Under assumptions and conditions of Theorem \ref{thm:extremeIQeps}, we have, as $n\to\infty$
\begin{equation}\label{CLTextcondQ}
\dfrac{\sqrt{n(1-\alpha_n)}}{\log\left((1-\alpha_n)/(1-\tau_n) \right)}\,\left(\dfrac{\widehat{\mu}_{\tau_n}(x)}{\mu_{\tau_n}(x)} - 1\right) \overset{d}
{\longrightarrow} \mathcal N\left(\dfrac{\lambda}{1-\rho}, \gamma^2\right).
\end{equation}
and
\begin{equation}\label{CLTcondES}
\dfrac{\sqrt{n(1-\alpha_n)}}{\log\left((1-\alpha_n)/(1-\tau_n) \right)}\,\left(\dfrac{\widehat{\text{CES}}_{\tau_n}(x)}{\text{CES}_{\tau_n}(x)} - 1\right) \overset{d}
{\longrightarrow} \mathcal N\left(\dfrac{\lambda}{1-\rho}, \gamma^2\right).
\end{equation}
\end{thm}

\begin{rem}
The limiting distributions in
\eqref{CLTextcondQ}--\eqref{CLTcondES} coincide with those obtained in
Theorem~\ref{thm:extremeIQeps} for the innovations alone, so that the
estimation of the functional operators $m(\cdot)$ and $\sigma(\cdot)$
carries no asymptotic cost. The mechanism is condition \eqref{condh},
which requires the uniform rates of Theorems~\ref{T1} and~\ref{T2},
namely $h^{\min(b_1,b_2)}$ and $\sqrt{\log n/(n\phi(h))}$, to vanish
faster than $\{n(1-\alpha_n)\}^{-1/2}$, the rate at which the
intermediate quantile and the tail index are estimated. The functional regression components therefore converge at a faster rate than the extreme-value component, and does not contribute to the limiting distribution. The uncertainty surrounding extreme conditional risk
measures is consequently governed by the tail-index estimator rather
than by the functional regression, and the finite-sample results of
Section~\ref{sec:simulation} display the same ordering: the gap between
the feasible and the oracle estimator narrows as the target level
becomes more extreme and as the tail index increases, so that in the
configurations closest to the asymptotic regime the two are of
comparable accuracy, whereas at moderate levels and lighter tails the
estimation of the operators still contributes materially to the total
error.
\end{rem}

\section{Selection of Tuning Parameters}\label{sec:tuning}

\subsection{Selection of the mean smoothing parameter}\label{choiceh_m}
The smoothing parameter controlling the conditional mean fit, denoted
$h_m>0$, is selected by $V$-fold cross-validation over a prespecified
candidate set $\Lambda_m$. Let $\{1,\ldots,n\}$ be partitioned into $V$
disjoint folds $\mathcal I_1,\ldots,\mathcal I_V$, and for any
$h_m\in\Lambda_m$ let $\widehat m^{(-v)}_{h_m}$ denote the local
estimator constructed from the training sample
$\{1,\ldots,n\}\setminus\mathcal I_v$ under the chosen semi-metric and
kernel. The cross-validation criterion is
\[
\mathrm{CV}_m(h_m)
=
\frac{1}{n}\sum_{v=1}^V \sum_{t\in\mathcal I_v}
\bigl\{Y_t-\widehat m^{(-v)}_{h_m}(X_t)\bigr\}^2 ,
\qquad
\widehat h_m \in \arg\min_{h_m\in\Lambda_m} \mathrm{CV}_m(h_m).
\]
We use $V=5$ throughout. This choice targets out-of-sample predictive
accuracy for the mean component and adapts to the scale and concentration
properties induced by the semi-metric: for a more concentrated
semi-metric, such as an index-based distance, the minimising
$\widehat h_m$ typically corresponds to a smaller neighbourhood in the
original function space, whereas for less concentrated distances, such as
$\mathbb L^2$ on a dense grid, a larger bandwidth may be required to
secure an adequate effective sample size
\citep{Ferraty2006,Cuevas2014}.

\subsection{Selection of the volatility smoothing parameter and numerical
stability}\label{choiceh_v}

The conditional variance is estimated by locally regressing the
\emph{log-variance} function
\[
\ell(x)=\log U(x), \qquad U(x)=\sigma^2(x),
\]
where $\sigma(x)$ denotes the conditional volatility. Working on the
logarithmic scale is convenient for two reasons. First, back-transformation
yields a strictly positive variance estimate, which avoids the negative
fitted values that can arise from direct local smoothing of variance
proxies; this is a genuine concern for the local linear estimator, whose
weights are not sign-constrained. Second, the log transformation improves
numerical stability in regions where squared residuals are close to zero.
We therefore use the stabilised proxy
\[
W_t=\log\bigl((Y_t-\widehat m(X_t))^2+c_{\log}\bigr),\qquad c_{\log}>0 .
\]
The constant $c_{\log}$ serves only as numerical regularisation: it
prevents undefined logarithms and attenuates the influence of extremely
small residuals. We fix $c_{\log}=10^{-6}$, negligible relative to the
typical magnitude of $(Y_t-\widehat m(X_t))^2$, and the estimates are
empirically insensitive to its exact value provided it is small.

Because $\mathbb E[\log(\cdot)]\neq\log\mathbb E[\cdot]$, the
back-transformed estimator $\exp\{\widehat \ell(x)\}$ is not unbiased for
$U(x)$. We therefore rescale the resulting standardized residuals to have
unit sample variance, which enforces the normalisation
$\mathbb E(\varepsilon_t^2)=1$ of Section~\ref{ESTCOMP} and removes this
distortion; the rescaling factor is absorbed into $\widehat\sigma(\cdot)$
and leaves the conditional risk measures unaffected up to that
normalisation.

The bandwidth governing the conditional variance, denoted $h_s$, is
selected \emph{separately} from $h_m$, since the smoothness of $U(\cdot)$
and its effective signal-to-noise ratio generally differ from those of
the conditional mean. We select $h_s$ by $V$-fold cross-validation with
the same partition, computing the regression targets from fold-specific
mean fits so as to avoid in-sample reuse of residuals. For each fold $v$,
let $\widehat m^{(-v)}_{\widehat h_m}$ be the mean estimator trained on
$\{1,\ldots,n\}\setminus\mathcal I_v$ with the previously selected
$\widehat h_m$, and set
\[
W_t^{(-v)}=\log\bigl((Y_t-\widehat m^{(-v)}_{\widehat h_m}(X_t))^2
+c_{\log}\bigr),\qquad t=1,\dots,n .
\]
For any candidate $h_s\in\Lambda_s$, let $\widehat \ell^{(-v)}_{h_s}$ be
the local estimator of $\ell(\cdot)$ fitted on
$\{1,\ldots,n\}\setminus\mathcal I_v$ using the responses
$\{W_j^{(-v)}:j\notin\mathcal I_v\}$. The criterion is
\[
\mathrm{CV}_U(h_s)
=
\frac{1}{n}\sum_{v=1}^V \sum_{t\in\mathcal I_v}
\bigl(W_t^{(-v)}-\widehat \ell^{(-v)}_{h_s}(X_t)\bigr)^2 ,
\qquad
\widehat h_s \in \arg\min_{h_s\in\Lambda_s} \mathrm{CV}_U(h_s).
\]
This isolates the smoothing level appropriate for the variance component
and avoids implicitly constraining $\widehat h_s$ by $\widehat h_m$, so
that the procedure accommodates settings in which $U(\cdot)$ varies more
smoothly than $m(\cdot)$, favouring a larger $\widehat h_s$, as well as
settings in which volatility is locally more irregular, favouring a
smaller one, all under the same semi-metric neighbourhood structure
\citep{Ferraty2006,Cuevas2014}.


\subsection{Choice of the semi-metric.}
A central tuning choice in functional regression is the semi-metric $d(\cdot,\cdot)$ used to define local neighborhoods, since it induces the topology under which ``closeness'' of trajectories is assessed and therefore determines which observations contribute non-negligible weight to a local estimator. In the general framework of nonparametric functional regression, the semi-metric is not required to arise from an inner product, and different choices can yield substantially different effective sample sizes and bias--variance behavior even when the smoothing rule is otherwise unchanged; see, e.g., the monograph-level discussions in \citet{Ferraty2006} and the review perspective in \citet{Cuevas2014}. Practically, the semi-metric should be aligned with the \emph{dominant sources of variability} (amplitude, phase, derivatives, shape, or distributional features) and with the \emph{regularity class} in which the functional covariate is believed to live (e.g., $\mathbb L^2$, Sobolev spaces, or quotient spaces modulo time-warping).

\subsection{Selection of the tail threshold and extrapolation
level}\label{choicek}

Let $\widehat{\varepsilon}_t$ denote the standardized residuals obtained
from the estimated location--scale representation. Since the objective is
to model extreme negative returns, tail estimation is applied to the loss
side of the innovation distribution. Define
\[
\xi_t = (-\widehat{\varepsilon}_t)^{+} = \max\{-\widehat{\varepsilon}_t,0\},
\qquad t=1,\dots,n,
\]
and write $\xi_{(1)} \leq \cdots \leq \xi_{(n_{\mathrm{tail}})}$ for the
order statistics of the strictly positive values among them, where
$n_{\mathrm{tail}}$ is the number of such values. Truncation at zero
removes the observations that carry no information about the loss tail.
All intermediate levels below are defined relative to
$n_{\mathrm{tail}}$ rather than to $n$; in the notation of
Section~\ref{sec:mainresults} the relevant quantity is the number $k$ of
upper order statistics retained, so that the rate $\sqrt{n(1-\alpha_n)}$
is read as $\sqrt{k}$.

Under the regular variation framework of Assumption~\ref{cgamma}, the
tail index $\gamma$ is estimated by the Hill estimator: for an
intermediate sequence $k=k_{n_{\mathrm{tail}}}$,
\begin{equation}\label{eq:hill-tuning}
\widehat{\gamma}(k)
=
\frac{1}{k}\sum_{j=1}^{k}
\Bigl(
\log \xi_{(n_{\mathrm{tail}}-j+1)}
-
\log \xi_{(n_{\mathrm{tail}}-k)}
\Bigr),
\end{equation}
where $k\to\infty$ and $k/n_{\mathrm{tail}}\to0$ as
$n_{\mathrm{tail}}\to\infty$, so that the estimator uses sufficiently many
exceedances while remaining within the asymptotic tail region. Following
common practice we fix the intermediate sequence at a conventional
fraction of the tail sample,
$k^\star=\lfloor 0.07\,n_{\mathrm{tail}}\rfloor$; the sensitivity of the
results to this choice is documented in Section~\ref{sec:insample}.

The associated intermediate probability level is
$\alpha_n=1-k^\star/n_{\mathrm{tail}}$ and the corresponding
empirical threshold is $\widehat{\mu}_{\alpha_n}(\xi)=
\xi_{(n_{\mathrm{tail}}-k^\star+1)}$. Extreme quantiles beyond the range
of the data are obtained by Weissman extrapolation: for
$\tau_n >\alpha_n$,
\begin{equation}\label{eq:weissman-tuning}
\widehat{\mu}_{\tau_n}(\xi)
=
\widehat{\mu}_{\alpha_n}(\xi)
\left(
\frac{1-\alpha_n}{1-\tau_n}
\right)^{\widehat{\gamma}} ,
\qquad
\widehat{\mathrm{ES}}_{\tau_n}(\xi)
=
\frac{\widehat{\mu}_{\tau_n}(\xi)}{1-\widehat{\gamma}} ,
\end{equation}
the second expression following from the asymptotic relation
\eqref{eq:VARESrelation} and requiring $\widehat\gamma<1$. The target level $\tau_n$ cannot be chosen arbitrarily far beyond
$\alpha_n$: condition \eqref{condalpha} of
Theorem~\ref{thm:extremeIQeps} requires
$\sqrt{k}/\log\{(1-\alpha_n)/(1-\tau_n)\}\to\infty$, so that a more
extreme target level demands a larger number of exceedances to preserve
the asymptotic approximation. Section~\ref{sec:insample} verifies that
the levels used in the empirical analysis satisfy this requirement.


\section{Simulation study}\label{sec:simulation}

This section examines the finite-sample performance of the proposed estimators
of the extreme conditional value-at-risk
(VaR), and the extreme conditional expected shortfall (ES) in a linear and nonlinear
functional location--scale framework. 


\subsection*{Functional covariate generation}

Let $[0,1]$ denote the functional domain and consider the orthonormal system
$\{\phi_j\}_{j=1}^4 \subset L^2[0,1]$ defined by
\[
\phi_1(u)=1,\qquad 
\phi_2(u)=\sqrt{2}\sin(\pi u),\qquad
\phi_3(u)=\sqrt{2}\cos(2\pi u),\qquad
\phi_4(u)=\sqrt{2}\sin(2\pi u),
\qquad u\in[0,1].
\]
For each $t=1,\ldots,n$, we generate latent scores
\[
a_{t1},a_{t2}\sim\mathrm{Unif}(0,1),
\qquad
a_{t3},a_{t4}\sim\mathrm{Beta}(2,2),
\]
independently across $t$ and components. The functional covariate is then
constructed as
\begin{equation}\label{eq:X-sim}
X_t(u)
=
\sum_{j=1}^4 a_{tj}\,\phi_j(u), 
\qquad u\in[0,1].
\end{equation}
This specification produces smooth random trajectories whose variability and
shape are governed by a low-dimensional latent structure, a setting that is
standard in functional regression simulations and allows for interpretable
signal-to-noise control.

\subsection*{Regression and volatility operators}

We define the coefficient vectors
\[
\beta=(1,1,2,2)^\top,
\qquad
\theta=(1,1,1,1)^\top,
\]
and the associated linear functionals
\[
\langle \beta, X_t\rangle_{L^2}
=
\sum_{j=1}^4 \beta_j a_{tj},
\qquad
\langle \theta, X_t\rangle_{L^2}
=
\sum_{j=1}^4 \theta_j a_{tj}.
\]

We consider the following two data-generating processes.

\medskip
\noindent\textit{Model~1 (Functional linear location--scale model).}
\begin{equation}\label{eq:FM1}
Y_t
=
\langle\beta,X_t\rangle_{L^2}
+
\Big(\langle\theta,X_t\rangle_{L^2}+1\Big)\varepsilon_t.
\end{equation}

\medskip
\noindent\textit{Model~2 (Functional nonlinear heteroscedastic model).}
\begin{equation}\label{eq:FM2}
Y_t
=
\exp\!\big(\langle\beta,X_t\rangle_{L^2}-2\big)+1
+
\Big[\exp\!\big(\langle\theta,X_t\rangle_{L^2}-1\big)+\tfrac12\Big]\varepsilon_t.
\end{equation}

Model~1 corresponds to a functional linear regression with multiplicative
heteroscedasticity, while Model~2 introduces nonlinear dependence in both the
conditional mean and scale. Together, these specifications allow us to assess
the robustness of the proposed estimators across varying degrees of functional
nonlinearity.

\subsection*{Heavy-tailed innovations and extreme risk benchmarks}

The innovations are constructed as $\varepsilon_t = S_t \delta_t/s_\gamma$,
where $S_t$ takes the values $\pm1$ with equal probability, 
$\delta_t=(1-\mathcal{U}_t)^{-\gamma}$ with $\mathcal{U}_t\sim \text{Uniform}(0,1)$ independent of
$S_t$, and $s_\gamma=(1-2\gamma)^{-1/2}$ ensures
$\mathbb E(\varepsilon_t)=0$ and $\mathbb E(\varepsilon_t^2)=1$ as
required by Assumption~\ref{cgamma}. Since $\delta_t$ is exactly standard
Pareto with tail index $1/\gamma$, the survival function of the
innovations is
$\overline F_\varepsilon(x)=\tfrac12 (s_\gamma x)^{-1/\gamma}$ for
$x\ge s_\gamma^{-1}$, and the true risk measures are available in closed
form,
\begin{equation}\label{eq:true-quantile}
\mu_{\tau_n}(\varepsilon)=\frac{\bigl(2(1-\tau_n)\bigr)^{-\gamma}}{s_\gamma},
\qquad
\mathrm{ES}_{\tau_n}(\varepsilon)
=\frac{\bigl(2(1-\tau_n)\bigr)^{-\gamma}}{(1-\gamma)\,s_\gamma}.
\end{equation}
The factor $2$ reflects the sign symmetrisation, which assigns half the
probability mass to the negative half-line. Two features of this design
are deliberate. The relation \eqref{eq:VARESrelation} holds here as an
exact identity rather than only in the limit, so the Expected-Shortfall
benchmark is not contaminated by second-order effects; and the
second-order function of Assumption~\ref{cgamma} vanishes identically,
so that $\lambda=0$ in condition \eqref{cond1} and the confidence
intervals implied by Theorem~\ref{thm:extremeQ} require no bias
correction. The design therefore isolates the error due to estimating
the functional operators from the second-order bias of the tail
extrapolation. The expressions in \eqref{eq:true-quantile} were verified
against empirical quantiles and tail means computed from $4\times10^{7}$
independent draws.

\subsection*{Conditional extreme risk measures at a target trajectory}

Estimation accuracy is evaluated at a fixed target trajectory
\[
x_0(u)
=
\sum_{j=1}^4 a_{0j}\phi_j(u),
\qquad
(a_{01},a_{02},a_{03},a_{04})
=
(0.5,0.3,0.5,0.4).
\]
For either model, the true extreme conditional VaR and ES at level $\tau_n$ are
given by
\[
\mathrm{VaR}_{\tau_n}(x_0)
=
m(x_0)
+
\sigma(x_0)\,\mu_{\tau_n}(\varepsilon),
\qquad
\mathrm{ES}_{\tau_n}(x_0)
=
m(x_0)
+
\sigma(x_0)\,\mathrm{ES}_{\tau_n}(\varepsilon),
\]
with $\mu_{\tau_n}(\varepsilon)$ and $\mathrm{ES}_{\tau_n}(\varepsilon)$ defined
in \eqref{eq:true-quantile}. These analytical expressions
serve as exact benchmarks for computing empirical bias and root mean squared
error of both the residual-based estimators and the oracle estimators that use
the true innovations and operators.

\subsection*{Performance measures}

To assess the finite-sample behaviour of the proposed estimators of the
conditional Value-at-Risk and Expected Shortfall, we consider four
criteria based on the relative estimation error. Let $T(x_0)$ denote the
true value of the risk measure at the target trajectory $x_0$, where $T$
stands for either $\mu_{\tau_n}$ or $\text{CES}_{\tau_n}$, and let
$\widehat{T}^{(b)}(x_0)$ be its estimate in the $b$-th Monte Carlo
replication, $b=1,\dots,B$. Writing
\[
e_b=\frac{\widehat{T}^{(b)}(x_0)-T(x_0)}{T(x_0)}
\]
for the relative error in replication $b$, we report the relative root
mean squared error and the relative median absolute deviation,
\[
\text{RMSE}=\Bigl(\frac1B\sum_{b=1}^{B}e_b^2\Bigr)^{1/2},
\qquad
\text{RMAD}=\operatorname*{median}_{1\le b\le B}|e_b| .
\]
Both criteria are reported because the sampling distribution of
$\widehat T$ inherits the heavy tail of the innovations, so that a small
number of replications can dominate a mean-based criterion; the
comparison of the two therefore indicates when this is occurring.

Because RMSE and RMAD are unsigned, they cannot reveal whether a
procedure systematically over- or under-states risk, a distinction that
matters here since the understatement of extreme losses is the costly
direction in risk management. We therefore also report the relative bias
\[
\text{RelBias}=\frac1B\sum_{b=1}^{B}e_b ,
\]
a positive value indicating that the risk measure is systematically
overstated. Finally, we report the empirical coverage of the asymptotic
confidence intervals implied by Theorem~\ref{thm:extremeQ}. Since the theorem delivers a limiting
distribution for the relative error, the interval is multiplicative:
\begin{equation}\label{eq:ci-sim}
\text{CR}
=\frac1B\sum_{b=1}^{B}
\1\Bigl\{
\widehat T^{(b)}(x_0)\,e^{-z_{1-\eta/2}\widehat\gamma^{(b)}/v_n}
\le T(x_0)\le
\widehat T^{(b)}(x_0)\,e^{+z_{1-\eta/2}\widehat\gamma^{(b)}/v_n}
\Bigr\},
v_n=\frac{\sqrt{n(1-\alpha_n)}}{\log\bigl((1-\alpha_n)/(1-\tau_n)\bigr)},
\end{equation}
at the nominal level $1-\eta=0.95$. The interval is centred at the
asymptotic mean zero, which is exact in the present design: because the
innovations are exactly Pareto in the upper tail, the second-order
auxiliary function vanishes identically and $\lambda=0$ in condition
\eqref{cond1}. Any departure of \eqref{eq:ci-sim} from the nominal level
is therefore a genuine finite-sample phenomenon and not an artefact of an
uncorrected second-order bias.

Each criterion is itself estimated from $B$ replications and is
therefore subject to simulation error. We accordingly report Monte Carlo
standard errors in Table~\ref{tab:mcse}, computed as
$\{B^{-1}\sum_b(e_b-\bar e)^2\}^{1/2}/\sqrt B$ for the relative bias and
by a nonparametric bootstrap over replications, with $500$ resamples, for
the RMSE and the RMAD. Throughout this simulation $B=500$ for $n\in\{200,500, 1000\}.$

\subsection*{Tuning parameters selection}
The estimators involve several tuning choices. The functional neighbourhood
is defined by the single-index semi-metric
$d_1(X_i,X_j)=|\langle\beta,\Xi_i-\Xi_j\rangle|$ with
$\rho(X_i,X_j)=\langle\beta,\Xi_i-\Xi_j\rangle$ the associated bilinear
form, evaluated at the true index direction $\beta=(1,1,2,2)$ of the
data-generating process. Treating $\beta$ as known isolates the
estimation of the location and scale operators and of the tail from the
estimation of the index itself; in the empirical application of
Section~\ref{sec:application} the direction is unknown and is estimated
from the functional principal components. 


The kernel is the Epanechnikov function
$K(u)=\tfrac34(1-u^2)\1\{|u|\le1\}$ throughout. Bandwidths are selected
separately for the location and the scale operators as described in Sections \ref{choiceh_m} and \ref{choiceh_v}.  The
scale operator is estimated on the logarithmic scale, by smoothing
$\log(\widehat r_i^{\,2}+c_{\log})$ with $c_{\log}=10^{-8}$ and back-transforming,
which guarantees a positive fitted variance; the standardized residuals
are then rescaled to unit variance, which enforces the normalisation
$\mathbb E(\varepsilon^2)=1$ of Assumption~\ref{cgamma} and removes the
Jensen bias introduced by the back-transformation. Finally, the intermediate sequence is fixed at a tenth of the positive
standardized residuals, giving an intermediate level
$\alpha_n\approx0.95$. Results are reported at the three target levels
$\tau_n\in\{0.99,0.995,0.999\}$, for which the extrapolation ratio
$(1-\alpha_n)/(1-\tau_n)$ equals approximately $5$, $10$ and $50$, so
that the design spans a range from moderate to substantial
extrapolation beyond the observable region.




\subsection*{Simulation results}

Tables~\ref{tab:model1} and~\ref{tab:model2} report the accuracy criteria
for Models~1 and~2, Table~\ref{tab:bias-coverage} the relative bias and
the coverage, Table~\ref{tab:gammahat} the mean of the estimated tail
index, and Table~\ref{tab:mcse} the Monte Carlo standard errors. Five
features deserve comment.

\begin{table}[!ht]
\centering
\footnotesize
\setlength{\tabcolsep}{3pt}
\renewcommand{\arraystretch}{1.1}
\caption{Monte Carlo performance of the proposed estimator in terms of relative RMSE and relative median absolute deviation (RMAD) for the conditional Value-at-Risk $\mu_{\tau_n}(x)$ and the conditional Expected Shortfall $\text{CES}_{\tau_n}(x)$ (in parentheses), compared with the oracle benchmark. Data generated according to Model 1.}
\label{tab:model1}
\begin{tabular}{|l|ll|ccc|ccc|ccc|}
\hline
\multicolumn{3}{|c}{} & \multicolumn{3}{|c|}{$\tau_n=0.99$}& \multicolumn{3}{c|}{$\tau_n=0.995$}& \multicolumn{3}{c|}{$\tau_n=0.999$} \\
\hline
\multicolumn{3}{|c|}{} & $\gamma=0.1$ & $\gamma=0.25$ & $\gamma=0.35$& $\gamma=0.1$ & $\gamma=0.25$ & $\gamma=0.35$& $\gamma=0.1$ & $\gamma=0.25$ & $\gamma=0.35$ \\
\hline
\multirow{4}{*}{$n=200$}
& \multirow{2}{*}{RMSE} & Full    & 0.135 & 0.132 & 0.179 & 0.178 & 0.175 & 0.240 & 0.339 & 0.313 & 0.435\\
&  &                    & (0.231) & (0.232) & (0.355) & (0.308) & (0.297) & (0.456) & (0.596) & (0.496) & (0.772)\\
&                       & Oracle  & 0.032 & 0.099 & 0.144 & 0.042 & 0.137 & 0.199 & 0.068 & 0.246 & 0.359\\
&  &                    & (0.048) & (0.179) & (0.285) & (0.059) & (0.229) & (0.363) & (0.088) & (0.370) & (0.586)\\
\cline{2-12}
& \multirow{2}{*}{RMAD} & Full    & 0.064 & 0.074 & 0.099 & 0.075 & 0.095 & 0.131 & 0.100 & 0.149 & 0.215\\
&  &                    & (0.080) & (0.115) & (0.168) & (0.089) & (0.136) & (0.207) & (0.121) & (0.198) & (0.287)\\
&                       & Oracle  & 0.023 & 0.067 & 0.093 & 0.029 & 0.088 & 0.125 & 0.047 & 0.139 & 0.203\\
&  &                    & (0.033) & (0.106) & (0.160) & (0.040) & (0.128) & (0.194) & (0.058) & (0.191) & (0.278)\\
\hline
\multirow{4}{*}{$n=500$}
& \multirow{2}{*}{RMSE} & Full    & 0.072 & 0.077 & 0.100 & 0.087 & 0.096 & 0.131 & 0.130 & 0.153 & 0.225\\
&  &                    & (0.098) & (0.117) & (0.179) & (0.116) & (0.143) & (0.223) & (0.170) & (0.211) & (0.344)\\
&                       & Oracle  & 0.019 & 0.059 & 0.087 & 0.025 & 0.079 & 0.120 & 0.040 & 0.134 & 0.212\\
&  &                    & (0.029) & (0.099) & (0.166) & (0.035) & (0.124) & (0.208) & (0.052) & (0.188) & (0.319)\\
\cline{2-12}
& \multirow{2}{*}{RMAD} & Full    & 0.040 & 0.049 & 0.066 & 0.043 & 0.060 & 0.084 & 0.055 & 0.097 & 0.139\\
&  &                    & (0.046) & (0.072) & (0.109) & (0.050) & (0.088) & (0.134) & (0.065) & (0.132) & (0.192)\\
&                       & Oracle  & 0.013 & 0.039 & 0.059 & 0.017 & 0.051 & 0.081 & 0.027 & 0.084 & 0.135\\
&  &                    & (0.019) & (0.063) & (0.107) & (0.023) & (0.077) & (0.131) & (0.035) & (0.113) & (0.187)\\
\hline
\multirow{4}{*}{$n=1000$}
& \multirow{2}{*}{RMSE} & Full    & 0.047 & 0.054 & 0.066 & 0.052 & 0.068 & 0.087 & 0.068 & 0.107 & 0.144\\
&  &                    & (0.056) & (0.081) & (0.114) & (0.062) & (0.099) & (0.140) & (0.081) & (0.143) & (0.206)\\
&                       & Oracle  & 0.014 & 0.043 & 0.062 & 0.018 & 0.057 & 0.085 & 0.028 & 0.094 & 0.146\\
&  &                    & (0.020) & (0.070) & (0.114) & (0.024) & (0.087) & (0.141) & (0.036) & (0.128) & (0.212)\\
\cline{2-12}
& \multirow{2}{*}{RMAD} & Full    & 0.034 & 0.040 & 0.045 & 0.033 & 0.048 & 0.058 & 0.038 & 0.074 & 0.099\\
&  &                    & (0.035) & (0.055) & (0.076) & (0.037) & (0.069) & (0.095) & (0.043) & (0.098) & (0.139)\\
&                       & Oracle  & 0.009 & 0.029 & 0.041 & 0.012 & 0.040 & 0.054 & 0.019 & 0.063 & 0.091\\
&  &                    & (0.014) & (0.046) & (0.071) & (0.017) & (0.058) & (0.087) & (0.024) & (0.087) & (0.124)\\
\hline
\end{tabular}
\end{table}

\begin{table}[!ht]
\centering
\footnotesize
\setlength{\tabcolsep}{3pt}
\renewcommand{\arraystretch}{1.1}
\caption{Monte Carlo performance of the proposed estimator in terms of relative RMSE and relative median absolute deviation (RMAD) for the conditional Value-at-Risk $\mu_{\tau_n}(x)$ and the conditional Expected Shortfall $\text{CES}_{\tau_n}(x)$ (in parentheses), compared with the oracle benchmark. Data generated according to Model 2.}
\label{tab:model2}
\begin{tabular}{|l|ll|ccc|ccc|ccc|}
\hline
\multicolumn{3}{|c}{} & \multicolumn{3}{|c|}{$\tau_n=0.99$}& \multicolumn{3}{c|}{$\tau_n=0.995$}& \multicolumn{3}{c|}{$\tau_n=0.999$} \\
\hline
\multicolumn{3}{|c|}{} & $\gamma=0.1$ & $\gamma=0.25$ & $\gamma=0.35$& $\gamma=0.1$ & $\gamma=0.25$ & $\gamma=0.35$& $\gamma=0.1$ & $\gamma=0.25$ & $\gamma=0.35$ \\
\hline
\multirow{4}{*}{$n=200$}
& \multirow{2}{*}{RMSE} & Full    & 0.173 & 0.179 & 0.172 & 0.215 & 0.247 & 0.231 & 0.356 & 0.538 & 0.423\\
&  &                    & (0.256) & (0.441) & (0.339) & (0.319) & (0.637) & (0.438) & (0.532) & (1.494) & (0.742)\\
&                       & Oracle  & 0.031 & 0.085 & 0.131 & 0.041 & 0.117 & 0.186 & 0.066 & 0.212 & 0.350\\
&  &                    & (0.047) & (0.153) & (0.275) & (0.058) & (0.196) & (0.356) & (0.086) & (0.317) & (0.595)\\
\cline{2-12}
& \multirow{2}{*}{RMAD} & Full    & 0.122 & 0.080 & 0.090 & 0.141 & 0.099 & 0.120 & 0.192 & 0.155 & 0.214\\
&  &                    & (0.157) & (0.116) & (0.161) & (0.178) & (0.146) & (0.205) & (0.235) & (0.207) & (0.298)\\
&                       & Oracle  & 0.020 & 0.056 & 0.083 & 0.026 & 0.074 & 0.113 & 0.043 & 0.118 & 0.187\\
&  &                    & (0.031) & (0.090) & (0.148) & (0.038) & (0.110) & (0.179) & (0.055) & (0.160) & (0.271)\\
\hline
\multirow{4}{*}{$n=500$}
& \multirow{2}{*}{RMSE} & Full    & 0.133 & 0.086 & 0.116 & 0.155 & 0.101 & 0.149 & 0.219 & 0.148 & 0.248\\
&  &                    & (0.171) & (0.117) & (0.198) & (0.199) & (0.138) & (0.244) & (0.277) & (0.198) & (0.372)\\
&                       & Oracle  & 0.018 & 0.051 & 0.079 & 0.023 & 0.069 & 0.109 & 0.036 & 0.116 & 0.193\\
&  &                    & (0.026) & (0.085) & (0.150) & (0.032) & (0.106) & (0.189) & (0.047) & (0.161) & (0.294)\\
\cline{2-12}
& \multirow{2}{*}{RMAD} & Full    & 0.106 & 0.055 & 0.068 & 0.117 & 0.066 & 0.088 & 0.150 & 0.092 & 0.134\\
&  &                    & (0.125) & (0.078) & (0.109) & (0.138) & (0.088) & (0.129) & (0.179) & (0.114) & (0.179)\\
&                       & Oracle  & 0.012 & 0.033 & 0.049 & 0.015 & 0.044 & 0.065 & 0.024 & 0.074 & 0.110\\
&  &                    & (0.017) & (0.054) & (0.083) & (0.021) & (0.068) & (0.105) & (0.031) & (0.099) & (0.157)\\
\hline
\multirow{4}{*}{$n=1000$}
& \multirow{2}{*}{RMSE} & Full    & 0.103 & 0.076 & 0.088 & 0.115 & 0.087 & 0.108 & 0.149 & 0.124 & 0.170\\
&  &                    & (0.124) & (0.100) & (0.138) & (0.138) & (0.116) & (0.166) & (0.175) & (0.162) & (0.243)\\
&                       & Oracle  & 0.013 & 0.040 & 0.056 & 0.017 & 0.054 & 0.078 & 0.027 & 0.092 & 0.138\\
&  &                    & (0.020) & (0.067) & (0.107) & (0.024) & (0.084) & (0.134) & (0.035) & (0.127) & (0.204)\\
\cline{2-12}
& \multirow{2}{*}{RMAD} & Full    & 0.094 & 0.054 & 0.055 & 0.106 & 0.057 & 0.059 & 0.132 & 0.070 & 0.096\\
&  &                    & (0.113) & (0.057) & (0.072) & (0.123) & (0.067) & (0.090) & (0.152) & (0.086) & (0.137)\\
&                       & Oracle  & 0.008 & 0.028 & 0.037 & 0.010 & 0.037 & 0.048 & 0.017 & 0.062 & 0.083\\
&  &                    & (0.012) & (0.046) & (0.064) & (0.015) & (0.057) & (0.080) & (0.023) & (0.082) & (0.120)\\
\hline
\end{tabular}
\end{table}

\begin{table}[!ht]
\centering
\footnotesize
\setlength{\tabcolsep}{3pt}
\renewcommand{\arraystretch}{1.1}
\caption{Relative bias and empirical coverage of the nominal $95\%$ asymptotic confidence intervals implied by Theorem~\ref{thm:extremeQ}, for the conditional Value-at-Risk and, in parentheses, the conditional Expected Shortfall.}
\label{tab:bias-coverage}
\begin{tabular}{|l|ll|ccc|ccc|ccc|}
\hline
\multicolumn{3}{|c}{} & \multicolumn{3}{|c|}{$\tau_n=0.99$}& \multicolumn{3}{c|}{$\tau_n=0.995$}& \multicolumn{3}{c|}{$\tau_n=0.999$} \\
\hline
\multicolumn{3}{|c|}{} & $\gamma=0.1$ & $\gamma=0.25$ & $\gamma=0.35$& $\gamma=0.1$ & $\gamma=0.25$ & $\gamma=0.35$& $\gamma=0.1$ & $\gamma=0.25$ & $\gamma=0.35$ \\
\hline
\multicolumn{12}{|l|}{\emph{Model 1}} \\
\hline
\multirow{4}{*}{$n=200$}
& \multirow{2}{*}{RelBias} & Full    & 0.074 & 0.023 & 0.017 & 0.091 & 0.026 & 0.021 & 0.141 & 0.042 & 0.043\\
&  &                    & (0.105) & (0.035) & (0.040) & (0.127) & (0.043) & (0.054) & (0.192) & (0.074) & (0.107)\\
&                       & Oracle  & 0.001 & 0.005 & 0.008 & 0.002 & 0.010 & 0.017 & 0.004 & 0.031 & 0.051\\
&  &                    & (0.003) & (0.020) & (0.039) & (0.004) & (0.029) & (0.057) & (0.007) & (0.061) & (0.115)\\
\cline{2-12}
& \multirow{2}{*}{CR} & Full    & 0.863 & 0.959 & 0.948 & 0.912 & 0.957 & 0.943 & 0.949 & 0.938 & 0.935\\
&  &                    & (0.695) & (0.789) & (0.799) & (0.775) & (0.827) & (0.812) & (0.846) & (0.848) & (0.829)\\
&                       & Oracle  & 0.978 & 0.974 & 0.956 & 0.982 & 0.972 & 0.948 & 0.980 & 0.968 & 0.940\\
&  &                    & (0.932) & (0.902) & (0.840) & (0.940) & (0.922) & (0.856) & (0.946) & (0.922) & (0.878)\\
\hline
\multirow{4}{*}{$n=500$}
& \multirow{2}{*}{RelBias} & Full    & 0.045 & 0.021 & 0.011 & 0.052 & 0.021 & 0.009 & 0.070 & 0.023 & 0.009\\
&  &                    & (0.057) & (0.022) & (0.011) & (0.065) & (0.024) & (0.012) & (0.086) & (0.029) & (0.019)\\
&                       & Oracle  & 0.001 & 0.002 & 0.003 & 0.001 & 0.004 & 0.007 & 0.001 & 0.010 & 0.022\\
&  &                    & (0.001) & (0.007) & (0.016) & (0.001) & (0.010) & (0.024) & (0.002) & (0.019) & (0.047)\\
\cline{2-12}
& \multirow{2}{*}{CR} & Full    & 0.766 & 0.968 & 0.975 & 0.852 & 0.975 & 0.970 & 0.909 & 0.966 & 0.950\\
&  &                    & (0.675) & (0.847) & (0.832) & (0.776) & (0.882) & (0.847) & (0.842) & (0.897) & (0.862)\\
&                       & Oracle  & 0.996 & 0.994 & 0.992 & 0.996 & 0.992 & 0.990 & 0.996 & 0.984 & 0.976\\
&  &                    & (0.968) & (0.902) & (0.840) & (0.978) & (0.916) & (0.868) & (0.984) & (0.932) & (0.882)\\
\hline
\multirow{4}{*}{$n=1000$}
& \multirow{2}{*}{RelBias} & Full    & 0.034 & 0.012 & 0.010 & 0.037 & 0.011 & 0.009 & 0.044 & 0.008 & 0.009\\
&  &                    & (0.038) & (0.010) & (0.010) & (0.042) & (0.009) & (0.010) & (0.049) & (0.007) & (0.013)\\
&                       & Oracle  & -0.001 & 0.002 & -0.001 & -0.001 & 0.003 & 0.001 & -0.001 & 0.006 & 0.008\\
&  &                    & (-0.001) & (0.004) & (0.005) & (-0.001) & (0.006) & (0.009) & (-0.001) & (0.011) & (0.021)\\
\cline{2-12}
& \multirow{2}{*}{CR} & Full    & 0.733 & 0.977 & 0.972 & 0.842 & 0.977 & 0.976 & 0.926 & 0.967 & 0.958\\
&  &                    & (0.644) & (0.809) & (0.835) & (0.762) & (0.842) & (0.844) & (0.871) & (0.860) & (0.858)\\
&                       & Oracle  & 0.992 & 0.984 & 0.992 & 0.992 & 0.984 & 0.984 & 0.992 & 0.976 & 0.968\\
&  &                    & (0.972) & (0.868) & (0.820) & (0.972) & (0.896) & (0.848) & (0.976) & (0.908) & (0.888)\\
\hline
\multicolumn{12}{|l|}{\emph{Model 2}} \\
\hline
\multirow{4}{*}{$n=200$}
& \multirow{2}{*}{RelBias} & Full    & 0.135 & 0.077 & 0.052 & 0.163 & 0.093 & 0.060 & 0.243 & 0.148 & 0.096\\
&  &                    & (0.185) & (0.122) & (0.085) & (0.219) & (0.151) & (0.105) & (0.317) & (0.255) & (0.174)\\
&                       & Oracle  & -0.001 & 0.007 & 0.010 & -0.001 & 0.013 & 0.020 & 0.001 & 0.033 & 0.056\\
&  &                    & (0.000) & (0.022) & (0.043) & (0.001) & (0.031) & (0.062) & (0.004) & (0.061) & (0.123)\\
\cline{2-12}
& \multirow{2}{*}{CR} & Full    & 0.678 & 0.958 & 0.974 & 0.801 & 0.958 & 0.974 & 0.902 & 0.965 & 0.954\\
&  &                    & (0.460) & (0.851) & (0.811) & (0.591) & (0.872) & (0.844) & (0.732) & (0.893) & (0.851)\\
&                       & Oracle  & 0.988 & 0.982 & 0.982 & 0.988 & 0.980 & 0.982 & 0.982 & 0.974 & 0.974\\
&  &                    & (0.938) & (0.920) & (0.876) & (0.950) & (0.926) & (0.892) & (0.954) & (0.936) & (0.900)\\
\hline
\multirow{4}{*}{$n=500$}
& \multirow{2}{*}{RelBias} & Full    & 0.114 & 0.049 & 0.052 & 0.132 & 0.049 & 0.060 & 0.177 & 0.051 & 0.083\\
&  &                    & (0.143) & (0.050) & (0.072) & (0.163) & (0.051) & (0.083) & (0.215) & (0.055) & (0.116)\\
&                       & Oracle  & -0.002 & -0.004 & 0.008 & -0.002 & -0.006 & 0.013 & -0.002 & -0.007 & 0.030\\
&  &                    & (-0.002) & (-0.006) & (0.022) & (-0.002) & (-0.006) & (0.031) & (-0.002) & (-0.006) & (0.056)\\
\cline{2-12}
& \multirow{2}{*}{CR} & Full    & 0.277 & 0.970 & 0.970 & 0.416 & 0.980 & 0.973 & 0.605 & 0.980 & 0.957\\
&  &                    & (0.178) & (0.887) & (0.848) & (0.277) & (0.920) & (0.875) & (0.476) & (0.937) & (0.900)\\
&                       & Oracle  & 0.992 & 0.984 & 0.990 & 0.994 & 0.984 & 0.988 & 0.994 & 0.978 & 0.980\\
&  &                    & (0.972) & (0.908) & (0.894) & (0.978) & (0.918) & (0.912) & (0.978) & (0.918) & (0.926)\\
\hline
\multirow{4}{*}{$n=1000$}
& \multirow{2}{*}{RelBias} & Full    & 0.097 & 0.056 & 0.052 & 0.109 & 0.060 & 0.055 & 0.139 & 0.069 & 0.064\\
&  &                    & (0.116) & (0.063) & (0.060) & (0.129) & (0.067) & (0.064) & (0.162) & (0.078) & (0.076)\\
&                       & Oracle  & -0.001 & 0.000 & 0.005 & -0.001 & 0.001 & 0.007 & -0.002 & 0.004 & 0.016\\
&  &                    & (-0.002) & (0.002) & (0.012) & (-0.002) & (0.004) & (0.016) & (-0.003) & (0.009) & (0.029)\\
\cline{2-12}
& \multirow{2}{*}{CR} & Full    & 0.075 & 0.929 & 0.959 & 0.134 & 0.974 & 0.966 & 0.313 & 0.968 & 0.966\\
&  &                    & (0.022) & (0.805) & (0.842) & (0.075) & (0.877) & (0.877) & (0.209) & (0.916) & (0.877)\\
&                       & Oracle  & 1.000 & 1.000 & 1.000 & 1.000 & 0.995 & 1.000 & 0.995 & 0.985 & 0.990\\
&  &                    & (0.975) & (0.925) & (0.885) & (0.985) & (0.930) & (0.890) & (0.990) & (0.940) & (0.920)\\
\hline
\end{tabular}
\end{table}

\begin{table}[!ht]
\centering
\footnotesize
\caption{Mean of the estimated tail index across replications.}
\label{tab:gammahat}
\begin{tabular}{|l|l|ccc|ccc|}
\hline
& & \multicolumn{3}{c|}{Model 1} & \multicolumn{3}{c|}{Model 2}\\
$n$ & Estimator & $\gamma=0.1$ & $\gamma=0.25$ & $\gamma=0.35$ & $\gamma=0.1$ & $\gamma=0.25$ & $\gamma=0.35$\\
\hline
$200$ & Full & 0.1245 & 0.2411 & 0.3340 & 0.1432 & 0.2575 & 0.3416\\
 & Oracle & 0.1005 & 0.2512 & 0.3490 & 0.1001 & 0.2538 & 0.3509\\
\hline
$500$ & Full & 0.1103 & 0.2447 & 0.3380 & 0.1279 & 0.2441 & 0.3507\\
 & Oracle & 0.1000 & 0.2500 & 0.3502 & 0.0992 & 0.2451 & 0.3531\\
\hline
$1000$ & Full & 0.1041 & 0.2441 & 0.3440 & 0.1194 & 0.2504 & 0.3478\\
 & Oracle & 0.0996 & 0.2501 & 0.3505 & 0.0989 & 0.2500 & 0.3512\\
\hline
\end{tabular}
\end{table}

\begin{table}[!ht]
\centering
\footnotesize
\setlength{\tabcolsep}{3pt}
\renewcommand{\arraystretch}{1.1}
\caption{Monte Carlo standard errors of the reported criteria for the feasible estimator, for the conditional Value-at-Risk and, in parentheses, the conditional Expected Shortfall. Standard errors for RMSE and RMAD are obtained by a nonparametric bootstrap over replications with $500$ resamples; that for the relative bias is $\{B^{-1}\sum_b (e_b-\bar e)^2\}^{1/2}/\sqrt{B}$.}
\label{tab:mcse}
\begin{tabular}{|l|ll|ccc|ccc|ccc|}
\hline
\multicolumn{3}{|c}{} & \multicolumn{3}{|c|}{$\tau_n=0.99$}& \multicolumn{3}{c|}{$\tau_n=0.995$}& \multicolumn{3}{c|}{$\tau_n=0.999$} \\
\hline
\multicolumn{3}{|c|}{} & $\gamma=0.1$ & $\gamma=0.25$ & $\gamma=0.35$& $\gamma=0.1$ & $\gamma=0.25$ & $\gamma=0.35$& $\gamma=0.1$ & $\gamma=0.25$ & $\gamma=0.35$ \\
\hline
\multicolumn{12}{|l|}{\emph{Model 1}} \\
\hline
\multirow{6}{*}{$n=200$}
& \multicolumn{2}{l|}{RMSE} & 0.0186 & 0.0068 & 0.0124 & 0.0323 & 0.0107 & 0.0193 & 0.0943 & 0.0268 & 0.0471\\
& \multicolumn{2}{l|}{} & (0.0567) & (0.0187) & (0.0409) & (0.0880) & (0.0271) & (0.0583) & (0.2221) & (0.0582) & (0.1233)\\
\cline{2-12}
& \multicolumn{2}{l|}{RMAD} & 0.0040 & 0.0044 & 0.0063 & 0.0070 & 0.0055 & 0.0061 & 0.0080 & 0.0081 & 0.0089\\
& \multicolumn{2}{l|}{} & (0.0068) & (0.0073) & (0.0059) & (0.0075) & (0.0078) & (0.0080) & (0.0070) & (0.0097) & (0.0179)\\
\cline{2-12}
& \multicolumn{2}{l|}{RelBias} & 0.0060 & 0.0068 & 0.0093 & 0.0082 & 0.0090 & 0.0125 & 0.0165 & 0.0162 & 0.0226\\
& \multicolumn{2}{l|}{} & (0.0110) & (0.0119) & (0.0184) & (0.0150) & (0.0153) & (0.0236) & (0.0301) & (0.0256) & (0.0399)\\
\hline
\multirow{6}{*}{$n=500$}
& \multicolumn{2}{l|}{RMSE} & 0.0041 & 0.0031 & 0.0058 & 0.0058 & 0.0042 & 0.0082 & 0.0110 & 0.0078 & 0.0172\\
& \multicolumn{2}{l|}{} & (0.0072) & (0.0057) & (0.0138) & (0.0095) & (0.0073) & (0.0188) & (0.0166) & (0.0126) & (0.0364)\\
\cline{2-12}
& \multicolumn{2}{l|}{RMAD} & 0.0025 & 0.0029 & 0.0029 & 0.0026 & 0.0035 & 0.0039 & 0.0047 & 0.0073 & 0.0061\\
& \multicolumn{2}{l|}{} & (0.0031) & (0.0046) & (0.0058) & (0.0038) & (0.0069) & (0.0059) & (0.0049) & (0.0104) & (0.0097)\\
\cline{2-12}
& \multicolumn{2}{l|}{RelBias} & 0.0028 & 0.0037 & 0.0050 & 0.0035 & 0.0047 & 0.0066 & 0.0054 & 0.0075 & 0.0113\\
& \multicolumn{2}{l|}{} & (0.0040) & (0.0057) & (0.0089) & (0.0048) & (0.0070) & (0.0112) & (0.0073) & (0.0104) & (0.0172)\\
\hline
\multirow{6}{*}{$n=1000$}
& \multicolumn{2}{l|}{RMSE} & 0.0026 & 0.0024 & 0.0034 & 0.0031 & 0.0029 & 0.0044 & 0.0050 & 0.0047 & 0.0074\\
& \multicolumn{2}{l|}{} & (0.0036) & (0.0036) & (0.0059) & (0.0044) & (0.0044) & (0.0074) & (0.0068) & (0.0067) & (0.0113)\\
\cline{2-12}
& \multicolumn{2}{l|}{RMAD} & 0.0026 & 0.0026 & 0.0025 & 0.0025 & 0.0041 & 0.0046 & 0.0031 & 0.0045 & 0.0072\\
& \multicolumn{2}{l|}{} & (0.0027) & (0.0049) & (0.0055) & (0.0030) & (0.0043) & (0.0069) & (0.0031) & (0.0079) & (0.0087)\\
\cline{2-12}
& \multicolumn{2}{l|}{RelBias} & 0.0023 & 0.0036 & 0.0045 & 0.0026 & 0.0046 & 0.0059 & 0.0036 & 0.0073 & 0.0099\\
& \multicolumn{2}{l|}{} & (0.0029) & (0.0055) & (0.0078) & (0.0033) & (0.0067) & (0.0096) & (0.0045) & (0.0098) & (0.0141)\\
\hline
\multicolumn{12}{|l|}{\emph{Model 2}} \\
\hline
\multirow{6}{*}{$n=200$}
& \multicolumn{2}{l|}{RMSE} & 0.0135 & 0.0203 & 0.0101 & 0.0208 & 0.0406 & 0.0148 & 0.0534 & 0.1559 & 0.0344\\
& \multicolumn{2}{l|}{} & (0.0327) & (0.1397) & (0.0293) & (0.0487) & (0.2321) & (0.0415) & (0.1166) & (0.6762) & (0.0868)\\
\cline{2-12}
& \multicolumn{2}{l|}{RMAD} & 0.0054 & 0.0070 & 0.0065 & 0.0075 & 0.0068 & 0.0075 & 0.0118 & 0.0126 & 0.0131\\
& \multicolumn{2}{l|}{} & (0.0093) & (0.0095) & (0.0118) & (0.0108) & (0.0129) & (0.0134) & (0.0170) & (0.0126) & (0.0209)\\
\cline{2-12}
& \multicolumn{2}{l|}{RelBias} & 0.0065 & 0.0095 & 0.0094 & 0.0085 & 0.0135 & 0.0128 & 0.0157 & 0.0305 & 0.0237\\
& \multicolumn{2}{l|}{} & (0.0107) & (0.0250) & (0.0189) & (0.0140) & (0.0365) & (0.0245) & (0.0258) & (0.0867) & (0.0416)\\
\hline
\multirow{6}{*}{$n=500$}
& \multicolumn{2}{l|}{RMSE} & 0.0050 & 0.0044 & 0.0062 & 0.0067 & 0.0058 & 0.0082 & 0.0127 & 0.0098 & 0.0150\\
& \multicolumn{2}{l|}{} & (0.0083) & (0.0073) & (0.0120) & (0.0110) & (0.0092) & (0.0154) & (0.0198) & (0.0144) & (0.0259)\\
\cline{2-12}
& \multicolumn{2}{l|}{RMAD} & 0.0038 & 0.0048 & 0.0044 & 0.0041 & 0.0060 & 0.0048 & 0.0062 & 0.0062 & 0.0064\\
& \multicolumn{2}{l|}{} & (0.0043) & (0.0049) & (0.0054) & (0.0062) & (0.0051) & (0.0058) & (0.0080) & (0.0076) & (0.0103)\\
\cline{2-12}
& \multicolumn{2}{l|}{RelBias} & 0.0037 & 0.0041 & 0.0057 & 0.0045 & 0.0051 & 0.0075 & 0.0070 & 0.0080 & 0.0129\\
& \multicolumn{2}{l|}{} & (0.0052) & (0.0061) & (0.0102) & (0.0063) & (0.0074) & (0.0127) & (0.0096) & (0.0110) & (0.0195)\\
\hline
\multirow{6}{*}{$n=1000$}
& \multicolumn{2}{l|}{RMSE} & 0.0030 & 0.0039 & 0.0059 & 0.0035 & 0.0050 & 0.0078 & 0.0050 & 0.0088 & 0.0136\\
& \multicolumn{2}{l|}{} & (0.0039) & (0.0064) & (0.0108) & (0.0045) & (0.0081) & (0.0136) & (0.0063) & (0.0130) & (0.0217)\\
\cline{2-12}
& \multicolumn{2}{l|}{RMAD} & 0.0026 & 0.0054 & 0.0047 & 0.0037 & 0.0052 & 0.0055 & 0.0051 & 0.0058 & 0.0074\\
& \multicolumn{2}{l|}{} & (0.0041) & (0.0061) & (0.0078) & (0.0046) & (0.0059) & (0.0075) & (0.0055) & (0.0088) & (0.0120)\\
\cline{2-12}
& \multicolumn{2}{l|}{RelBias} & 0.0029 & 0.0041 & 0.0059 & 0.0033 & 0.0051 & 0.0077 & 0.0046 & 0.0083 & 0.0131\\
& \multicolumn{2}{l|}{} & (0.0036) & (0.0062) & (0.0103) & (0.0042) & (0.0076) & (0.0127) & (0.0057) & (0.0115) & (0.0191)\\
\hline
\end{tabular}
\end{table}

\medskip\noindent\emph{Consistency and the role of the sample size.}
Both criteria decline monotonically in $n$ in every configuration. For
Model~1 at $\tau_n=0.99$ and $\gamma=0.25$, the relative RMSE of the
conditional VaR falls from $0.132$ at $n=200$ to $0.077$ at $n=500$ and
$0.054$ at $n=1000$, and the corresponding RMAD from $0.074$ to $0.049$
and $0.040$. The rate of decay is consistent with the
$\sqrt{n(1-\alpha_n)}$ normalisation of
Theorem~\ref{thm:extremeQ}, under which a doubling of the sample size at
a fixed intermediate fraction should reduce the error by a factor of
about $\sqrt2$. Accuracy deteriorates as the level $\tau_n$ becomes more
extreme, which is expected since the extrapolation range
$(1-\alpha_n)/(1-\tau_n)$ then widens and the estimation error in
$\widehat\gamma$ is magnified through the exponent of the Weissman
formula.

\medskip\noindent\emph{The cost of estimating the functional operators
declines as the problem becomes more extreme.} The ratio of the feasible
to the oracle RMSE is the natural measure of what is lost by not
observing $m(\cdot)$ and $\sigma(\cdot)$. For Model~1 at $n=1000$ and
$\tau_n=0.99$, this ratio equals $3.47$ at $\gamma=0.1$, $1.26$ at
$\gamma=0.25$ and $1.06$ at $\gamma=0.35$; at $\tau_n=0.999$ it equals
$2.43$, $1.13$ and $0.99$ respectively. Two regularities are visible. The
ratio decreases sharply with the tail index, and it decreases further as
the probability level becomes more extreme. Both are precisely the
ordering predicted by Theorem~\ref{thm:extremeQ}: the estimation error of
the functional regression is asymptotically negligible relative to that
of the extreme-value step, so the closer a configuration lies to the
regime in which the extreme-value error dominates --- heavy tails and
very extreme levels --- the smaller the penalty for not observing the
true operators. Conversely, at $\gamma=0.1$ and moderate levels the
extreme-value step is comparatively accurate and the estimation of the
operators contributes the larger part of the total error. The asymptotic
equivalence established in Section~\ref{sec:mainresults} should therefore
not be read as a finite-sample statement: it describes a limit that the
simulations approach from a direction that is itself informative.

\medskip\noindent\emph{Expected Shortfall is estimated less accurately
than Value-at-Risk, and increasingly so as the tail thickens.} For
Model~1 the ratio of the ES to the VaR relative RMSE rises from $1.19$ at
$\gamma=0.1$ to $1.73$ at $\gamma=0.35$ when $n=1000$ and
$\tau_n=0.99$. This reflects the structure of the estimator
\eqref{estimatorcES} rather than any deficiency of the extrapolation: the
conditional ES multiplies the extrapolated quantile by
$(1-\widehat\gamma)^{-1}$, so that the estimation error in
$\widehat\gamma$ enters twice, once through the extrapolation exponent
and once through this multiplier. Since $(1-\gamma)^{-1}$ is increasing
and convex, the amplification grows with the tail index. That the two
estimators nevertheless share the same limiting distribution in
Theorem~\ref{thm:extremeQ} indicates that this is a finite-sample
amplification which vanishes asymptotically, and indeed the ES-to-VaR
ratio declines with $n$ in every configuration.

\medskip\noindent\emph{The feasible estimator is conservative.}
Table~\ref{tab:bias-coverage} shows that the relative bias of the
feasible estimator is positive throughout, ranging from $0.008$ to
$0.141$ for Model~1 and reaching $0.243$ for Model~2 at $n=200$, whereas
the oracle is essentially unbiased. The source is visible in
Table~\ref{tab:gammahat}: the tail index estimated from the standardized
residuals is biased upward --- $0.1245$ against a true value of $0.10$ at
$n=200$ for Model~1 --- while the same estimator applied to the true
innovations recovers $0.1005$. The upward distortion arises because the
residuals $\widehat\varepsilon_i$ carry the estimation error of both
$m(\cdot)$ and $\sigma(\cdot)$, which inflates their dispersion in the
region where the local sample is thin, and it decays with $n$, falling to
$0.1041$ at $n=1000$. Since an overstated tail index produces an
overstated extreme quantile, the procedure errs on the side of caution, a
direction of error that is the less damaging one in risk management but
which should nonetheless be reported.

\medskip\noindent\emph{Coverage is adequate for heavy tails and
unsatisfactory for light ones.} The empirical coverage of the intervals
\eqref{eq:ci-sim} is close to the nominal $95\%$ when $\gamma\in\{0.25,
0.35\}$, lying between $0.93$ and $0.98$ for the conditional VaR across
both models and all sample sizes. At $\gamma=0.1$, by contrast, coverage
is poor and, more strikingly, deteriorates as $n$ grows: for Model~1 at
$\tau_n=0.99$ it falls from $0.863$ at $n=200$ to $0.733$ at $n=1000$,
and for Model~2 from $0.678$ to $0.075$. The mechanism follows from the
two preceding observations. The half-width of the interval is
proportional to $\widehat\gamma/v_n$ and therefore shrinks at the rate
$\{n(1-\alpha_n)\}^{-1/2}$, whereas the residual-induced bias in
$\widehat\gamma$ documented above decays more slowly; as $n$ increases,
the interval contracts around a point estimate whose systematic error has
not contracted at the same rate. Since the second-order bias of the
data-generating process is exactly zero by construction, this is
unambiguously an effect of the plug-in of estimated operators into the
tail step rather than of the tail extrapolation itself. Two implications
follow. First, the asymptotic intervals of
Theorem~\ref{thm:extremeQ} should be used with caution when the
innovations are only moderately heavy-tailed and the sample is small,
which is the configuration in which the functional estimation error is
relatively most important. Second, the difficulty is one of centring
rather than of scale, and could in principle be addressed by a
bias-corrected tail-index estimator; we do not pursue this here, since
the tail indices estimated in the empirical application of
Section~\ref{sec:application} (we obtained $\widehat{\gamma} = 0.297$) lie in the range where coverage is
satisfactory.

\medskip\noindent\emph{Model specification and the role of the median
criterion.} Comparison of Tables~\ref{tab:model1} and~\ref{tab:model2}
shows that the nonlinear specification is uniformly harder: at $n=1000$,
$\tau_n=0.99$ and $\gamma=0.1$ the relative RMSE of the feasible VaR is
$0.047$ under Model~1 and $0.103$ under Model~2, and the
feasible-to-oracle ratio rises correspondingly from $3.47$ to $7.76$. The
oracle rows are almost identical across the two models, confirming that
the difference originates in the estimation of the operators and not in
the tail step. A final observation concerns the two accuracy criteria
themselves. In a small number of configurations they diverge markedly ---
for Model~2 at $n=200$, $\gamma=0.25$ and $\tau_n=0.999$ the ES relative
RMSE is $1.494$ while the corresponding RMAD is $0.207$ --- which
indicates that a handful of replications, in which the local variance
estimate is close to degenerate, dominate the mean-based criterion. The
median-based RMAD is the more reliable summary in such cases, and the
divergence between the two is itself a useful diagnostic of the stability
of the procedure.


\section{Application to iShares S\&P500 Value ETF}\label{sec:application}
\subsection{Financial setting and data}

This section investigates whether intraday price dynamics contain predictive information about next-day downside risk in equity markets. While conventional risk models rely primarily on daily returns, intraday price movements reflect richer information about order flow, liquidity conditions, volatility clustering, and market stress. Sharp intraday drawdowns, persistent negative drifts, or elevated volatility bursts may signal latent risk accumulation that is not fully incorporated into the closing price. Understanding whether such intraday patterns translate into heightened next-day tail risk is of direct relevance for portfolio risk management, volatility forecasting, and regulatory capital assessment.

We focus on the iShares S\&P 500 Value ETF (IVE), a liquid exchange-traded fund tracking the S\&P 500 Value Index. Value-oriented portfolios are particularly interesting from a risk perspective because their performance can exhibit nonlinear sensitivity during market downturns, especially when liquidity conditions deteriorate. By concentrating on a style-based ETF rather than the broad market index, the analysis evaluates whether intraday return geometry contains economically meaningful signals for the conditional distribution of next-day returns in a factor-oriented portfolio.

The intraday data consist of $1{,}367{,}714$ one-minute bid and ask price records for IVE covering $4{,}127$ trading days from September 28, 2009 to February 24, 2026. To mitigate microstructure noise and bid--ask bounce effects, we construct the midquote price at minute $m$ of day $t$ as
\[
\text{Mid}_{t,m}
=
\frac{\text{BidClose}_{t,m} + \text{AskClose}_{t,m}}{2}.
\]
Each trading day is then represented by a cumulative intraday log-return curve
\[
X_t(u)
=
\log\!\bigl(\text{Mid}_{t,u}\bigr)
-
\log\!\bigl(\text{Mid}_{t,0}\bigr),
\qquad u \in [0,1],
\]
where $u$ denotes normalized intraday time over the regular trading session (09:30--16:00), discretized on a common grid of $391$ one-minute points; missing minutes are filled by the last available quote, and trading days with fewer than $300$ observed quote-minutes (mostly half-days and sparsely quoted sessions early in the sample) are excluded, leaving $3{,}232$ complete intraday curves. Isolated stub quotes, defined as minutes whose midquote deviates from the daily median by more than $25\%$ in log terms, are removed prior to curve construction; this filter discards nine observations in the entire sample, all of them on May 6, 2010, and is required because such quotes would otherwise dominate the covariance operator of the functional covariate. The functional covariate $X_t(\cdot)$ captures the entire evolution of price movements throughout the trading session, including early declines, late-day reversals, volatility spikes, and persistent trends. By construction $X_t(0)=0$, so that each curve measures the cumulative move relative to the same-day opening midquote.

\begin{figure}[h]
  \centering
  \includegraphics[width=0.9\linewidth]{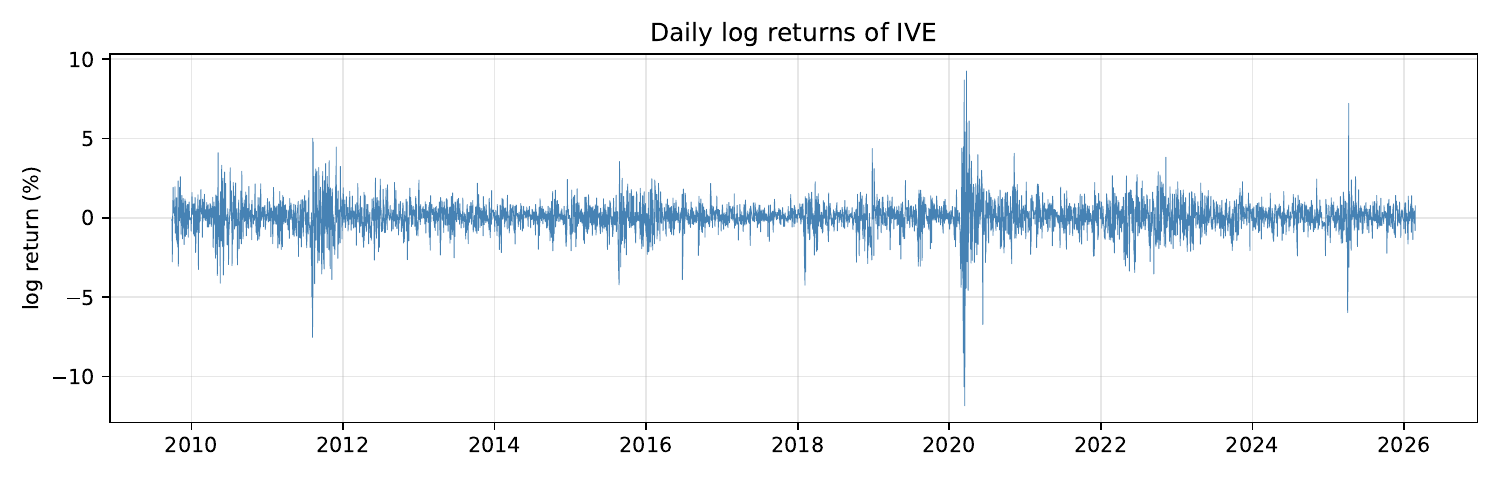}
  \caption{Daily log returns of IVE, September 2009 -- February 2026.}
  \label{fig:ive-logreturn}
\end{figure}

The response variable is the next-day close-to-close log return,
\[
Y_{t+1}
=
\log\!\left(
\frac{\text{Close}_{t+1}}{\text{Close}_{t}}
\right),
\]
constructed from $4{,}125$ daily closing prices of IVE over the same period. The daily return series is heavy-tailed and asymmetric, with skewness $-0.70$ and excess kurtosis $13.7$. Figure~\ref{fig:ive-logreturn} displays the daily log returns over the sample period. The series fluctuates around zero and exhibits several pronounced volatility bursts, particularly during episodes of market stress. Large negative realizations occur intermittently, reflecting the heavy-tailed nature of financial returns and the presence of volatility clustering, two well-documented stylized facts of financial markets.

The estimation sample includes all trading days for which both the full intraday curve $X_t(\cdot)$ and the subsequent daily return $Y_{t+1}$ are observed, yielding $n=3{,}229$ aligned pairs $(X_t, Y_{t+1})$. This temporal alignment ensures that the functional covariate strictly precedes the response variable and preserves the predictive interpretation of the model. 

\begin{figure}[t]
\centering
\includegraphics[width=\linewidth]{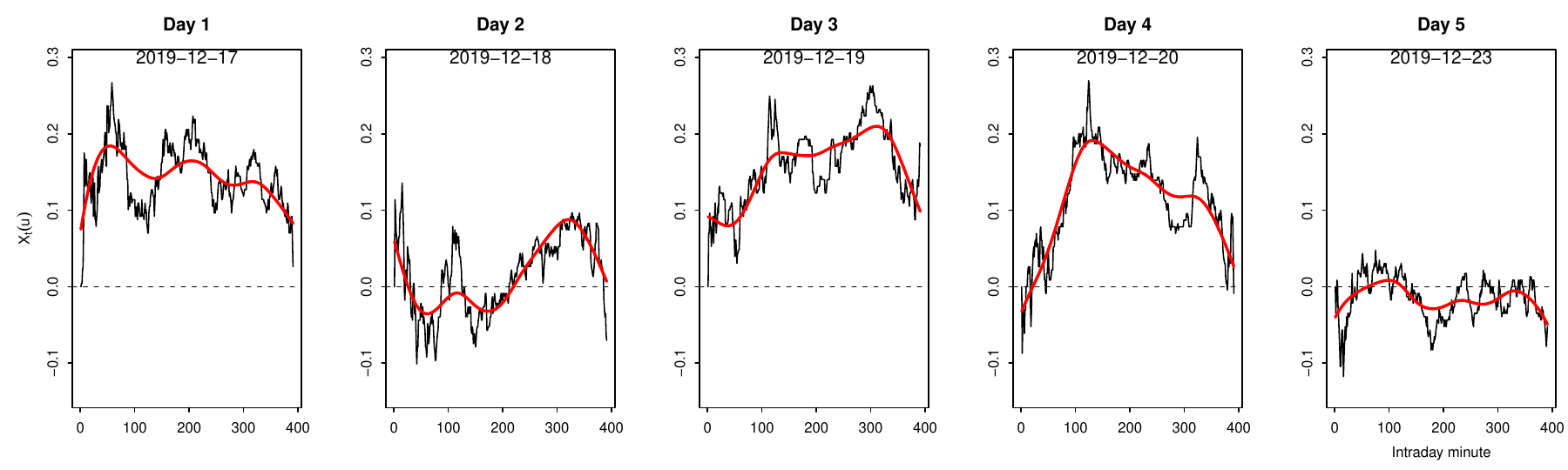}
\caption{Five consecutive intraday functional covariate curves $X_t(u)$ for IVE, plotted over the regular trading session grid $u$ (intraday minute). The black curves are the observed $X_t(u)$, and the red curves are their projections onto the subspace spanned by the first three functional principal components, providing a low-dimensional representation of the dominant intraday variation.}
\label{fig:IVE_consecutive_Xt}
\end{figure}

Figure~\ref{fig:IVE_consecutive_Xt} displays five consecutive intraday curves together with their projections onto the subspace spanned by the first three functional principal components. The observed curves (black) exhibit diverse shapes, including gradual trends, intraday reversals, and localized volatility spikes. The projected curves (red) closely track the original trajectories, indicating that the leading functional principal components capture the dominant modes of variation in the intraday data. Indeed, the first component alone explains $83.2\%$ of the total variation and the first five components together account for $96.1\%$; the leading mode corresponds to the overall intraday drift, while the second ($7.9\%$) captures the within-day reversal pattern. This low-dimensional representation provides a parsimonious summary of intraday dynamics while preserving the key information relevant for predicting future risk; accordingly, the score-based semi-metric used throughout the empirical analysis is constructed from the first $J=5$ functional principal component scores.

From an economic perspective, the functional covariate $X_t(\cdot)$ summarizes the structure of intraday market activity, whereas $Y_{t+1}$ captures the subsequent adjustment in daily returns. Persistent intraday sell-offs may reflect liquidity withdrawal or the gradual incorporation of information into prices, both of which can increase uncertainty and amplify volatility in the following trading session. In contrast, late-day recoveries may reflect temporary order imbalances or short-lived liquidity shocks and therefore contain weaker predictive content for subsequent returns.

To capture these effects, we embed $X_t(\cdot)$ into the nonparametric location--scale model of Section~\ref{ESTCOMP},
\[
Y_{t+1}
=
m(X_t)
+
\sigma(X_t)\varepsilon_{t+1},
\]
which allows both the conditional mean and the conditional volatility of returns to depend flexibly on the entire intraday trajectory. Conditional Value-at-Risk and Expected Shortfall are then obtained through extreme value methods applied to the standardized residuals. The resulting representation
\[
\mathrm{VaR}_{\tau}(X_t)
=
m(X_t)
+
\sigma(X_t)\,q_{\tau}(\varepsilon)
\]
provides an economically interpretable decomposition of tail risk into drift-driven and volatility-driven components. In particular, nonlinearities in $\sigma(\cdot)$ reveal whether specific intraday patterns disproportionately increase next-day crash risk.

The central empirical question is therefore whether the shape of the intraday return curve contains incremental information about extreme next-day losses beyond what is captured by traditional daily volatility measures. By exploiting the full functional structure of high-frequency price paths, the proposed framework links intraday market dynamics to daily tail risk in a flexible and data-driven manner.

\subsection{Extreme conditional VaR and ES estimation}\label{sec:insample}

\begin{figure}[h]
  \centering
  \includegraphics[width=0.92\textwidth]{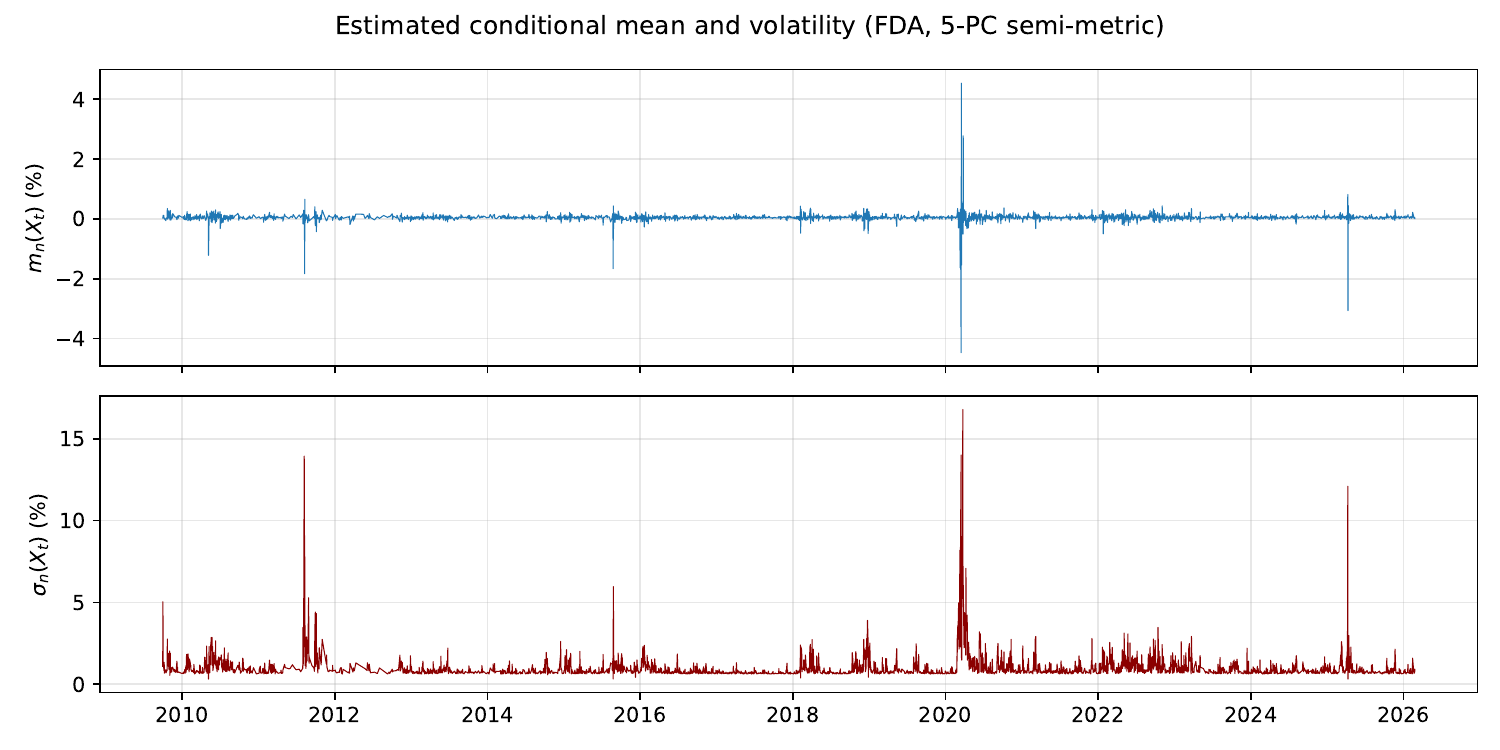}
  \caption{Estimated conditional mean and volatility functions over time:
  $m_n(X_t)$ and $\sigma_n(X_t)$.}
  \label{fig:ive-m-sigma}
\end{figure}

The conditional mean and log-variance operators are estimated by local linear regression in the space of standardized functional principal component scores, with bandwidths selected by $V$-fold cross-validation as described in Section~\ref{sec:tuning}. Figure~\ref{fig:ive-m-sigma} reports the estimated conditional mean and conditional volatility functions over time. The estimated mean component $m_n(X_t)$ remains close to zero throughout the sample, indicating that the conditional location effect plays only a minor role in explaining next-day returns. In contrast, the estimated scale function $\sigma_n(X_t)$ exhibits pronounced time variation, with clear volatility spikes during periods of market stress, most visibly around the February 2018 volatility event, the COVID-19 crash of early 2020, the 2022 bear market, and the tariff-related turbulence of spring 2025. These episodes correspond to days characterized by large intraday fluctuations and are followed by substantially higher predicted volatility for the next trading day. The results therefore confirm that the functional covariate contains economically meaningful information about future risk conditions. In particular, the conditional volatility function captures the clustering behavior that is typically observed in financial returns (see Figure~\ref{fig:ive-logreturn}), while allowing this dependence to be driven by the entire intraday trajectory rather than by lagged squared returns alone.

\begin{figure}[h]
  \centering
  \includegraphics[width=9cm, height = 5cm]{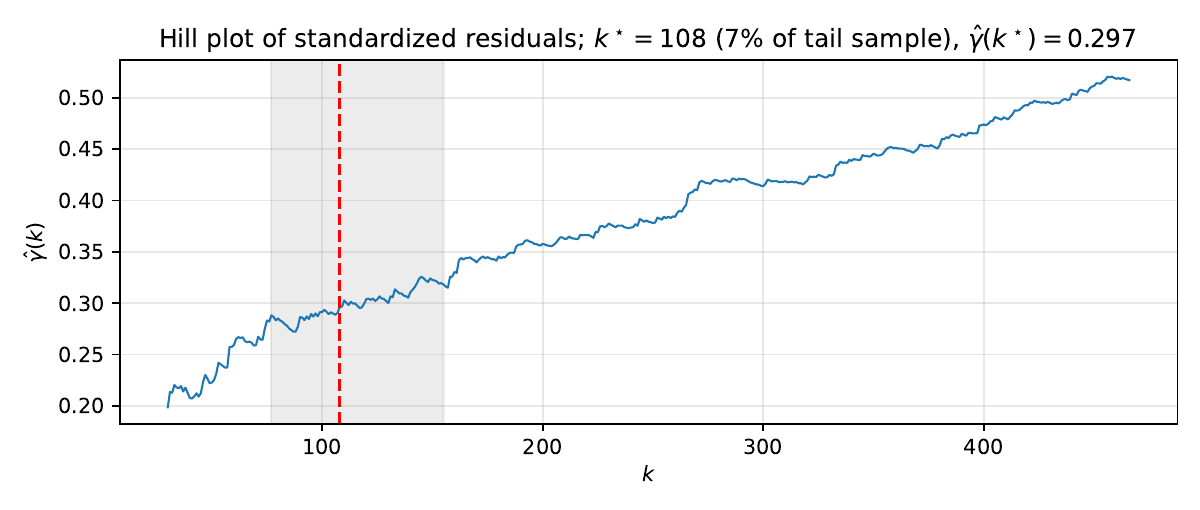}
  \caption{Tail index estimation and stability selection: Hill-type estimates
  $k \mapsto \widehat{\gamma}(k)$ with the selected threshold $k^\star = 108$, corresponding to $7\%$ of the tail sample (dashed line; the shaded band marks the $5\%$--$10\%$ range examined in Table~\ref{tab:threshold}), and $\widehat{\gamma}(k^\star) = 0.297.$}
  \label{fig:ive-tail-index-stability}
\end{figure}

\begin{figure}[h]
  \centering
  \includegraphics[width=9cm, height = 5cm]{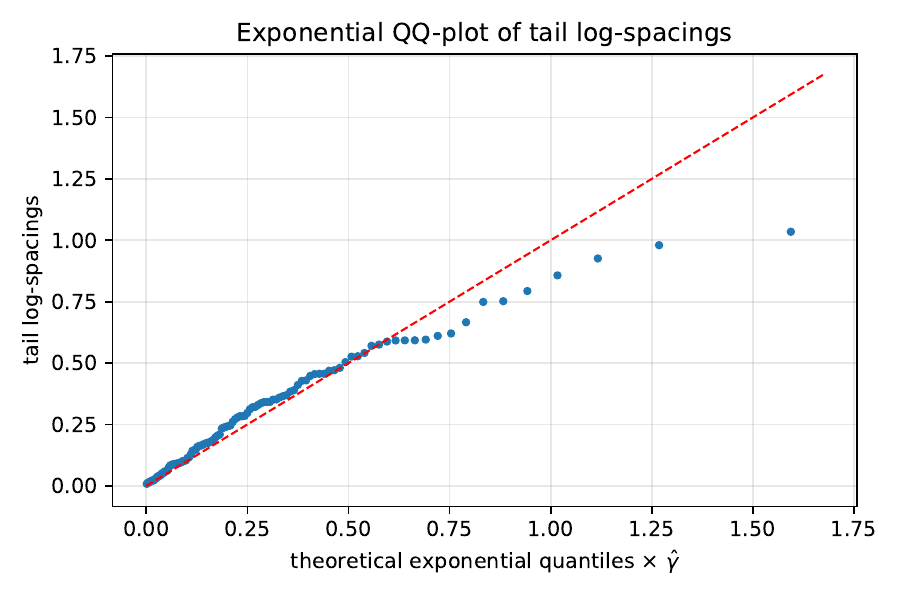}
  \caption{Exponential QQ-plot of tail log-spacings
  $\log\!\big(\widehat{\epsilon}_{n-i+1,n}/\widehat{\epsilon}_{n-k^\star,n}\big)$
  for $1\le i\le k^\star = 108.$}
  \label{fig:ive-expqq-logspacings}
\end{figure}

Figures~\ref{fig:ive-tail-index-stability} and~\ref{fig:ive-expqq-logspacings} present diagnostic tools supporting the extreme value modeling of the residuals. Since the objective is the left tail of the return distribution, tail estimation is based on the negative standardized residuals, of which $n_{\mathrm{tail}}=1{,}557$ are strictly positive after sign reversal. Figure~\ref{fig:ive-tail-index-stability} shows the Hill-type estimates of the tail index as a function of the number of upper order statistics $k$. The path rises slowly and monotonically with $k$, a pattern indicative of second-order bias at deep thresholds, and flattens over the region $k\in[80,130]$, where $\widehat{\gamma}(k)$ varies only between $0.285$ and $0.304$. Because this path has no genuine plateau over the full admissible range, automatic flatness rules, whether based on the local variance or the local slope of $k\mapsto\widehat\gamma(k)$, are numerically unstable and can select deep, bias-dominated thresholds. We therefore fix the intermediate sequence at a conventional fraction of the tail sample, $k_n=\lfloor 0.07\,n_{\mathrm{tail}}\rfloor$, in the spirit of \citet{McNeilFrey2000}, which yields $k^\star=108$ and $\widehat{\gamma}(k^\star)=0.297$; Table~\ref{tab:threshold} documents the sensitivity of the results to this choice. This estimate indicates moderately heavy tails in the residual distribution: the innovations possess a finite variance but an infinite fourth moment, which is consistent with the stylized fact that financial return innovations often display Pareto-type tail behavior, and lies comfortably within the region $0<\gamma<1$ required by the asymptotic theory of Section~\ref{sec:mainresults}. Figure~\ref{fig:ive-expqq-logspacings} provides an exponential QQ-plot of the tail log-spacings used in the Hill estimation. The empirical points align closely with the theoretical reference line, supporting the assumption that the tail of the residual distribution is well approximated by the extreme value model. Taken together, these diagnostics provide empirical justification for applying extreme value methods to estimate tail risk measures.

Table~\ref{tab:threshold} reports the sensitivity of the FDA results to the tail fraction defining the intermediate sequence. Over the range $5\%$--$10\%$ the estimated tail index moves only from $0.288$ to $0.318$ and the in-sample exceedance rates are essentially unchanged, so the conclusions are not driven by the particular value adopted. For fractions of $15\%$ and above the Hill estimator becomes visibly bias-dominated, $\widehat\gamma$ drifts towards $0.418$, and the implied risk measures become markedly over-conservative---which is precisely the failure mode that motivates fixing the fraction in the stable region rather than letting an automatic rule search the full range. The results below are based on tail fraction equals to $7\%$ (i.e. $\widehat\gamma = 0.297$).

\begin{table}[H]
\centering
\caption{Sensitivity of the FDA tail estimates and in-sample exceedance rates to the tail fraction $k_n/n_{\mathrm{tail}}$ used for the Hill estimator ($n_{\mathrm{tail}}=1{,}557$). }
\label{tab:threshold}
\begin{tabular}{lccccc}
\hline
Tail fraction (\%) & 5 & 7 & 10 & 15 & 20 \\
\hline
$k_n$                                & 77 & 108 & 155 & 233 & 311 \\
$\widehat{\gamma}$                   & 0.288 & 0.297 & 0.318 & 0.374 & 0.418 \\
Exceedance rate, $\tau=0.99$ (\%)    & 1.115 & 1.115 & 1.053 & 0.960 & 0.805 \\
Exceedance rate, $\tau=0.995$ (\%)   & 0.526 & 0.526 & 0.465 & 0.217 & 0.217 \\
\hline
\end{tabular}
\end{table}

\begin{figure}[h]
  \centering
  \includegraphics[width=0.92\textwidth]{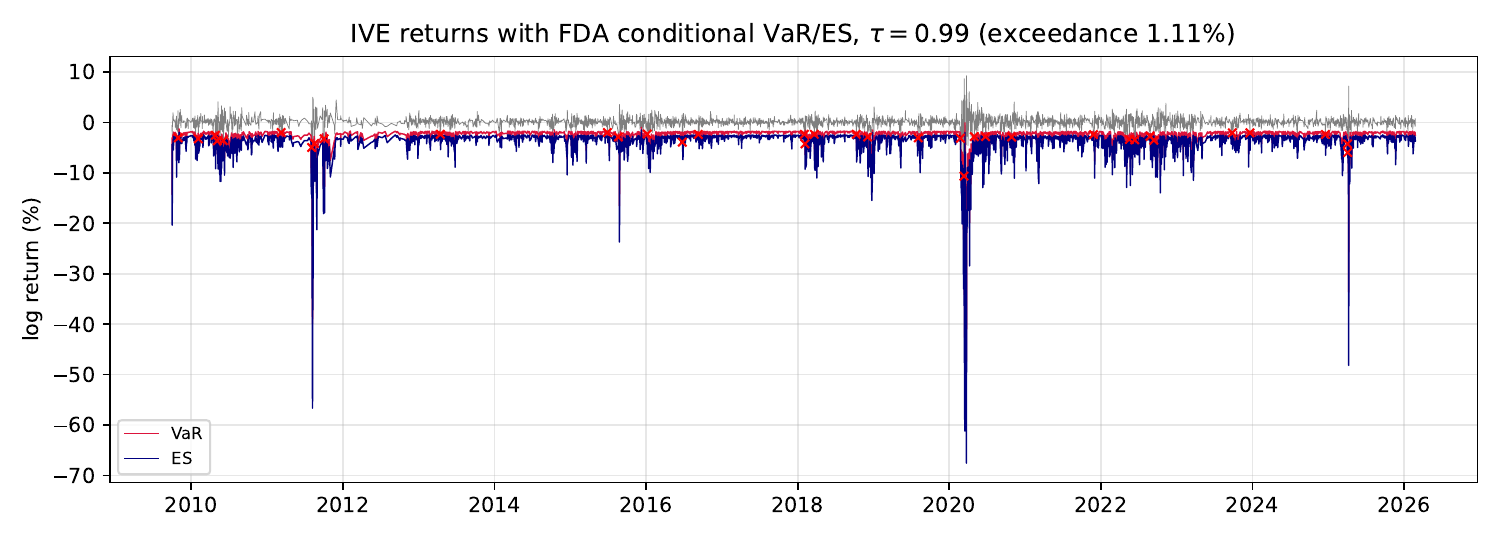}
  \caption{Daily returns with conditional risk measures at level $\tau=0.99.$}
  \label{fig:ive-returns-var-es_099}
\end{figure}

\begin{figure}[h]
  \centering
  \includegraphics[width=0.92\textwidth]{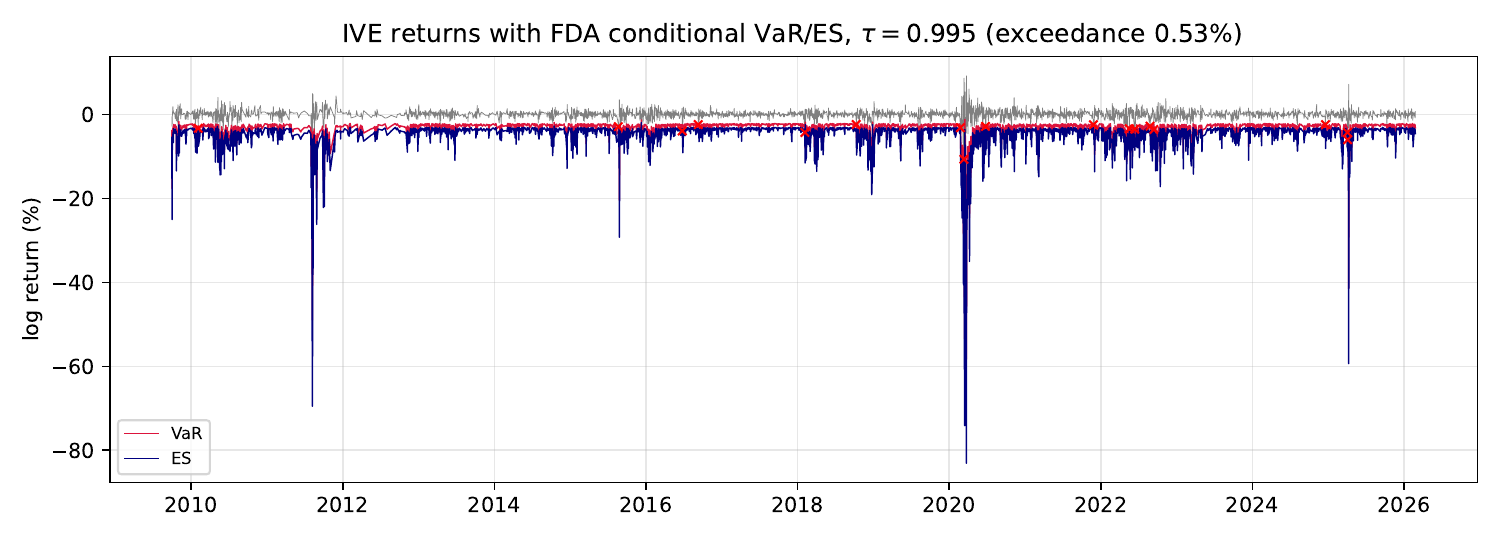}
  \caption{Daily returns with conditional risk measures at level $\tau=0.995.$}
  \label{fig:ive-returns-var-es_0995}
\end{figure}

Figures~\ref{fig:ive-returns-var-es_099} and~\ref{fig:ive-returns-var-es_0995} report the estimated conditional Value-at-Risk (VaR) and Expected Shortfall (ES) together with the observed daily returns for two confidence levels, $\tau=0.99$ and $\tau=0.995$. Several features are noteworthy. First, both risk measures exhibit substantial time variation and react strongly to periods of elevated market volatility. Episodes of large negative returns are associated with pronounced downward movements of the estimated VaR and ES curves, reflecting an increase in predicted downside risk. Second, the frequency of violations of the VaR threshold is broadly consistent with the nominal confidence levels: the observed exceedance rate is $1.11\%$ for $\tau=0.99$ and $0.53\%$ for $\tau=0.995$, against theoretical probabilities of $1\%$ and $0.5\%$, respectively. Third, the ES curves lie systematically below the corresponding VaR curves, as expected since ES reflects the expected magnitude of losses beyond the VaR threshold.

Table~\ref{tab:insample_backtesting} formalizes the in-sample comparison at $\tau=0.995$ across the proposed FDA model and the four benchmarks described in Section~\ref{sec:backtesting} below, each fitted once on the full aligned sample. Four of the five specifications deliver violation rates statistically indistinguishable from the nominal level, the exception being the GARCH-$t$ model, whose in-sample unconditional coverage (UC) is clearly rejected ($\text{Rate} = 0.96\%$, UC $p=0.001$); a single Student-$t$ quantile multiplier applied to a GARCH scale is evidently too rigid to track the extreme tail over a sixteen-year sample. Among the remaining models the FDA and GARCH-EVT specifications attain the lowest joint VaR--ES scores, which are essentially tied. It should be emphasized that in-sample diagnostics are reported for completeness only: all five models are fitted to the same data used for evaluation, so these figures measure goodness of fit rather than predictive ability, and the out-of-sample exercise of Section~\ref{sec:backtesting} is the basis for our conclusions.

\begin{table}[H]
\centering
\caption{In-sample VaR backtesting and joint VaR--ES evaluation results for $\tau=0.995$ ($n=3{,}229$; nominal rate $0.5\%$). UC, IND and CC denote the $p$-values of the unconditional coverage, independence and conditional coverage tests of \citet{Christoffersen1998}; DQ is the $p$-value of the dynamic quantile test of \citet{EngleManganelli2004} with four lags; FZ0 is the mean of the zero-degree homogeneous Fissler--Ziegel joint loss \citep{FisslerZiegel2016,PattonZiegelChen2019} (lower is better).}
\label{tab:insample_backtesting}
\begin{tabular}{lccccccc}
\hline
Model & Exceed. & Rate (\%) & UC & IND & CC & DQ & Mean FZ0 \\
\hline
FDA        & 17 & 0.526 & 0.833 & 0.078 & 0.207 & 0.068 & 1.298 \\
GARCH-$t$  & 31 & 0.960 & 0.001 & 0.305 & 0.003 & 0.000 & 1.361 \\
GARCH-EVT  & 15 & 0.465 & 0.772 & 0.708 & 0.894 & 0.034 & 1.283 \\
NP-Y       & 15 & 0.465 & 0.773 & 0.708 & 0.894 & 0.034 & 1.491 \\
CAViaR     & 17 & 0.526 & 0.833 & 0.671 & 0.894 & 0.000 & --    \\
\hline
\end{tabular}
\end{table}

\subsection{Backtesting and comparison with benchmark methods}\label{sec:backtesting}

The predictive performance of the proposed FDA-based model is evaluated using standard backtesting procedures widely used in financial risk management. We assess the adequacy of the VaR forecasts by comparing the observed violation frequency with the nominal probability $1-\tau$ and employ the unconditional coverage (UC), independence (IND), and conditional coverage (CC) tests proposed by \citet{Christoffersen1998}, complemented by the dynamic quantile (DQ) test of \citet{EngleManganelli2004}, which detects predictability of violations from a constant and four lagged violation indicators. The joint accuracy of the VaR and ES forecasts is evaluated using the strictly consistent scoring function of \citet{FisslerZiegel2016}; we adopt its zero-degree homogeneous parametrization \citep{PattonZiegelChen2019}, whose mean value is directly comparable across models and subsamples, with lower values indicating better joint forecasts.

The forecasting exercise is conducted using a rolling-window scheme. At each forecasting date, the models are estimated using the most recent $1{,}000$ observations and one-step-ahead VaR and ES forecasts are produced; the estimation window is then rolled forward throughout the evaluation sample and model parameters are re-estimated every $20$ observations. For the FDA model, the functional principal component representation is obtained from the intraday curves observed within each estimation window; the conditional location and scale functions are then estimated by local linear regression using the semi-metric defined on the first five functional principal component scores, with bandwidths selected by $V$-fold cross-validation within each estimation window. Extreme tail risk is estimated by applying the Hill estimator to the left tail of the standardized residuals, with the intermediate threshold selected automatically by the stability criterion, and the resulting tail index is combined with the Weissman extrapolation formula to obtain extreme conditional quantiles and the associated Expected Shortfall.

The proposed model is compared with four benchmarks. The first is a parametric GARCH$(1,1)$ model with Student-$t$ innovations following \citet{Bollerslev1986}. The second is a GARCH$(1,1)$--EVT model in which the conditional mean and variance are first filtered through a Gaussian GARCH$(1,1)$ specification and the left tail of the standardized residuals is subsequently modeled by the same Hill--Weissman procedure, as in \citet{McNeilFrey2000}. The third is a nonparametric location--scale model with scalar predictor, denoted NP-Y, defined by $Y_{t+1}=m(Y_t)+\sigma(Y_t)\varepsilon_{t+1}$, where $m(\cdot)$ and $\sigma(\cdot)$ are estimated by local linear smoothing with cross-validated bandwidths and EVT is applied to the standardized residuals; NP-Y thus isolates the marginal contribution of the functional covariate, since it shares the entire estimation pipeline of the FDA model but conditions only on the previous daily return. Finally, we consider the symmetric absolute value specification of the Conditional Autoregressive Value-at-Risk (CAViaR) model of \citet{EngleManganelli2004}, estimated by minimizing the check loss over a large set of randomly drawn parameter vectors followed by simplex refinement of the best candidates, as recommended by the original authors.

Because the FDA forecasts require a complete intraday curve on the day preceding each response, the out-of-sample comparison is conducted on the common evaluation sample on which all five models produce forecasts, comprising $2{,}229$ trading days from February 11, 2016 to February 24, 2026, aligned by response date. This decade-long evaluation window is particularly demanding, as it contains several distinct stress episodes: the February 2018 volatility spike, the fourth quarter of 2018, the COVID-19 crash, the 2022 inflation-driven bear market, the August 2024 carry-trade unwinding, and the tariff shock of spring 2025.

\begin{figure}[h]
  \centering
  \includegraphics[width=\textwidth]{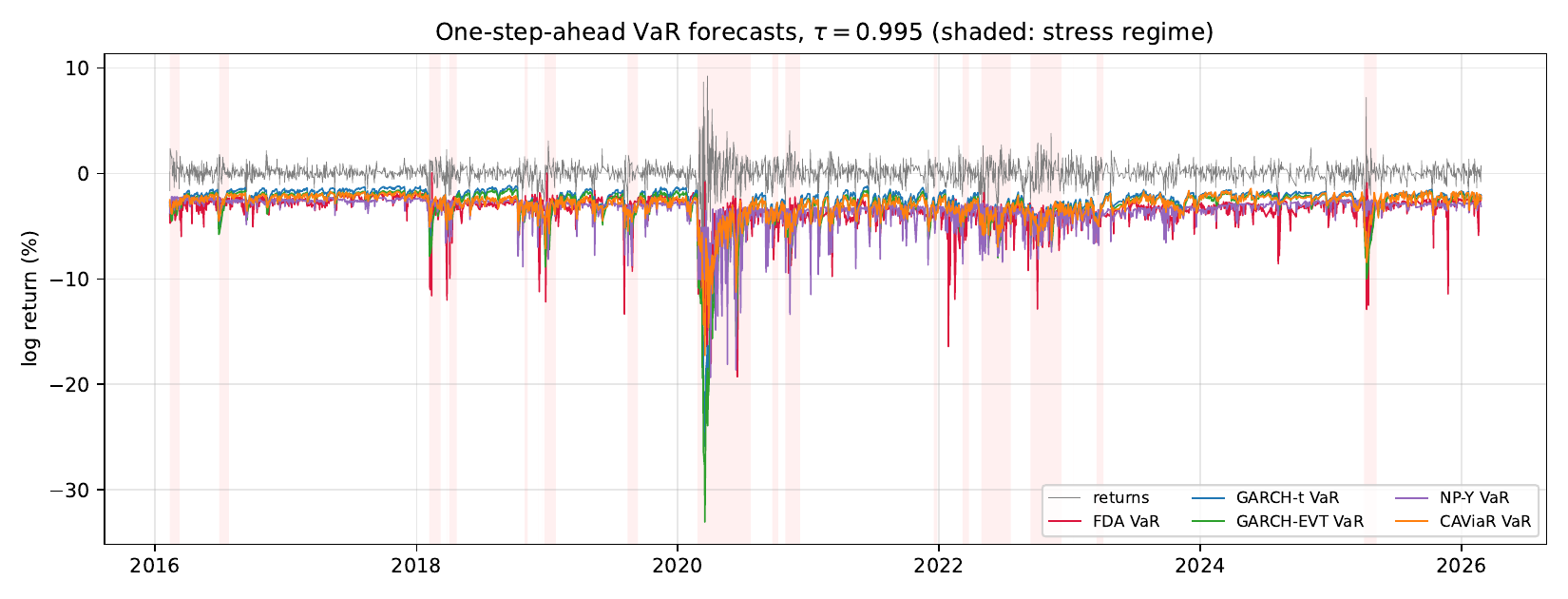}
  \caption{One-step-ahead VaR forecasts of the five competing models at $\tau=0.995$ over the common evaluation sample, together with realized daily returns. Shaded bands indicate the stress regime defined in Section~\ref{sec:subperiods}.}
  \label{fig:oos-var-all}
\end{figure}

Tables~\ref{tab:oos_backtesting_99} and~\ref{tab:oos_backtesting} report the out-of-sample results for $\tau=0.99$ and $\tau=0.995$, and Figure~\ref{fig:oos-var-all} displays the corresponding VaR forecast paths of all five models together with the realized returns. At the extreme level $\tau=0.995$, the FDA model attains an exceedance rate of $0.54\%$ against the nominal $0.5\%$---the most accurate unconditional coverage of any specification (UC $p=0.800$)---and passes the conditional coverage test (CC $p=0.148$). The GARCH-EVT model performs comparably ($0.63\%$, CC $p=0.147$), whereas GARCH-$t$ is markedly too aggressive ($0.94\%$) and is rejected by all three Christoffersen tests, and NP-Y and CAViaR attain acceptable frequencies but generate strongly clustered violations (IND $p=0.002$ and $p=0.004$, respectively). The picture at $\tau=0.99$ is similar in ranking: GARCH-EVT, CAViaR and FDA deliver the exceedance rates closest to nominal, while GARCH-$t$ over-shoots.

Two qualifications should be stated plainly. First, the DQ test, which is more powerful than the Christoffersen statistics because it exploits the magnitude and autocorrelation of the violation indicators jointly, rejects every model on the full evaluation sample at conventional levels. This is not surprising: the sample spans a decade containing several major liquidity events, and no one-step-ahead model conditioned on information available at the previous close absorbs all of them. Second, under the joint scoring function the ordering differs from the coverage-based ordering: GARCH-EVT achieves the lowest mean FZ0 loss ($1.325$ at $\tau=0.995$), followed by GARCH-$t$, with FDA and NP-Y trailing. The two families of criteria reward different properties, and the divergence is informative rather than contradictory, a point we develop below and, with the regime breakdown of Section~\ref{sec:subperiods}, that is where the contribution of the intraday information becomes visible.

\begin{table}[!ht]
\centering
\caption{Out-of-sample VaR backtesting and joint VaR--ES evaluation results for $\tau=0.99$ on the common evaluation sample ($N=2{,}229$ one-step-ahead forecasts, February 2016 -- February 2026; nominal rate $1\%$).}
\label{tab:oos_backtesting_99}
\begin{tabular}{lccccccc}
\hline
Model & Exceed. & Rate (\%) & UC & IND & CC & DQ & Mean FZ0 \\
\hline
FDA        & 28 & 1.256 & 0.242 & 0.000 & 0.000 & 0.000 & 1.393 \\
GARCH-$t$  & 36 & 1.615 & 0.007 & 0.136 & 0.009 & 0.006 & 1.233 \\
GARCH-EVT  & 27 & 1.211 & 0.332 & 0.043 & 0.080 & 0.010 & 1.170 \\
NP-Y       & 32 & 1.436 & 0.052 & 0.010 & 0.006 & 0.000 & 1.390 \\
CAViaR     & 27 & 1.211 & 0.332 & 0.043 & 0.080 & 0.000 & --    \\
\hline
\end{tabular}
\end{table}

\begin{table}[!ht]
\centering
\caption{Out-of-sample VaR backtesting and joint VaR--ES evaluation results for $\tau=0.995$ on the common evaluation sample ($N=2{,}229$ one-step-ahead forecasts, February 2016 -- February 2026; nominal rate $0.5\%$).}
\label{tab:oos_backtesting}
\begin{tabular}{lccccccc}
\hline
Model & Exceed. & Rate (\%) & UC & IND & CC & DQ & Mean FZ0 \\
\hline
FDA        & 12 & 0.538 & 0.800 & 0.053 & 0.148 & 0.009 & 1.621 \\
GARCH-$t$  & 21 & 0.942 & 0.008 & 0.015 & 0.002 & 0.000 & 1.432 \\
GARCH-EVT  & 14 & 0.628 & 0.410 & 0.075 & 0.147 & 0.021 & 1.325 \\
NP-Y       & 14 & 0.628 & 0.410 & 0.002 & 0.007 & 0.000 & 1.547 \\
CAViaR     & 16 & 0.718 & 0.171 & 0.004 & 0.007 & 0.000 & --    \\
\hline
\end{tabular}
\end{table}

The contrast between the coverage tests and the joint scoring function is instructive. The scoring function rewards forecasts that track the conditional tail closely on average, which the parsimonious GARCH filter does well; the coverage and independence tests reward correct \emph{dynamics} of the violation process. The economic mechanism behind the divergence is that the intraday information embedded in $X_t(\cdot)$ does not primarily deepen the estimated tail---the average level of the FDA risk forecasts is comparable to that of its competitors---but rather re-times it. GARCH-type filters update their volatility state only after a large daily return has materialized, so the first losses of a stress episode arrive before the risk forecast has adjusted. The functional covariate, by contrast, responds to the intraday drawdown pattern, the within-day volatility bursts, and the late-session order imbalance of the current day, allowing $\sigma_n(X_t)$ to rise before the daily return series reveals the regime change. Consistent with this interpretation, the twelve FDA violations at $\tau=0.995$ fall on isolated and economically identifiable surprise events---the Brexit referendum (June 24, 2016), the Volmageddon episode (February 5, 2018), the October--December 2018 correction, the onset of the COVID-19 crash (February 24--25, 2020), and the tariff announcement of April 4, 2025---rather than in extended runs. From a regulatory perspective this distinction matters, since supervisory backtesting frameworks penalize clustered exceedances, which signal a systematically late model, more severely than isolated ones.

\subsection{Performance across market regimes and stress episodes}\label{sec:subperiods}

Aggregate backtests may mask economically important heterogeneity in model performance across market conditions. To investigate this, we partition the evaluation sample into three volatility regimes constructed from the annualized 21-day realized volatility of daily returns: \emph{calm} days, on which realized volatility lies below its evaluation-sample median of $11.8\%$; \emph{stress} days, on which it lies in the top quantile (at or above $18.0\%$); and \emph{normal} days in between. This yields $1{,}114$ calm, $669$ normal and $446$ stress days. The stress regime captures, among others, the February 2018 and Q4-2018 corrections, the COVID-19 crash, the most turbulent phases of the 2022 bear market, and the 2025 tariff shock. Table~\ref{tab:regimes} reports the backtesting results by regime at $\tau=0.995$, and Table~\ref{tab:episodes} complements the regime analysis with VaR exceedance rates during six named stress episodes.

\begin{table}[!ht]
\centering
\caption{Out-of-sample backtesting by volatility regime at $\tau=0.995$ (nominal rate $0.5\%$). Regimes are defined by the annualized 21-day realized volatility of daily returns: Calm (below the evaluation-sample median of $11.8\%$), Stress (top quintile, $\geq 18.0\%$), Normal (in between).}
\label{tab:regimes}
\begin{tabular}{llccccccc}
\hline
Regime & Model & Exceed. & Rate (\%) & UC & IND & CC & DQ & Mean FZ0 \\
\hline
Calm    & FDA        & 0  & 0.000 & 0.001 & 1.000 & 0.004 & --    & 1.124 \\
($N=1{,}114$) & GARCH-$t$  & 7  & 0.628 & 0.559 & 0.766 & 0.807 & 0.988 & 1.000 \\
        & GARCH-EVT  & 5  & 0.449 & 0.805 & 0.832 & 0.948 & 1.000 & 0.985 \\
        & NP-Y       & 0  & 0.000 & 0.001 & 1.000 & 0.004 & --    & 1.131 \\
        & CAViaR     & 2  & 0.180 & 0.081 & 0.932 & 0.216 & 0.808 & --    \\
\hline
Normal  & FDA        & 6  & 0.897 & 0.191 & 0.742 & 0.402 & 0.766 & 1.631 \\
($N=669$) & GARCH-$t$  & 9  & 1.345 & 0.010 & 0.620 & 0.033 & 0.049 & 1.660 \\
        & GARCH-EVT  & 6  & 0.897 & 0.191 & 0.742 & 0.402 & 0.766 & 1.440 \\
        & NP-Y       & 2  & 0.299 & 0.425 & 0.913 & 0.723 & 0.990 & 1.507 \\
        & CAViaR     & 5  & 0.747 & 0.398 & 0.784 & 0.674 & 0.956 & --    \\
\hline
Stress  & FDA        & 6  & 1.345 & 0.037 & 0.685 & 0.104 & 0.188 & 2.846 \\
($N=446$) & GARCH-$t$  & 5  & 1.121 & 0.110 & 0.039 & 0.033 & 0.000 & 2.171 \\
        & GARCH-EVT  & 3  & 0.673 & 0.623 & 0.010 & 0.033 & 0.000 & 2.005 \\
        & NP-Y       & 12 & 2.691 & 0.000 & 0.035 & 0.000 & 0.000 & 2.648 \\
        & CAViaR     & 9  & 2.018 & 0.001 & 0.165 & 0.001 & 0.000 & --    \\
\hline
\end{tabular}
\end{table}

\begin{table}[!ht]
\centering
\caption{VaR exceedance rates (in \%) during named stress episodes at $\tau=0.995$ (nominal rate $0.5\%$). $N$ denotes the number of evaluation days within each episode.}
\label{tab:episodes}
\begin{tabular}{lcccccc}
\hline
Episode & $N$ & FDA & GARCH-$t$ & GARCH-EVT & NP-Y & CAViaR \\
\hline
Volmageddon 2018      & 65  & 3.08 & 4.62 & 3.08 & 1.54  & 6.15 \\
Q4-2018 sell-off      & 61  & 4.92 & 3.28 & 3.28 & 0.00  & 0.00 \\
COVID-19 crash        & 50  & 4.00 & 2.00 & 2.00 & 16.00 & 8.00 \\
2022 bear market      & 193 & 0.52 & 0.52 & 0.52 & 0.52  & 0.00 \\
Carry unwind 2024     & 25  & 0.00 & 4.00 & 0.00 & 0.00  & 0.00 \\
Tariff shock 2025     & 58  & 1.72 & 3.45 & 3.45 & 5.17  & 3.45 \\
\hline
\end{tabular}
\end{table}

Three findings emerge. First, in calm markets the FDA and NP-Y models are markedly conservative---neither records a single violation over $1{,}114$ calm days, so their unconditional coverage is formally rejected---whereas the GARCH-based models are almost perfectly calibrated and attain the lowest FZ0 losses. The functional model's errors in tranquil conditions are therefore concentrated in the economically inexpensive direction of excess prudence rather than in under-estimation of risk, but they are errors nonetheless, and they explain a substantial part of the FDA model's unfavourable full-sample FZ0 score. Second, in the normal regime the FDA and GARCH-EVT specifications are indistinguishable and both are adequate on every diagnostic, while the GARCH-$t$ model already over-shoots ($1.35\%$ against $0.5\%$ nominal, CC $p=0.033$). Third, and most importantly, the stress regime separates the models sharply. Violation frequencies rise for all models, and the mean FZ0 loss roughly doubles relative to calm markets, quantifying the systematic cost of stress episodes for joint VaR--ES accuracy. But the FDA model is the only specification that passes both the conditional coverage test (CC $p=0.104$) and the dynamic quantile test (DQ $p=0.188$) within the stress regime; every competitor is rejected by DQ at any conventional level, and GARCH-$t$, GARCH-EVT and NP-Y are additionally rejected on independence. In other words, when markets are turbulent the daily-data models fail not only in the number of their exceedances but in their timing, concentrating violations in runs at the onset of each episode, whereas the intraday-conditioned forecasts distribute them unpredictably. Since the stress regime is precisely where capital adequacy is tested, this is the economically relevant comparison, and it is the clearest evidence in our results that intraday price curves carry incremental information for extreme risk.

The episode-level results in Table~\ref{tab:episodes} refine this picture and also caution against over-claiming. No one-day-ahead model conditioned on the previous close absorbs a sequence of successive record-breaking jumps: during the COVID-19 crash the FDA model records two violations in fifty days, better than NP-Y and CAViaR but worse than the GARCH filters, and during the Q4-2018 correction it is the weakest of the five. Conversely, in the two most recent episodes the ranking is reversed and the margin is substantial: in the August 2024 carry-trade unwinding the FDA model records no violation while GARCH-$t$ exceeds at a $4\%$ rate, and in the spring 2025 tariff shock it attains the lowest exceedance rate of all five models ($1.72\%$ against $3.45$--$5.17\%$). These are episodes characterized by pronounced intraday drawdown-and-reversal patterns in the days preceding the largest daily losses, which is exactly the configuration in which the shape of the intraday curve is informative relative to the magnitude of past daily returns. With between $25$ and $193$ observations per episode, however, these comparisons rest on a handful of violations each and should be read as illustrative rather than as formal tests.

Overall, the empirical results support that intraday price curves do improve extreme VaR and ES forecasting, but along a specific dimension: rather than uniformly tightening the estimated tail, the functional covariate corrects the \emph{timing} of tail risk forecasts. On the full evaluation sample the FDA model delivers the most accurate unconditional coverage at $\tau=0.995$ and passes conditional coverage; within the stress regime, where the comparison matters most for capital adequacy, it is the only specification whose violations are neither too clustered nor predictable from their own past. These gains come at two identifiable costs: additional conservatism in tranquil markets, and a joint VaR--ES score that remains inferior to that of the parsimonious GARCH-EVT filter, which tracks the average level of the tail more tightly. Since conditional coverage and the absence of violation clustering are the properties targeted by supervisory backtesting frameworks, whereas the scoring function summarizes average accuracy, the two criteria are best read as complementary. Taken together, the evidence supports incorporating functional high-frequency information into the econometric modeling of financial tail risk, and points to hybrid specifications---a GARCH-type filter whose innovations or conditional scale are modulated by intraday functional covariates---as a natural way to combine the average accuracy of the former with the superior timing of the latter.

\section{Conclusion}\label{sec:conclusion}

This paper develops a semiparametric framework for estimating extreme
Value-at-Risk and Expected Shortfall when the distribution of asset
returns depends on a functional covariate derived from high-frequency
intraday prices. Under a functional location--scale specification, the
conditional mean and volatility of next-day returns are estimated by
local linear regression on a score-based semi-metric, and extreme risk
measures are obtained by applying extreme value methods to the resulting
standardized residuals. Under stationarity and ergodicity, we establish
the consistency and asymptotic normality of the residual-based Hill
estimator and of the extrapolated conditional VaR and ES, extending the
asymptotic theory of functional nonparametric regression to the
extreme-value setting. Monte Carlo experiments show that the efficiency loss attributable to
estimating the location and scale operators diminishes as the
probability level becomes more extreme and as the tail becomes heavier,
so that the feasible and oracle estimators are of comparable accuracy in
the regime for which the asymptotic theory is developed; at moderate
levels, by contrast, operator estimation remains a material source of
error.

The empirical application to the IVE ETF, based on a decade-long
out-of-sample evaluation using one-minute bid--ask data, refines rather
than simply confirms this theoretical picture. The functional model
attains the most accurate unconditional coverage among five competing
specifications at the $99.5\%$ level, but the source of this advantage is
specific rather than general: it operates through the timing of extreme
losses rather than their average magnitude. A GARCH-EVT filter achieves a
tighter average joint VaR--ES loss, and the functional model is markedly
conservative in calm markets, recording no violations over more than a
thousand low-volatility trading days. The advantage is concentrated in,
and clearest during, periods of market stress, where the functional
model is the only specification whose violations are neither too
frequent nor predictable from their own history; competing models
conditioned on daily returns alone are rejected in this regime, most
commonly because their violations cluster in the days following a large
loss rather than anticipating it. Since supervisory backtesting
frameworks penalize clustered exceedances more severely than isolated
ones, this timing advantage, rather than a uniformly deeper estimated
tail, constitutes the paper's central empirical finding.

More broadly, these results suggest that the value of high-frequency
functional information for tail risk is conditional on the market regime
and should be assessed jointly with, rather than as a substitute for,
measures of average forecast accuracy. The intraday return curve
registers a shift in market conditions before that shift is reflected in
the closing price, which is precisely the information a filter driven
only by past daily returns cannot access until the following
observation.

Three directions appear promising for future research. First, the
analysis is conducted under a stationary framework, whereas financial
markets exhibit evolving dynamics over time; extending the methodology to
locally stationary processes would allow both the conditional operators
and the tail behavior of the innovations to vary smoothly, capturing
regime changes in volatility and extreme risk directly rather than
through a rolling re-estimation scheme. Second, the functional covariate
enters the model through a fixed number of leading principal components,
ranked by the variance they explain in the intraday curve; our empirical
results show that this ranking need not align with predictive relevance,
since a single additional component explaining little variance is
necessary for adequate performance in stress periods while further
components of even lower rank are actively harmful. This non-monotonic
sensitivity motivates a functional single-index structure, in which the
relevant direction of variation is estimated jointly with the conditional
mean and scale operators rather than fixed in advance by a
variance-based truncation rule. Third, because the timing advantage
documented here is complementary to the average-accuracy advantage of
parametric volatility filters, a hybrid specification in which a
GARCH-type filter is modulated by an intraday functional covariate may
combine the strengths of both approaches and merits formal development.

\section*{Proofs}
\section*{Appendix A: Preliminary results}
Throughout this section, let us denote by  $\phi(r_1, r_2) = \mathbb{P}(r_2\leq \rho(X, x)\leq r_1)$ , the small ball probability function and we will denote by $C$ and $C_0$ some strictly positive generic constants
In order to prove the main results of this paper, we introduce additional notations and necessary Lemmas. Let $j \in \{1, 2 \} $ and for $l\in \{1, 2 \}$. Let
\begin{eqnarray}
  m^{[l]}_{n,j}(x) &=& \frac{1}{n\mathbb{E}(K_{1}(x)\Gamma_{1,l}(x))}\sum_{s=1}^{n}Y^{j-1}_{s} K_{s}(x)\Gamma_{s,l}(x),\label{*} \\
  \overline{m}^{[l]}_{n,j}(x) &=& \frac{1}{n\mathbb{E}(K_{1}(x)\Gamma_{1,l}(x))}\sum_{s=1}^{n}\mathbb{E}[Y^{j-1}_{s} K_{s}(x)\Gamma_{s,l}(x)|\F_{s-1}],\notag \label{**}
\end{eqnarray}
and
\begin{equation}\label{Vn0}
V^{[l]}_{n}(x)=\frac{1}{n \mathbb{E}\left[K_{1}(x)\Gamma_{1,l}(x)\right]} \sum_{s=1}^{n} \sqrt{U(X_{s})}\epsilon_{s}K_{s}(x)\Gamma_{s,l}(x)\text{.}\color{black}
\end{equation}
where $K_s(x)$ and $\Gamma_{s,l}(x)$  are defined in (\ref{rK}) and (\ref{VS}), respectively. 
\color{black}
\begin{lem} [\cite{LL16}]\label{Lemma 4.4}
Let $(Z_{n})_{n\geq 1} $ be a sequence of martingale differences with respect to the sequence of $\sigma$-fields $\F_{n}=\{\sigma(Z_{1},\ldots, Z_{n}):
 n \geq 1 \} $, where $\sigma (Z_{1},\ldots, Z_{n})$ is the $\sigma$-field generated by the random variables $Z_{1},\ldots, Z_{n}$. Set $S_{n}=\displaystyle\sum_{t=1}^{n}Z_{t}$. Assume that there exist for any $ t \in \mathbb{N}$ some nonnegative constants $C$ and $d_{t}$ such that 
 $\mathbb{E}(M_{t}^{p}| \F_{t-1}) \leq M^{p-1}p!d_{t}^{2} $
almost surely.\\ Then, for any $\epsilon >0$,
\begin{equation*}
\mathbb{P}(|S_{n}|>\epsilon)\leq 2\exp\left\lbrace\frac{-\epsilon^{2}}{(4D_{n}+2C\epsilon)}\right\rbrace,
\end{equation*}
where $D_{n}=\sum_{t=1}^{n}d_{t}^{2}.$
\end{lem}
\color{black}

\begin{lem}\label{Lemma 4.5}
Suppose that Assumptions~\ref{A1}, \ref{SBP}(iii), and \ref{SMC}(i) hold true. For all $(k,i) \in \mathbb{N}^{*}\times \mathbb{N}$, $1 \leq j \leq 2 + \kappa$ and $l \in \{0, 1\}$, we have:
\begin{itemize}
    \item[(i)] $\mathbb{E}[K_s^j |\rho_s|^i | \mathcal{F}_{s-1}] = M_j h^i \phi(h) + o_{\text{a.s.}}(h^i \phi(h)),$
    \item[(ii)] $\mathbb{E}[\Gamma_{s,l}^j K_s^j | \mathcal{F}_{s-1}] = M_j (n h^2)^j \phi(h) + o_{\text{a.s.}}((n h^2)^j \phi(h)),$
    \item[(iii)] $\mathbb{E}[\Gamma_{1,l} K_1] = M_1 (n h^2) \phi(h) + o_{\text{a.s.}}(n h^2 \phi(h)).$
\end{itemize}
\end{lem}
\begin{proof}
{Proof of part (i):}  Based on Assumptions~\ref{A1}(i) and \ref{SMC}(i), we have
\begin{equation*}
\mathbb{E}\left[K_s^j |\rho_s|^i \mid \mathcal{F}_{s-1}\right] \leq \mathbb{E}\left[K_s^j h^i \mid \mathcal{F}_{s-1}\right] \leq C h^i \mathbb{E}\left[K_s^j \mid \mathcal{F}_{s-1}\right].
\end{equation*}
By applying \cite{LL10},  we obtain
\begin{equation*}
\mathbb{E}\left[K_s^j |\rho_s|^i \mid \mathcal{F}_{s-1}\right] = M_j h^i \phi(h) + \mathcal{O}_{\text{a.s.}}(h^i \phi(h)).
\end{equation*}
{Proof of part (ii):}  First, observe that:
\begin{equation*}
\mathbb{E}\left[K_s^j \Gamma_{s,l}^j \mid \mathcal{F}_{s-1}\right] \leq 
 \mathbb{E} \left(\left( \sum_{t=1}^n \rho_t^2 K_t \right)^j K_s^j \mid \mathcal{F}_{s-1} \right)   +  \mathbb{E} \left(\left( \sum_{t=1}^n |\rho_t| K_t \right)^j \rho_s^j K_s^j \mid \mathcal{F}_{s-1} \right).
\end{equation*}
In view of Assumptions~\ref{A1} and \ref{SMC}(i), we obtain
\begin{equation*}
\mathbb{E}\left[K_s^j \Gamma_{s,l}^j \mid \mathcal{F}_{s-1}\right] \leq C' n^j h^{2j} \mathbb{E}\left[K_s^j \mid \mathcal{F}_{s-1}\right] + C' n^j h^j \mathbb{E}\left[\rho_s^j K_s^j \mid \mathcal{F}_{s-1}\right].
\end{equation*}
Making use of the result from part (i)  it follows that
\begin{equation}
\mathbb{E}\left[K_s^j \Gamma_{s,l}^j \mid \mathcal{F}_{s-1}\right] = M_j (n h^2)^j \phi(h) + \mathcal{O}_{\text{a.s.}}((n h^2)^j \phi(h)).
\end{equation}
{Proof of part (iii):} 
By combining the Assumption~\ref{SBP}(iii)  with the result obtained in part (ii) for $j=1$, it follows
\begin{equation*}
\mathbb{E}\left[\Gamma_{1,l} K_1\right] = M_1 (n h^2) \phi(h) + \mathcal{O}_{\text{a.s.}}(n h^2 \phi(h)).
\end{equation*}
This completes the proof of the Lemma.
\end{proof} 
\color{black}
\begin{lem}\label{Lemma 7.1.}
Suppose that Assumptions~\ref{A1}, \ref{SBP}(i)-(iii), \ref{SMC}, \ref{bdw} hold true. Moreover, assume that $\eta=\eta_n=o(h/n)$ and condition (\ref{condunifconv}) are satisfied.   
Then, 
\begin{equation}\label{sup1}
\underset{n\rightarrow\infty}{\lim}\sup_{x\in \mathcal{C}}\left|m^{[l]}_{n,1}(x)-1\right|=0 \quad  a.s. \quad \text{for} \quad l\in \{0,1\}. 
\end{equation}
\end{lem}

\begin{proof}
Observe that
\begin{equation}\label{sup****}
\sup_{x\in \mathcal{C}}| m^{[l]}_{n,1}(x)-1|\leq\sup_{x\in \mathcal{C}}|m^{[l]}_{n,1}(x)-\overline{m}^{[l]}_{n,1}(x)\color{black}|+\sup_{x\in \mathcal{C}}|\overline{m}^{[l]}_{n,1}(x)-1|.
\end{equation}
Then, the first term in right-hand side of (\ref{sup****}) could be decomposed as follows.  For $\eta >0$ and $B(c_{k}, \eta):=\{x \in \mathcal{C}: \varrho(x, c_{k})<\eta \}$, we have
\begin{eqnarray}\label{decomp1}
\displaystyle{\sup_{x\in \mathcal{C}}}\left|m^{[l]}_{n,1}(x)-\overline{m}^{[l]}_{n,1}(x)\right|&\leq& \displaystyle{\max_{1\leq k \leq N(\eta, \mathcal{C}, \varrho)}\sup_{x \in B(c_{k}, \eta)}}\left|m^{[l]}_{n,1}(x)-\overline{m}^{[l]}_{n,1}(x) \right| \nonumber\\
& \leq&  \displaystyle{\max_{1\leq k \leq N(\eta, \mathcal{C}, \varrho)}\sup_{x \in B(c_{k}, \eta)}}\left|m^{[l]}_{n,1}(x)-m^{[l]}_{n,1}(c_{k}) \right|+\displaystyle{\max_{1\leq k \leq N(\eta, \mathcal{C}, \varrho)}}\left|m^{[l]}_{n,1}(c_{k})-\overline{m}^{[l]}_{n,1}(c_{k}) \right|\nonumber\\
& & \quad +\displaystyle{\max_{1\leq k \leq N(\eta, \mathcal{C}, \varrho)}\sup_{x \in B(c_{k}, \eta)}}\left|\overline{m}^{[l]}_{n,1}(x)-\overline{m}^{[l]}_{n,1}(c_{k}) \right|  \nonumber\\
&=: &\Upsilon
_{1}+\Upsilon
_{2}+\Upsilon
_{3}.
\end{eqnarray}
We start with studying the term $\Upsilon
_{2}.$
Observe that, for any $x \in B(c_{k}$, $\eta)$ and making use of the definition of $m^{[l]}_{n,1}(x)$ in \eqref{*}, one gets
$$\begin{aligned}
m^{[l]}_{n,1}(x)-m^{[l]}_{n,1}(c_{k})&=\frac{1}{n \mathbb{E}(K_{1}(x)\Gamma_{1,l}(x))}\displaystyle{\sum _{s=1}^{n}}K_{s}(x)\Gamma_{s,l}(x)-\frac{1}{n \mathbb{E}(K_{1}(c_{k})\Gamma_{1,l}(c_{k}))}\displaystyle{\sum _{s=1}^{n}}K_{s}(c_{k})\Gamma_{s,l}(c_{k})\end{aligned}$$ 
By using Lemma \ref{Lemma 4.5}, it follows that  
$$\begin{aligned}
\Upsilon_{1}&\leq \frac{C}{n^2h^2\phi(h)}\sup_{x \in B(c_{k}, \eta)}\displaystyle{\sum _{s=1}^{n}}|\Gamma_{s,l}(x)||K_{s}(x)-K_{s}(c_{k})|+\frac{C}{n^2h^2\phi(h)}\sup_{x \in B(c_{k}, \eta)}\displaystyle{\sum _{s=1}^{n}}|K_{s}(c_{k})||\Gamma_{s,l}(x)-\Gamma_{s,l}(c_{k})|\\
& := \mathcal{Q}_1+\mathcal{Q}_2.
\end{aligned}$$
Firstly, let us check the term $\mathcal{Q}_1$. Based on the definition \ref{VS} with the use of Lemma \ref{Lemma 4.5}, it yields 

\begin{equation}\label{BondGamma}
|\Gamma_{s,l}(x)|\leq Cnh^2.
\end{equation}
Then, in view of condition \ref{A1} which suppose that the kernel K is Lipschitzian and by using the above equation, one gets
\begin{eqnarray}
\mathcal{Q}_1 &\leq \displaystyle \frac{Cnh^2}{n^2h^2\phi(h)}\sup_{x \in B(c_{k}, \eta)}\displaystyle{\sum _{s=1}^{n}}|K_{s}(x)-K_{s}(c_{k})| \notag \\
& \leq  \displaystyle\frac{Cnh^2n\eta}{n^2h^2\phi(h)h}:= \frac{C\eta}{\phi(h)h}.\label{Q1}
\end{eqnarray}
Regarding the term $\mathcal{Q}_2$. Based on the definition \ref{VS}, we have  \\
\begin{small}
$\begin{aligned}
\Gamma_{s,l}(x)-\Gamma_{s,l}(c_k)&=\displaystyle  \left(\sum_{t=1}^n  \rho^2_t(x)K_t(x)\right)-\left(\sum_{t=1}^n \rho_t(x)K_t(x)\displaystyle\right)\displaystyle\rho_s(x) -\left(\sum_{t=1}^n  \rho^2_t(c_k)K_t(c_k)\right)\\
&+\left(\sum_{t=1}^n \rho_t(c_k)K_t(c_k)\displaystyle\right)\displaystyle\rho_s(c_k).  
\end{aligned}$
\end{small}
\\
Thus
$$\begin{aligned}
|\Gamma_{s,l}(x)-\Gamma_{s,l}(c_k)|&
\leq \displaystyle \sum_{t=1}^n \left|\rho^2_t(x)K_t(x)- \rho^2_t(c_k)K_t(c_k) \right|+\displaystyle \sum_{t=1}^n \left|\rho_t(c_k)K_t(c_k) \rho_s(c_k) -\rho_t(x)K_t(x)\displaystyle \displaystyle\rho_s(x)
\right|\\
&\leq |\rho_s(x)| \sum_{t=1}^n |K_t(x)| |\rho_t(x)-\rho_t(c_k)| + |\rho_s(x)-\rho_s(c_k)| \sum_{t=1}^n |\rho_t(c_k)| |K_t(x)| \\
&\quad + \sum_{t=1}^n \left( |\rho_s(c_k)||\rho_t(c_k)|+ \rho^2_t(c_k)\right)|K_t(x)-K_t(c_k)| + \sum_{t=1}^n |K_t(x)| |\rho^2_t(x)-\rho^2_t(c_k)| \\
\\&:=\mathcal{J}_{1}+\mathcal{J}_{2}+\mathcal{J}_{3}+\mathcal{J}_{4}.
\end{aligned}$$

Firstly, by using the boundedness of $K$, Assumption~\ref{SMC}(i) with the Lipschitz condition on $\rho$, it follows that,    
\begin{equation}\label{J1}
\mathcal{J}_{1}\leq Cn \eta h
\end{equation}
 and 
\begin{equation}\label{J2}
\mathcal{J}_{2} \leq Cn \eta h.
\end{equation}
 Then, making use of Assumption~\ref{SMC}(i) with the fact that the kernel $K$ is Lipschitz, it yields 
\begin{equation}\label{J3}
\mathcal{J}_{3} \leq Cn \eta h.
\end{equation}
Regarding the term $\mathcal{J}_{4}$, in view of Assumptions~\ref{A1} and \ref{SMC}(i), we obtain, 
\begin{equation}\label{J4}
\mathcal{J}_{4} \leq Cn \eta h.
\end{equation}
Therefore, we get 
\begin{equation}\label{termA}
|\Gamma_{s,l}(x)-\Gamma_{s,l}(c_k)|\leq Cnh\eta.
\end{equation}

In addition, by  the the boundedness of the kernel $K$, it follows that 
\begin{eqnarray}
\mathcal{Q}_2 & \displaystyle\leq \frac{C}{n^2h^2\phi(h)}\sup_{x \in B(c_{k}, \eta)}\displaystyle{\sum _{s=1}^{n}}|\Gamma_{s,l}(x)-\Gamma_{s,l}(c_{k})| \notag \\
& \displaystyle \leq \frac{Cnhn\eta}{n^2h^2\phi(h)}:= \frac{C\eta}{\phi(h)h}.\label{Q2*}
\end{eqnarray}
Hence, by taking $\eta=\eta_n=(h/n)$ with the use Assumption~\ref{bdw}(i), it yields
\begin{equation}\label{ups1}
\Upsilon_1=o(1).
\end{equation}
Concerning the term $\Upsilon_3$, we follow the same steps used in studying the term $\Upsilon_1$, it results
\begin{equation}\label{ups3}
\Upsilon_3=o(1).
\end{equation}
Now, we turn to the term $\Upsilon_2$. Let

\begin{equation*}
 m^{[l]}_{n,1}(c_{k})-\overline{m}^{[l]}_{n,1}(c_{k})=\frac{1}{n\mathbb{E}(K_{1}(c_{k})\Gamma_{1,l}(c_{k}))}\sum _{s=1}^{n}L_{n,s}(c_{k}),
\end{equation*}
where $L_{n,s}(c_{k})=K_{s}(c_{k})\Gamma_{s,l}(c_{k})-\mathbb{E}(K_{s}(c_{k})\Gamma_{s,l}(c_{k})|\F_{s-1})$ is a martingale difference with respect to the $\sigma$-field $\F_{s-1}$, for $s= 1, 2,\ldots, n.$. Then, we have
$$\begin{aligned}
\mathbb{E}(L^{2}_{n,s}(c_{k})|\mathcal{F}_{s-1})
&\leq 2\mathbb{E}(K^{2}_{s}(c_{k})\Gamma^2_{s,l}(c_{k})|\F_{s-1}) \\
& \leq 2 \mathbb{E}(K^{2}_{s}(c_{k})\Gamma^2_{s,l}(c_{k})|\F_{s-1}). \\
\end{aligned}$$
Then, making use of Lemma (\ref{Lemma 4.5}), it follows that
\begin{equation}
\mathbb{E}(L^{2}_{n,s}(c_{k})|\mathcal{F}_{s-1}) \leq 2 Cn^2 h^4 \phi_s(h)=:d^2_s.
\end{equation}
In addition, the use of Assumption~\ref{SBP}(iii) permits us to find $D_n=2 C n^3 h^4 \phi(h)$. \\  
Now, we can use Lemma~\ref{Lemma 4.4}, it follows that 
$$\begin{aligned}
\mathbb{P}\left(\Upsilon_2>\overline{\lambda}_{n}\right)& \leq \displaystyle{\sum_{k=1}^{N(\eta, \mathcal{C}, \varrho)}}\mathbb{P}\left(\left|m^{[l]}_{n,1}(c_{k})-\overline{m}^{[l]}_{n,1}(c_{k})\right|>\overline{\lambda}_{n}\right)\\
& \leq \displaystyle{\sum_{k=1}^{N(\eta, \mathcal{C}, \varrho)}}\mathbb{P}\left(\left|\sum_{s=1}^n L_{n, s}(c_{k})\right|> \overline{\lambda}_{n} n\mathbb{E}[K_{1}(c_{k})\Gamma_{1,l}(c_{k})]\right)\\
&\leq 2N(\eta, \mathcal{C}, \varrho) \exp \left( -\frac{\left(\overline{\lambda}_{n} n\mathbb{E}[K_{1}(c_{k})\Gamma_{1,l}(c_{k})]\right)^{2}}{2D_{n}+2C \overline{\lambda}_{n} n\mathbb{E}[K_{1}(c_{k})\Gamma_{1,l}(c_{k})] } \right) \\
&\leq 2N(\eta, \mathcal{C}, \varrho) \exp \left( -\frac{Cn^4h^4\phi^{2}(h)\overline{\lambda}_{n}^{2}}{ 
2Cn^3h^4\phi(h)+2Cn^2h^2\phi(h)\overline{\lambda}_{n} } \right)\\
& \leq 2N(\eta, \mathcal{C}, \varrho) \exp \left( -\frac{Cn^4h^4\phi^{2}(h)\overline{\lambda}_{n}^{2}}{2Cn^3h^4\phi(h)\left(1+\frac{\overline{\lambda}_{n}}{nh^2}\right)} \right)\\
& \leq 2\exp \left( -\mathcal{O}(n\phi(h))\overline{\lambda}_{n}^{2} \left[1-\frac{\log N(\eta, \mathcal{C}, \varrho) }{\mathcal{O}(n\phi(h))\overline{\lambda}_{n}^{2}}\right]\right),
\end{aligned}
$$ 
where $\overline{\lambda}_{n}=\sqrt{\frac{\log n}{n\phi(h)}}\ell_n^{-1}$, $\lambda_{n}$ and $\ell_n$ are defined in Theorem \ref{T1}. Hence,  using condition (\ref{condunifconv}), one gets
$$
\displaystyle{\sum_{n\geq 1}}\mathbb{P}\left( \Upsilon_2>\overline{\lambda}_{n}\right)<\infty.
$$ 
Then, using Borel-Cantelli Lemma we have 
\begin{equation}\label{sup21++}
 \Upsilon_2= \displaystyle{\max_{1\leq k \leq N(\eta, \mathcal{C}, \varrho)}}\left|m_{n,1}^{[l]}(c_{k})-\overline{m}_{n,1}^{[l]}(c_{k})\right|= \mathcal{O}_{a.s.}(\overline{\lambda}_{n})=o_{a.s.}(1).
 \end{equation}
Combining \eqref{decomp1} with \eqref{ups1}, \eqref{ups3} and \eqref{sup21++}, one deduces that the first term in the right-hand side of inequality \eqref{sup****} converges to zero as $n$ goes to infinity. 
\\
Regarding the second term $\displaystyle\sup_{x\in \mathcal{C}}|\overline{m}^{[l]}_{n,1}(x)-1|$, we have
$$\begin{aligned}
\overline{m}^{[l]}_{n,1}(x)-1 &=\frac{1}{n \mathbb{E}( K_1(x)\Gamma_{1,l}(x))} \sum_{s=1}^n \mathbb{E}\left[ K_s(x)\Gamma_{s,l}(x) \mid \F_{s-1}\right]-1 \\
&=\frac{1}{n \mathbb{E} (K_{1}(x)\Gamma_{1,l}(x))} \sum_{s=1}^n \left( \mathbb{E}\left[K_{s}(x)\Gamma_{s,l}(x) \mid \F_{s-1}\right] - \mathbb{E}[K_{1}(x)\Gamma_{1,l}(x)] \right).
\end{aligned}$$
Finally, Lemma \ref{Lemma 4.5} (ii)-(iii) along with Assumption~\ref{SBP}(iii) allow to conclude the proof of this Lemma.
 \end{proof}

 \begin{lem}\label{Lemmacvn}
Suppose that Assumptions~\ref{A1}, \ref{SBP}(i)-(iii), \ref{errors}(i), \ref{SMC}, \ref{bdw}(i) and condition (\ref{condunifconv}) are satisfied. Then, we have 
 \begin{equation*}
\displaystyle{\sup_{x\in \mathcal{C}}}\left|V^{[l]}_{n}(x)\right|=\mathcal{O}_{a.s.}\left(\sqrt{\frac{\log n}{n\phi(h)}}\right) \,\ \hbox{as} \quad n \rightarrow \infty \text{,} \quad \text{for} \quad l \in \{0,1\}.
\end{equation*}
\end{lem}
 \begin{proof} Let 
\begin{equation}\label{DecVn0}
V^{[l]}_{n}(x)=\frac{1}{n \mathbb{E}\left[K_{1}(x)\Gamma_{1,l}(x)\right]} \sum_{s=1}^{n} \mathcal{L}_{s} K_{s}(x)\Gamma_{s,1}(x)= \left(  V_{n}(x)-\overline{V}_{n}^{[l]}(x)  \right)  + \widetilde{V}_{n}(x) + \widehat{V}_{n}^{[l]}(x),
\end{equation}
where \\
 $ \overline{V}_{n}^{[l]}(x)=\displaystyle{\frac{1}{n \mathbb{E}\left[K_{1}(x)\Gamma_{1,l}(x)\right]} \sum_{s=1}^{n} \mathcal{L}_{s} \1_{\left(|\mathcal{L}_{s}|\leq\ell_{n}\right)} K_{s}(x)\Gamma_{s,l}(x)}, \\
\widetilde{V}_{n}^{[l]}(x)= \displaystyle{\frac{1}{n \mathbb{E}\left[K_{1}(x)\Gamma_{1,l}(x)\right]} \sum_{s=1}^{n}\mathbb{E}( \mathcal{L}_{s} \1_{\left(|\mathcal{L}_{s}|\leq\ell_{n}\right)}\mid \mathcal{F}_{s-1}) K_{s}(x)\Gamma_{s,l}(x)},
$ \\
$\widehat{V}_{n}^{[l]}(x)=\displaystyle{\frac{1}{n \mathbb{E}\left[K_{1}(x)\Gamma_{1,l}(x)\right]} \sum_{s=1}^{n} \left(  \mathcal{L}_{s} \1_{\left(|\mathcal{L}_{s}|\leq\ell_{n} \right)}-\mathbb{E}( \mathcal{L}_{s} \1_{\left(|\mathcal{L}_{s}|\leq\ell_{n}\right)}\mid \mathcal{F}_{s-1})\right)\color{black} K_{s}(x)\Gamma_{s,l}(x)},$ \\ such that $\mathcal{L}_{s}=\sqrt{U(X_{s})}\epsilon_{s}$  and $\ell_{n}$ is as defined in Theorem \ref{T1}.
\\
Lemma \ref{LemmaA} allows to conclude that the first term in (\ref{DecVn0}) equals zero almost surely as $n\rightarrow \infty.$ Moreover, making use of Lemma \ref{LemmaB+}  and Lemma \ref{LemmaB} and below, the second term is $\mathcal{O}\left(\{\ell^{\chi-1}_{n}n^{\nu}h^{2\nu}\phi(h)^{\nu}\}^{-1}\right)$ where $\nu=(\chi-1)/\chi,$ 
$\nu$ is defined in \eqref{T1}  and the third term is $\mathcal{O}_{a.s.}\left(\sqrt{\frac{\log n}{n\phi(h)}}\right),$ respectively. Finally, since 
$\left(\sqrt{\frac{\log n}{n\phi(h)}}\ell^{\chi-1}_{n}n^{\nu}h^{2\nu}\phi(h)^{\nu}\right)^{-1}
=o(1)$ for n large enough, the proof of this lemma is achieved.
 \end{proof}
 \begin{lem}\label{LemmaA}  Assume that $\left(X_{s}, \epsilon_{s}\right)$ is a strictly stationary ergodic process and suppose that Assumption~\ref{errors}(i)  hold. Then, for each $\omega$ outside a null set $D$, there exists a positive integer $n_{0}(\omega)$ such that $V^{[l]}_{n}(x)=\overline{V}_{n}^{[l]}(x)$ for $n \geq n_{0}(\omega)$, $l\in \{0,1\} $ and $x \in \mathcal{E}$.
\end{lem}
\begin{proof}
The proof follows along the same lines as the proof of Proposition 3.1 in \citep{Roussas1990}.
\end{proof}
\begin{lem}\label{LemmaB+} Suppose that Assumptions~\ref{A1}, \ref{SBP}(i)-(iii), \ref{errors}(i), 
\ref{SMC}, \ref{bdw}(i) and condition (\ref{condunifconv}) are satisfied. Then, we have, as $n \rightarrow \infty,$
\begin{equation*}
\displaystyle{\sup_{x\in \mathcal{C}}}|\widehat{V}_{n}^{[l]}(x)| =\mathcal{O}_{a.s.}\left(\sqrt{\frac{\log n}{n\phi(h)}}\right).
\end{equation*}
\end{lem}
\begin{proof}
Observe that

$$
\sup _{x \in \mathcal{C}}\left|\widehat{V}_{n}^{[l]}(x)\right| \leq \max_{1 \leq k \leq N\left(\eta, \mathcal{C}, \varrho\right)} \displaystyle\sup_{x \in B\left(c_{k}, \eta\right)}\left|\widehat{V}_{n}^{[l]}(x)-\widehat{V}_{n}^{[l]}\left(c_{k}\right)\right|+\max _{1 \leq k \leq N\left(\eta, \mathcal{C}, \varrho\right)}\left|\widehat{V}_{n}^{[l]}\left(c_{k}\right)\right| .
$$

We can easily show that

$$\begin{aligned}
\widehat{V}_{n}^{[l]}(x)-\widehat{V}_{n}^{[l]}\left(c_{k}\right)&=\frac{1}{n \mathbb{E}(K_{1}(x)\Gamma_{1,l}(x))}\displaystyle{\sum _{s=1}^{n}}\mathcal{L}_{s}^{-}K_{s}(x)\Gamma_{s,l}(x)-\frac{1}{n \mathbb{E}(K_{1}(c_{k})\Gamma_{1,l}(c_{k}))}\displaystyle{\sum _{s=1}^{n}}\mathcal{L}_{s}^{-}K_{s}(c_{k})\Gamma_{s,l}(c_{k})\end{aligned}$$ 
Making use of Lemmas \ref{Lemma 4.5} and \ref{LemmaA}, it follows that  
\begin{small}
$$\begin{aligned}
|\widehat{V}_{n}^{[l]}(x)-\widehat{V}_{n}^{[l]}\left(c_{k}\right)|&\leq \frac{2C\ell_n}{n^2h^2\phi(h)}\sup_{x \in B(c_{k}, \eta)}\displaystyle{\sum _{s=1}^{n}}|\Gamma_{s,l}(x)||K_{s}(x)-K_{s}(c_{k})|
\\
&+\frac{2C\ell_n}{n^2h^2\phi(h)}\sup_{x \in B(c_{k}, \eta)}\displaystyle{\sum _{s=1}^{n}}|K_{s}(c_{k})||\Gamma_{s,l}(x)-\Gamma_{s,l}(c_{k})|\\
& \leq 2\ell_n (\mathcal{{Q}}_1+\mathcal{{Q}}_2)
\end{aligned}$$
\end{small}
Then, using results given in (\ref{Q1}) and (\ref{Q2*}), it follows that
$$\widehat{V}_{n}^{[l]}(x)-\widehat{V}_{n}^{[l]}\left(c_{k}\right) \leq \frac{C\ell_n\eta}{\phi(h)h}. \label{Q2}   $$
Therefore, by taking $\eta=\eta_n=(h/n\ell_n)$  with the use Assumption~\ref{regularityconditions}(i), it results
$$
\max _{1 \leq k \leq N\left(\eta, \mathcal{C}, \varrho\right)} \sup _{x \in B\left(c_{k}, \eta\right)}\left|\widehat{V}_{n}^{[l]}(x)-\widehat{V}_{n}^{[l]}\left(c_{k}\right)\right|=o(1) .
$$

Now, we turn our attention to the term $\widehat{V}_{n}^{[l]}\left(c_{k}\right)$. Observe that

$$
\widehat{V}_{n}^{[l]}\left(c_{k}\right)=\frac{1}{n} \sum_{s=1}^{n} \frac{K_{s}\left(c_{k}\right)\Gamma_{s,l}(c_{k}) \mathcal{L}_{s}^{-}}{\mathbb{E}\left\{K_{1}\left(c_{k}\right)\Gamma_{1,l}(c_{k})\right\}}=\frac{1}{n} \sum_{s=1}^{n} T_{s}\left(c_{k}\right),
$$

where $T_{s}\left(c_{k}\right)=K_{s}\left(c_{k}\right)\Gamma_{s,l}(c_{k}) \mathcal{L}_{s}^{-} / \mathbb{E}\left\{K_{1}\left(c_{k}\right)\Gamma_{1,l}(c_{k})\right\}$ is a martingale difference, for any $1 \leq k \leq N\left(\eta, \mathcal{C}, d_{\mathcal{C}}\right)$. The convergence of the term $\widehat{V}_{n}^{[l]}\left(c_{k}\right)$ will be treated by the exponential inequality for a sum of martingale differences; see Lemma \ref{Lemma 4.4}. Now, let us check the condition under which we can apply the exponential inequality. For this purpose, we observe that

$$
\begin{aligned}
\mathbb{E}\left\{T_{s}^{2}\left(c_{k}\right) \mid \mathcal{F}_{s-1}\right\} & =\frac{1}{\left[\mathbb{E}\left\{K_{1}\left(c_{k}\right)\Gamma_{1,l}(c_{k})\right\}\right]^{2}} \mathbb{E}\left\{K_{s}^{2}\left(c_{k}\right)\Gamma^2_{s,l}(c_{k})\left(\mathcal{L}_{s}^{-}\right)^{2} \mid \mathcal{F}_{s-1}\right\} 
\end{aligned}
$$
By applying Lemmas \ref{Lemma 4.5} and \ref{LemmaA}, it follows that 
$$
\begin{aligned}
\mathbb{E}\left\{T_{s}^{2}\left(c_{k}\right) \mid \mathcal{F}_{s-1}\right\} & \leq \frac{2 \ell_{n}^{2}\mathcal{O}\left( n^2h^4\phi(h)\right)}{\mathcal{O}\left(n^2h^4\phi^2\left(h\right)\right)} \\
&\leq C'\displaystyle \frac{2 \ell_{n}^{2}}{\phi\left(h\right)}.
\end{aligned}
$$
Consequently, we obtain $d_{s}^{2}=2C'\ell_{n}^{2}/\phi\left(h\right).$ Therefore, we are now able  to apply Lemma \ref{Lemma 4.4}.\\
Thus, for $n^{-1}Dn=2C'\ell_{n}^{2}/\phi\left(h\right)$, we have

$$
\begin{aligned}
\operatorname{Pr}\left\{\max _{1 \leq k \leq N\left(\eta, \mathcal{C}, \varrho\right)}\left|\widehat{V}_{n}^{[l]}\left(c_{k}\right)\right|>  \sqrt{\frac{\log n}{n\phi(h)}}  \right\}& \leq 2 N\left(\eta, \mathcal{C}, \varrho\right) \exp \left\{ \frac{-n \left(\frac{\log n}{n\phi(h)}\right)}{ (2D_n)/n +2C\lambda  }\right\} \\
&\leq 2 N\left(\eta, \mathcal{C}, \varrho\right) \exp \left\{ \frac{-n \left(\frac{\log n}{n\phi(h)}\right)}{ (2D_n)/n}\right\} 
\\ & \leq 2 N\left(\eta, \mathcal{C}, \varrho\right) \exp \left\{- \frac{\log n}{n\phi(h)} \mathcal{O}(n\phi\left(h\right)) \ell_{n}^{-2} \right\} \\
& =2 \exp \left\{ -\log n \ell_{n}^{-2} C+\ln N\left(\eta, \mathcal{C}, \varrho\right)\right\} \\
& =2 \exp \left[ -\log n \ell_{n}^{-2}\left\{C-\frac{\ln N\left(\eta, \mathcal{C}, \varrho\right)}{\log n \ell_{n}^{-2}}\right\}\right] .
\end{aligned}
$$
Therefore, by (\ref{condunifconv}), we get
\begin{equation}
\displaystyle{\sum_{n\geq 1}}\mathbb{P}\left( \max _{1 \leq k \leq N\left(\eta, \mathcal{C}, \varrho\right)}\left|\widehat{V}_{n}^{[l]}\left(c_{k}\right)\right|>\sqrt{\frac{\log n}{n\phi(h)}}\right)<\infty.
\end{equation}
Using Borel-Cantelli Lemma, we can deduce that


$$
\max _{1 \leq k \leq N\left(\eta, \mathcal{C}, \varrho\right)}\left|\widehat{V}_{n}^{[l]}\left(c_{k}\right)\right|=\mathcal{O}_{\text {a.s. }}\left(\sqrt{\frac{\log n}{n\phi(h)}}\right) .
$$

This completes the proof.
\end{proof}

\color{black}
\begin{lem}\label{LemmaB}
Under Assumptions~\ref{A1}, \ref{A1}(i)-(iii), \ref{errors}(i), \ref{SMC}(i) and \ref{bdw}(i), we have, as $n \rightarrow \infty,$
\begin{equation*}
\displaystyle{\sup_{x\in \mathcal{C}}}|\widetilde{V}_{n}^{[l]}(x)| = \mathcal{O}\left(\{\ell^{\chi-1}_{n}\phi(h)^{\chi-1/\chi}\}^{-1}\right).
\end{equation*}
\end{lem}
\color{black}
\begin{proof}
Let $a>1$ and $b>1$ be real numbers such that $1 / a+1 / b=1$ and the H\"{o}lder's and Markov's inequalities, we can get
$$
\begin{aligned}
\left|\mathbb{E}\left\{\mathcal{L}_{s} \1_{\left(\left|\mathcal{L}_{s}\right|  \leq \ell_{n}\right)} \mid \mathcal{F}_{s-1}\right\}\right| & \leq \mathbb{E}^{1 / a}\left(\left|\mathcal{L}_{s}\right|^{a} \mid \mathcal{F}_{s-1}\right) \times \mathbb{E}^{1 / b}\left\{\1_{\left(\left|\mathcal{L}_{s}\right| \leq \ell_{n}\right)} \mid \mathcal{F}_{s-1}\right\} \\
& \leq \mathbb{E}^{1 / a}\left(\left|\mathcal{L}_{s}\right|^{a} \mid \mathcal{F}_{s-1}\right) \times \frac{\mathbb{E}^{1 / b}\left(\left|\mathcal{L}_{s}\right| \mid \mathcal{F}_{s-1}\right)}{\ell_{n}^{1 / b}} \\
& \leq \frac{\mathbb{E}\left(\left|\mathcal{L}_{s}\right|^{a} \mid \mathcal{F}_{s-1}\right)}{\ell_{n}^{a / b}}=\frac{\mathbb{E}\left(\left|\sqrt{U\left(X_{s}\right)}\right|^{a}\left|\epsilon_{s}\right|^{a} \mid \mathcal{F}_{s-1}\right)}{\ell_{n}^{a / b}}.
\end{aligned}$$

Then, making use of Assumption~\ref{errors}(i),  the ergodicity of $X_t$ and by setting $a=\chi$ and $b=\chi /(\chi-1)$ such that $1/a + 1/b = 1$, we obtain
\color{black}
$$
\mathbb{E}\left\{\mathcal{L}_{s} \1_{\left(\left|\mathcal{L}_{s}\right| \leq \ell_{n}\right)} \mid \mathcal{F}_{s-1}\right\}=\mathcal{O}\left[\mathbb{E}\left\{ \left|\sqrt{U\left(X_{s}\right)}\right|^{\chi} \mid \mathcal{F}_{s-1}\right\} / \ell_{n}^{\chi-1}\right].
$$
Observe that another use of the H\"{o}lder inequality allows us to get
\color{black} 
$$\begin{small}
\begin{aligned}
\left|\widetilde{V}_{n}^{[l]}(x)\right| & \leq \frac{\ddot{C}}{\ell_{n}^{\chi-1\color{black}}} \times\left|\frac{1}{n \mathbb{E}\left\{K_{1}(x)\Gamma_{1,l}(x)\right\}} \sum_{s=1}^{n} K_{s}(x)\Gamma_{s,l}(x) \mathbb{E}\left\{\left|\sqrt{U\left(X_{s}\right)}\right|^{\chi} \mid \mathcal{F}_{s-1}\right\}\right| \\
& \leq \frac{\ddot{C}}{\ell_{n}^{\chi-1} }\times\left[\frac{1}{n \mathbb{E}\left\{K_{1}(x)\Gamma_{1,l}(x)\right\}} \sum_{s=1}^{n} \mathbb{E}^{b}\left\{\left|\sqrt{U\left(X_{s}\right)}\right|^{\chi} \mid \mathcal{F}_{s-1}\right\}\right]^{1 / b}\left\{\frac{1}{n \mathbb{E}\left\{K_{1}(x)\Gamma_{1,l}(x)\right\}}\left|\sum_{s=1}^{n} K_{s}^{a}(x)\Gamma_{s,l}^a(x)\right|\right\}^{1 / a}
\\
& \leq \frac{\ddot{C}}{\ell_{n}^{\chi-1\color{black}} \mathbb{E}^{1 / b}\left\{K_{1}(x)\Gamma_{1,l}(x)\right\}}\left[\frac{1}{n} \sum_{s=1}^{n} \mathbb{E}^{b}\left\{\left|\sqrt{U\left(X_{s}\right)}\right|^{\chi}  \mid \mathcal{F}_{s-1}\right\}\right]^{1 / b} \times\left[\frac{1}{n \mathbb{E}\left\{K_{1}(x)\Gamma_{1,l}(x)\right\}} \sum_{s=1}^{n} K_{s}^{a}(x)\Gamma_{s,l}^a(x)\right]^{1 / a} \\
& \leq \frac{\ddot{C}}{\ell_{n}^{\chi-1} \mathbb{E}^{1 / b}\left\{K_{1}(x)\Gamma_{1,l}(x)\right\}}\left[\frac{1}{n} \sum_{s=1}^{n} \mathbb{E}\left\{\left|\sqrt{U\left(X_{s}\right)}\right|^{b \chi}  \mid \mathcal{F}_{s-1}\right\}\right]^{1 / b} \times\left[\frac{1}{n \mathbb{E}\left\{K_{1}(x)\Gamma_{1,l}(x)\right\}} \sum_{s=1}^{n} K_{s}^{a}(x)\Gamma_{s,l}^a(x)\right]^{1 / a}.  (A.1)
\end{aligned}\end{small}
$$

By the ergodicity of the $X_{s},$ the first sum in braces converges a.s to $\mathbb{E}\left\{\mathbb{E}\left(\left|U\left(X_{s}\right)\right|^{b \chi} \mid \mathcal{F}_{s-1}\right)\right\}= \mathbb{E}\left\{|\sqrt{U(X)}|^{(\chi^{2} /(\chi-1)}  \right\}$, which is finite by the first part of Assumption \ref{errors}(i).\\
Now, we have to show that the second term in (A.1) is bounded uniformly in $x$. Observe that
\begin{equation}\label{A.2}
\begin{aligned}
\frac{1}{n \mathbb{E}\left\{K_{1}(x)\Gamma_{1,l}(x)\right\}} \sum_{s=1}^{n} K_{s}^{a}(x)\Gamma^{a}_{s,l}(x)&=\frac{1}{n \mathbb{E}\left\{K_{1}(x)\Gamma_{1,l}(x)\right\}} \sum_{s=1}^{n}\left[K_{s}^{a}(x)\Gamma^{a}_{s,l}(x)-\mathbb{E}\left\{K_{s}^{a}(x)\Gamma^{a}_{s,l}(x) \mid \mathcal{F}_{s-1}\right\}\right]\\ & +\frac{1}{n \mathbb{E}\left\{K_{1}(x)\Gamma_{1,l}(x)\right\}} \sum_{s=1}^{n} \mathbb{E}\left\{K_{s}^{a}(x)\Gamma^{a}_{s,l}(x) \mid \mathcal{F}_{s-1}\right\}.
\end{aligned}
\end{equation}
By using the martingale property, $\sum_{s=1}^{n} \mathbb{E}\left\{K_{s}^{a}(x)\Gamma_{s,l}^a(x) \mid \mathcal{F}_{s-1}\right\} /\left[n \mathbb{E}\left\{K_{1}(x)\Gamma_{1,l}(x)\right\}\right]$ converges to  $\mathbb{E}\left\{K_{1}^{a}(x)\Gamma_{1,l}^a(x)\right\} / \mathbb{E}\left\{K_{1}(x)\Gamma_{1,l}(x)\right\}$ which is equal, by Lemma \ref{Lemma 4.5}, to $ \mathcal{O}(n^{a-1} h^{2(a-1)}) $. Because of assumption \ref{bdw}(i), the last term is negligible as $n \rightarrow \infty$. Concerning the first term on the right-hand side of (A.1), denote

$$
G^{[l]}_{n, 1}(x)=\frac{1}{n \mathbb{E}\left\{K_{1}(x)\Gamma_{1,l}(x)\right\}} \sum_{s=1}^{n}\left[K_{s}^{a}(x)\Gamma_{s,l}^a(x)-\mathbb{E}\left\{K_{s}^{a}(x)\Gamma_{s,l}^a(x) \mid \mathcal{F}_{s-1}\right\}\right] .
$$

Using the Burkholder and Jensen inequalities, we deduce that, for any $\epsilon>0$, there exists a constant $c_{0}>$ such that
$$
\operatorname{Pr}\left(\left|G^{[l]}_{n, 1}(x)\right|>\epsilon\right) \leq c_{0} \frac{\mathbb{E}\left\{K_{1}^{2 a}(x)\Gamma_{1,l}^{2a}(x)\right\}}{\epsilon^{2} n\left[\mathbb{E}\left\{K_{1}(x)\Gamma_{1,l}(x)\right\}\right]^{2}}=c_{0} \frac{\mathcal{O}(n^{2a} h^{4a}) \phi(h)}{\epsilon^2 n \mathcal{O}(n^{2} h^{4}) \phi^2(h)} .
$$
Using Assumption~\ref{bdw}(i) permits us to conclude that $G_{n, 1}(x)=o_{ \mathbb{P}}(1)$. Finally, by considering  $b=\chi /(\chi-1)$, and applying Lemma \ref{Lemma 4.5} ,it yields
$$
\left[\ell_{n}^{\chi-1} \mathbb{E}^{(\chi-1) / \chi}\left\{K_{1}(x)\Gamma_{1,l}(x)\right\}\right]^{-1} \leq \ddot{C} \ell_{n}^{(1-\chi)}n^{\chi /(\chi-1)}h^{2\chi /(\chi-1)} \phi\left(h\right)^{\chi /(\chi-1)}.$$
\end{proof}

The following Lemma presents the asymptotic normality of  $\vartheta^{[l]}_{n,3}(x)$ needed to prove Theorem \ref{T3}.
\begin{lem}\label{Lemma-T3}
Suppose that Assumptions~\ref{A1}(i), \ref{SBP}(i)-(iv), (vi), \ref{errors}(ii)-(iii), \ref{SMC}(ii)-(ii)  are satisfied. Then, as $n \to \infty$, we have, 
\begin{equation*}
\sqrt{n \phi(h)}\vartheta^{[l]}_{n,3}(x)  \overset{\mathcal{D}}
{\longrightarrow}  \mathcal{N}\left(0, \sigma^{2}(x)\right).
\end{equation*} 
where $\sigma^{2}(x)=\frac{M_2}{M_1^2} \omega \color{black} U^{2}(x)$ and $l \in \{0,1\} $. 
\end{lem}

\begin{proof}
Define
$$\xi _{n, s}=\{\sqrt{\phi\left(h\right) / n}\} \left[\frac{U\left(X_{s}\right) K_s(x){\Gamma_{s,l}(x)}\left(\epsilon_{s}^{2}-1\right)}{\mathbb{E}\left\{K_1(x){\Gamma_{1,l}(x)}\right\}}\right], \quad s=1,...,n.$$
Further, observe that
\begin{equation}\label{36}
\sqrt{n \phi\left(h\right) }\vartheta^{[l]}_{n,3}(x)=\sum_{s=1}^{n} \xi_{n, s}.
\end{equation}
The summands $\xi_{n,s}$ form a triangular array of stationary martingale differences with respect to the $\sigma$-field $\mathcal{F}_{s-1}$. 
To establish the asymptotic normality, we must prove that
\begin{itemize}
\item[(i)] $\underset{n \rightarrow \infty}{\lim} \sum_{s=1}^{n} \mathbb{E}\left(\xi_{n, s}^{2} \mid \mathcal{F}_{s-1}\right) \stackrel{\mathbb{P}}{=} \sigma^{2}(x),$
\item[(ii)] $n \mathbb{E}\left\{\xi_{n, s}^{2} \1_{\left(\left|\xi_{n, s}\right|>\zeta\right)}\right\}=o(1)$ holds for any $\zeta>0$.
\end{itemize}
\textbf{Proof of part (i).}  Since $\epsilon_s$ is independent of $X_s$ (and hence of $\mathcal{F}_{s-1}$), Using Assumption \ref{errors}(ii), we have \color{black}
\begin{align*}
\sum_{s=1}^{n} \mathbb{E}\left(\xi_{n, s}^{2} \mid \mathcal{F}_{s-1}\right)
& = \frac{\phi(h)}{n[\mathbb{E}(K_{1}\Gamma_{1,l})]^{2}} \sum_{s=1}^{n} \mathbb{E}\left[U^{2}(X_{s}) \omega K^{2}_{s}(x)\Gamma^{2}_{s,l}(x) \mid \mathcal{F}_{s-1}\right]\\
&= \frac{\omega\phi(h)}{n[\mathbb{E}(K_{1}\Gamma_{1,l})]^{2}} \sum_{s=1}^{n} \mathbb{E}\left[\{U^{2}(X_s) - U^2(x) \} K^{2}_{s}\Gamma^{2}_{s,l} \mid \mathcal{F}_{s-1}\right] \\
& \quad + \frac{U^{2}(x)\omega \phi(h)}{n[\mathbb{E}(K_{1}\Gamma_{1,l})]^{2}} \sum_{s=1}^{n} \mathbb{E}\left[K^{2}_{s}\Gamma^{2}_{s,l} \mid \mathcal{F}_{s-1}\right] \\
&\equiv \mathcal{H}_1 + \mathcal{H}_2.
\end{align*}
\textit{Regarding the term $\mathcal{H}_1$.} By using Lemma \ref{Lemma 4.5} and  Assumptions~\ref{A1}(i), \ref{SBP}(iii)-(iv), (vi),  \ref{regularityconditions}(ii), and \ref{bdw}(ii) we can show that $\mathcal{H}_1 \longrightarrow 0 \,\ \hbox{as} \,\ n \to \infty.$

\textit{Regarding the term $\mathcal{H}_2$.} Under the same conditions, we obtain
\begin{equation*}
\mathcal{H}_2 = U^{2}(x)\omega\times \frac{\phi(h) \cdot n \phi(h) M_2}{n (M_1 \phi(h))^2} + o_P(1) \longrightarrow \frac{M_2}{M_1^2} U^2(x)\omega,
\end{equation*}
Combining these results, we get
\begin{equation*}
\lim_{n\longrightarrow \infty} \sum_{s=1}^{n} \mathbb{E}\left(\xi_{n, s}^{2} \mid \mathcal{F}_{s-1}\right)= \frac{M_2}{M_1^2} U^{2}(x) \,\omega = \sigma^{2}(x) \quad \text{a.s.}
\end{equation*}
Combining these results, we get
\begin{equation*}
\lim_{n\longrightarrow \infty} \sum_{s=1}^{n} \mathbb{E}\left(\xi_{n, s}^{2} \mid \mathcal{F}_{s-1}\right)= \frac{M_2}{M_1^2} U^{2}(x) \omega(x) = \sigma^{2}(x) \quad \text{a.s.}
\end{equation*}

\textbf{Proof of part (ii).}
Consider $a>1$ and $b >1$ such that $\displaystyle{\frac{1}{a}+\frac{1}{b}=1}$, by using  H\"{o}lder's and Markov's inequalities one can
have, for all $\zeta > 0$,
$$
\mathbb{E}\left\{\xi_{n, s}^{2} \1_{\left(\left|\xi_{n, s}\right|>\zeta\right)}\right\}
\leq \frac{\mathbb{E}|\xi_{n, s}|^{2a}}{\zeta^{2a/b}}\text{.}
$$
Taking $2a=\kappa + 2$ (where $\kappa$ is given in Assumption~\ref{errors}(iii)) and let $C_{0}$ be a positive constant. Under Assumptions~\ref{A1}, \ref{SBP}(iii), \ref{errors}(iii),  \ref{regularityconditions}(iii), \ref{bdw}(iv),  and by using  Lemma 
\ref{Lemma 4.5}, we obtain
$$\begin{aligned}
n\mathbb{E}\left\{\xi_{n, s}^{2} {1}_{\left(\left|\xi_{n, s}\right|>\zeta\right)}\right\}& \leq C_{0}\left\{\phi\left(h\right) / n\right\}^{(2+\kappa) / 2} \times \frac{n}{\left[\mathbb{E}\left( 
    K_{1}(x)\Gamma_{1,l}(x)   \right\}\right]^{2+\kappa}} \mathbb{E}\left\{U^{2+\kappa}\left(X_{s}\right) K^{2+\kappa}_{s}(x)\Gamma^{2+\kappa}_{s,l}(x)\left(\epsilon_{s}^{2}-1\right)^{2+\kappa}\right\}\\
&\leq C_{0}\left\{n \phi\left(h\right)\right\}^{-\kappa / 2} \frac{M_{2+k}}{M_{1}^{2+k}}\left\{U^{2+\kappa}(x)+o(1)\right\}  \\
&=\mathcal{O}\left[\left\{n \phi\left(h\right)\right\}^{-\kappa / 2}\right]\text{.}
\end{aligned}$$ \\
Finally,  since $n \phi(h)\longrightarrow \infty$ as $n\longrightarrow \infty$, we get
\begin{equation*} 
n \mathbb{E} \left\{ \xi_{n, s}^{2} \1_{\left(\left|\xi_{n, s} \right| > \zeta \right)}\right\}=o(1).
\end{equation*}
Therefore,  Lemma \ref{Lemma-T3} holds.
\end{proof}
\color{black}

\begin{lem}\label{cor:residual-approx}
Let the assumptions of Theorems~\ref{T1} and~\ref{T2} hold, and let
$\widehat\varepsilon_t$ be the residuals defined in~\eqref{epsilonhat}. Moreover, suppose that $\log n/(n\phi(h)) \to 0$, as $n\to \infty.$ Then, for every $t\in\{1,\dots,n\}$, the approximation error satisfies
\[
|\widehat\varepsilon_t - \varepsilon_t|
\;\le\;
R_{n,t}\,\bigl(1+|\varepsilon_t|\bigr),
\]
where
\[
\max_{1\le t\le n} R_{n,t}
\;=\;
\mathcal{O}_{\mathbb{P}}\!\left(
h^{\min(b_1,b_2)} + \sqrt{\dfrac{\log n}{n\phi(h)}}
\right)
=o_{\mathbb{P}}(1).
\]
\end{lem}

\begin{proof} 

For all $t\in \{1, \dots, n\}$ we write
\[
|\widehat\varepsilon_t-\varepsilon_t|
=
\biggl|
\frac{Y_t-m_n(X_t)}{\sigma_n(X_t)}
-
\frac{Y_t-m(X_t)}{\sigma(X_t)}
\biggr|
=
\biggl|
\frac{m(X_t)-m_n(X_t)}{\sigma_n(X_t)}
+
\varepsilon_t\Bigl(
\frac{\sigma(X_t)}{\sigma_n(X_t)}-1
\Bigr)
\biggr|.
\]
Hence
\begin{align*}
|\widehat\varepsilon_t-\varepsilon_t|
&\le
\frac{\sigma(X_t)}{\sigma_n(X_t)}
\Bigl(
\Bigl|\frac{m_n(X_t)-m(X_t)}{\sigma(X_t)}\Bigr|
+
\Bigl|\frac{\sigma_n(X_t)-\sigma(X_t)}{\sigma(X_t)}\Bigr|\,
|\varepsilon_t|
\Bigr)
\\[0.2cm]
&\le
\frac{\sigma(X_t)}{\sigma_n(X_t)}
\max\bigl\{|\xi^{(m)}_{t,n}|,|\xi^{(\sigma)}_{t,n}|\bigr\}
\bigl(1+|\varepsilon_t|\bigr)
=: R_{n,t}\bigl(1+|\varepsilon_t|\bigr),
\end{align*}
where
\[
\xi^{(m)}_{t,n}
=
\frac{m_n(X_t)-m(X_t)}{\sigma(X_t)},\quad
\xi^{(\sigma)}_{t,n}
=
\frac{\sigma_n(X_t)-\sigma(X_t)}{\sigma(X_t)}.
\]

We now bound $\max_{1\leq t\leq n}R_{n,t}$. By definition,
\[
\max_{1\leq t\leq n}R_{n,t}
\le
\Bigl(\max_{1\leq t\leq n}\frac{\sigma(X_t)}{\sigma_n(X_t)}\Bigr)
\max\Bigl\{\max_{1\leq t\leq n}|\xi^{(m)}_{t,n}|,\;
           \max_{1\leq t\leq n}|\xi^{(\sigma)}_{t,n}|\Bigr\}.
\]

\medskip\noindent
\emph{Step 1: rate for $\xi^{(m)}_{t,n}$ and $\xi^{(\sigma)}_{t,n}$.}
Since $X_t\in\mathcal C$ a.s., we have
\[
\max_{1\leq t\leq n}|\xi^{(m)}_{t,n}|
\le
\sup_{x\in\mathcal C}
\frac{|m_n(x)-m(x)|}{\sigma(x)},
\qquad
\max_{1\leq t\leq n}|\xi^{(\sigma)}_{t,n}|
\le
\sup_{x\in\mathcal C}
\frac{|\sigma_n(x)-\sigma(x)|}{\sigma(x)}.
\]
By the boundedness assumption $c\le U(x)\le C$ and the definition
$\sigma(x)=\sqrt{U(x)}$, we have $\sigma(x)\ge \sqrt c$ on $\mathcal C$; hence
\[
\max_{1\leq t\leq n}|\xi^{(m)}_{t,n}|
\le
\frac{1}{\sqrt c}\sup_{x\in\mathcal C}|m_n(x)-m(x)|,
\qquad
\max_{1\leq t\leq n}|\xi^{(\sigma)}_{t,n}|
\le
\frac{1}{\sqrt c}\sup_{x\in\mathcal C}|\sigma_n(x)-\sigma(x)|.
\]

Theorems~\ref{T1} and~\ref{T2} yield 
\[
\max\Bigl\{\max_{1\leq t\leq n}|\xi^{(a)}_{t,n}|,\;
           \max_{1\leq t\leq n}|\xi^{(b)}_{t,n}|\Bigr\}
=
\mathcal O_{\mathbb P}(h^{b_1}+h^{b_2}+\sqrt{\log n/(n\phi(h))}).
\]

\medskip\noindent
\emph{Step 2: control of $\sigma(X_t)/\sigma_n(X_t)$.}
We now show that
\[
\max_{1\leq t\leq n}\frac{\sigma(X_t)}{\sigma_n(X_t)}
=\mathcal O_{\mathbb P}(1).
\]
For any $M>0$,
\[
\mathbb{P}\Bigl(\max_{1\leq t\leq n}\tfrac{\sigma(X_t)}{\sigma_n(X_t)}\ge M\Bigr)
\le
\mathbb{P}\Bigl(
\sup_{x\in\mathcal C}
\Bigl|\frac{\sigma_n(x)-\sigma(x)}{\sigma(x)}\Bigr|
\ge 1-\tfrac1M
\Bigr).
\]
Since $\sigma(x)\ge\sqrt c$ on $\mathcal C$ and
$\sup_{x\in\mathcal C}|\sigma_n(x)-\sigma(x)|\to0$ in probability,
the right-hand side tends to $0$ as $M\to\infty$. Hence
$\max_{1\leq t\leq n}\sigma(X_t)/\sigma_n(X_t)=\mathcal O_{\mathbb P}(1)$.

\medskip\noindent
Finally, putting the bounds from \emph{Steps~1} and\emph{~2} together yields
\[
\max_{1\leq t\leq n}R_{n,t}
\le
\Bigl(\max_{1\leq t\leq n}\tfrac{\sigma(X_t)}{\sigma_n(X_t)}\Bigr)
\max\Bigl\{\max_{1\leq t\leq n}|\xi^{(a)}_{t,n}|,\;
           \max_{1\leq t\leq n}|\xi^{(b)}_{t,n}|\Bigr\}
=o_{\mathbb P}(1).
\]
This concludes the proof of the lemma.
\end{proof}

\section*{Appendix B: Proof of main results}



\subsection*{Proof of Theorem \ref{T1}}
By Model (\ref{model}), we have
\begin{equation*}
m_{n}(x)-m(x)=\frac{1}{m^{[l]}_{n,1}(x)}\left[\frac{1}{n \mathbb{E} (K_1(x)\Gamma_{1,l}(x))}\displaystyle{\sum_{s=1}^{n}}
\{m(X_{s})-m(x)+\sqrt{U(X_{s})}\epsilon_{s}\}K_{s}(x)\Gamma_{s,l}(x)\right].
\end{equation*}
Then,
\begin{equation}\label{Decosup2}
\displaystyle{\sup_{x\in \mathcal{C}}}|m_{n}(x)-m(x)|\leq \displaystyle{\sup_{u\in B(x,h)}}|m(u)-m(x)|+\{\displaystyle{\inf_{x\in \mathcal{C}}}m^{[l]}_{n,1}(x)\}^{-1}\times \displaystyle{\sup_{x\in \mathcal{C}}}|V^{[l]}_{n}(x)|.
\end{equation}
Further, observe that
\begin{equation*}
\displaystyle{\inf_{x\in \mathcal{C}}}\left|m^{[l]}_{n,1}(x)\right|>1-\displaystyle{\sup_{x\in \mathcal{C}}}\left|m^{[l]}_{n,1}(x)-1\right|.
\end{equation*}
Thus, making use of Lemma \ref{Lemma 7.1.} and  Assumption~\ref{SBP}(iv), we get
 \begin{equation}\label{***+}
 \displaystyle{\inf_{x\in \mathcal{C}}}\left|m^{[l]}_{n,1}(x)\color{black}\right|>1\,\ a.s.
\end{equation}
Finally, using Lemma \ref{Lemmacvn}, combined with  Assumption~\ref{regularityconditions}(i), equations (\ref{Decosup2})-(\ref{***+}), one concludes the proof of the theorem.  
 \subsection*{Proof of Theorem \ref{T2}}
 \quad  Let 
 \begin{equation}\label{TU}
U_{n}(x)-U(x)=\left(\vartheta_{n,1}(x)+\vartheta_{n,2}(x)+\vartheta_{n,3}(x)+\vartheta_{n,4}(x)\right)/m^{[l]}_{n,1}(x),
 \end{equation}
 where $m^{[l]}_{n,1}(x)$ is defined in (\ref{*}). 

 \begin{eqnarray}
\vartheta^{[l]}_{n,1}(x)&=&\displaystyle\frac{1}{n \mathbb{E}\left[K_{1}(x)\Gamma_{1,l}(x)\right]} \sum_{s=1}^{n}\left(m\left(X_{s}\right)-m_{n}\left(X_{s}\right)\right)^{2} K_{s}(x)\Gamma_{s,l}(x), \label{Vn1}\\
\vartheta^{[l]}_{n,2}(x)&=&\displaystyle\frac{2}{n \mathbb{E}\left[K_{1}(x)\Gamma_{1,l}(x)\right]} \sum_{s=1}^{n}\left(m\left(X_{s}\right)-m_{n}\left(X_{s}\right)\right) \sqrt{U\left(X_{s}\right)} \epsilon_{s} K_{s}(x)\Gamma_{s,l}(x), \label{Vn2} \\
\vartheta^{[l]}_{n,3}(x)&=&\displaystyle\frac{1}{n \mathbb{E}\left[K_{1}(x)\Gamma_{1,l}(x)\right]} \sum_{s=1}^{n} U\left(X_{s}\right) K_{s}(x)\Gamma_{s,l}(x)\left(\epsilon_{s}^{2}-1\right), \label{Decoimp}\\
\vartheta^{[l]}_{n,4}(x)&=&\displaystyle\frac{1}{n \mathbb{E}\left[K_{1}(x)\Gamma_{1,l}(x)\right]} \sum_{s=1}^{n}\left(U\left(X_{s}\right)-U(x)\right) K_{s}(x)\Gamma_{s,l}(x), \notag\\
\end{eqnarray}

 Our main objective is to establish the uniform consistency rate, with respect to $x,$ of $U_{n}(x)$. For this purpose, let us consider
\begin{equation}\label{15*}
\displaystyle{\sup_{x\in \mathcal{C}}}\left| U_{n}(x)-U(x)\right| \leq  \big\{\displaystyle{\sup_{x\in \mathcal{C}}}|\vartheta_{n,1}(x)|+\displaystyle{\sup_{x\in \mathcal{C}}}|\vartheta_{n,2}(x)|+\displaystyle{\sup_{x\in \mathcal{C}}}|\vartheta_{n,3}(x)|+\displaystyle{\sup_{x\in \mathcal{C}}}|\vartheta_{n,4}(x)| \big\} \times \big\{\displaystyle{\inf_{x\in \mathcal{C}}}|m^{[l]}_{n,1}(x)|\big\}^{-1}.
 \end{equation}
Note that 
$$\begin{aligned}\left|\vartheta^{[l]}_{n,1}(x)\right| &\leq \frac{1}{n \mathbb{E}\left[K_{1}(x)\Gamma_{1,l}(x)\right]} \sum_{s=1}^{n}\left|m\left(X_{s}\right)-m_{n}\left(X_{s}\right)\right|^{2} K_{s}(x)\Gamma_{s,l}(x)\\ &\leq \sup _{x \in \mathcal{C}}\left|m_{n}(x)-m(x)\right|^{2} \times \frac{1}{n \mathbb{E}\left[K_{1}(x)\Gamma_{1,l}(x)\right]} \sum_{s=1}^{n} K_{s}(x)\Gamma_{s,l}(x)\text{.}
\end{aligned}
$$
Then, in view of Theorem \ref{T1} and  Lemma \ref{Lemma 7.1.}, we get
\begin{equation}\label{32}
\displaystyle{\sup_{x\in \mathcal{C}}}\left|\vartheta^{[l]}_{n,1}(x)\right|=\mathcal{O}_{a.s.}(h^{2b_1})+
\mathcal{O}_{a.s.}\left(\frac{\log n}{n\phi(h)}\right)\text{.}
\end{equation}
\textit{Study of the term $\vartheta^{[l]}_{n,2}(x)$.} Observe that
$$
\vartheta^{[l]}_{n,2}(x) \leq 2 \sup _{x \in \mathcal{C}}\left|m_{n}(x)-m(x)\right| \times V^{[l]}_{n}(x),\,\
$$
where $V^{[l]}_{n}(x)$ is defined in (\ref{Vn0})
Making use of Theorem \ref{T1} with Lemma \ref{Lemmacvn}, it yields
\begin{equation}\label{33}
\displaystyle{\sup_{x\in \mathcal{C}}}\left|\vartheta^{[l]}_{n,2}(x)\right|=\mathcal{O}_{a.s.}\left\{\left(h^{b_1}+\sqrt{\frac{\log n}{n\phi(h)}}\right)\sqrt{\frac{\log n}{n\phi(h)}}\right\}\text{.}
\end{equation}
 \textit{Study of the term $\vartheta^{[l]}_{n,3}(x).$} 
Observe that $\vartheta^{[l]}_{n,3}(x)$ has the same form as $V^{[l]}_{n}(x)$  when $\epsilon$ and $\sqrt{U}$ are replaced by $\epsilon^{2}-1$ and $U,$ respectively.
 Similar to the proof of Lemma \ref{Lemmacvn},  under Assumptions~\ref{cgamma}, \ref{A1}(i)-(iii), \ref{SBP}(iv)-(v), \ref{errors}, \ref{regularityconditions}(i)  and condition (\ref{condunifconv}), we obtain
\begin{equation}\label{34}
\displaystyle{\sup_{x\in \mathcal{C}}}\left|\vartheta^{[l]}_{n,3}(x)\right|=\mathcal{O}_{a.s.}\left(\sqrt{\frac{\log n}{n\phi(h)}}\right)\text{.}
\end{equation}  \color{black}
\textit{Study of the term $\vartheta^{[l]}_{n,4}(x).$} One can easily show, using Assumption~\ref{SMC}(ii) with Lemma \ref{Lemma 7.1.}, that
\begin{equation}\label{35}
\displaystyle{\sup_{x\in \mathcal{C}}}\left|\vartheta^{[l]}_{n,4}(x)\right|=\mathcal{O}_{a.s.}\left(h^{b_2}\right)\text{.}
\end{equation}
Finally,  using  equation  (\ref{15*}) combining   with  the results obtained in (\ref{***+}) and (\ref{32})-(\ref{35}), we conclude the proof of the theorem.


\subsection*{Proof of Theorem \ref{T3}.}
The proof is based on the decomposition (\ref{TU}). By defining the bias term as:
\begin{equation}
B_n(x) = \frac{\vartheta_{n,4}^{[l]}(x)}{ U(x)},
\end{equation}
we have
\begin{equation}
\sqrt{n\phi(h)} \left( \frac{U_n(x)}{U(x)} - 1 - B_n(x) \right) = \sqrt{n\phi(h)} \left[ \frac{\vartheta_{n,1}^{[l]}(x) + \vartheta_{n,2}^{[l]}(x) + \vartheta_{n,3}^{[l]}(x)}{ U(x)} \right].
\end{equation}
By combining equations (\ref{32}), (\ref{33}), and (\ref{34}) with the conditions stated in (\ref{BI}) and applying  Lemma \ref{Lemma-T3}, we obtain the result for part (i) of the theorem. \\
Furthermore, by considering the bias condition (\ref{BI2}) together with the asymptotic result (\ref{35}), it follows that the bias term $B_n(x)$ becomes asymptotically negligible:
\begin{equation}
\sqrt{n\phi(h)} B_n(x) \xrightarrow{\mathbb{P}} 0, \quad \text{as } n \to \infty.
\end{equation}
This leads directly to the second result in part (ii) of the theorem.

\subsection*{Proof of Corollary \ref{Cor}}

Consider the following decomposition
\begin{equation*}
\frac{\widehat{M}_1}{\sqrt{\widehat{M}_2}}\sqrt{\dfrac{n \phi(h)}{\omega_n U_{n}(x)}} \left(\frac{U_{n}(x)}{U(x)}-1 \right)
= {\left( \frac{\widehat{M}_1}{\sqrt{\widehat{M}_2}} \frac{\sqrt{M_2}}{M_1} \right) \sqrt{\dfrac{U(x)\omega}{\omega_n U_{n}(x)}}} \times {\left[ \frac{M_1}{\sqrt{M_2}}\sqrt{\dfrac{n \phi(h)}{\omega U(x)}}\left(\frac{U_{n}(x)}{U(x)}-1 \right) \right]}.
\end{equation*}
According to Theorem \ref{T3}, we have:
\begin{equation*}
\frac{M_1}{\sqrt{M_2}}\sqrt{\dfrac{n \phi(h)}{\omega U(x)}}\left(\frac{U_{n}(x)}{U(x)}-1 \right) \xrightarrow{\mathcal{D}} \mathcal{N}(0, 1) \quad \text{as } n \to \infty.
\end{equation*}

Following the same arguments as in the proof of Corollary 1 in \cite{LL10}, we have $\widehat{M}_j \xrightarrow{\mathbb{P}} M_j$ for $j=1,2$. By Theorem \ref{T2}, we also have $U_n(x) \xrightarrow{\mathbb{P}} U(x)$. Therefore, it remains to prove that
\begin{equation}\label{P22_proof}
\omega_n \xrightarrow{\mathbb{P}} \omega \quad \text{as } n \to \infty.
\end{equation}
Let
\begin{equation*}
\omega_n-\omega =\mathfrak{T}_{1,n}+\mathfrak{T}_{2,n},
\end{equation*}
 where $\mathfrak{T}_{1,n}= \frac{1}{n}\displaystyle\sum_{s=1}^{n}\left[(\widehat{\epsilon}_{s}^{2}-1)^{2}-(\epsilon_{s}^{2}-1)^{2}\right]$ and   $\mathfrak{T}_{2,n}=\frac{1}{n}\displaystyle\sum_{s=1}^{n}(\epsilon_{s}^{2}-1)^{2}-\omega.$

\textit{Regarding $\mathfrak{T}_{2,n}$.} Since $(\epsilon_s)_{s=1,\dots,n}$ is an i.i.d. sequence with Assumption \ref{errors}(ii), the  strong law of large numbers gives directly $\mathfrak{T}_{2,n} \xrightarrow{a.s.} 0$.

\textit{Regarding $\mathfrak{T}_{1,n}$.} Let 
\begin{equation*}
(\widehat\epsilon_s^2-1)^2-(\epsilon_s^2-1)^2 = \delta_s\,(\widehat\epsilon_s+\epsilon_s)\,(\widehat\epsilon_s^2+\epsilon_s^2-2),
\end{equation*}
where $\delta_s:=\widehat\epsilon_s-\epsilon_s.$\\
By Lemma \ref{cor:residual-approx}, we have $|\delta_s|\le R_n(1+|\epsilon_s|)$ and  for $n$ large enough that $R_n \le 1$ (which holds with probability tending to one, since $R_n=o_{\mathbb{P}}(1)$), we have
\begin{equation*}
|\widehat\epsilon_s+\epsilon_s| = |2\epsilon_s+\delta_s| \le 2|\epsilon_s|+|\delta_s| \le 2|\epsilon_s|+(1+|\epsilon_s|) \le C(1+|\epsilon_s|),
\end{equation*}
In the same manner, we have $|\widehat\epsilon_s|\le|\epsilon_s|+|\delta_s|\le1+2|\epsilon_s|$, then using some algebraic manipulations, we get
\begin{equation*}
|\widehat\epsilon_s^2+\epsilon_s^2-2| \le \widehat\epsilon_s^2+\epsilon_s^2+2  \le C(1+\epsilon_s^2).
\end{equation*}

Combining the three factors, using  elementary inequalities, we obtain
\begin{equation*}
\left|(\widehat\epsilon_s^2-1)^2-(\epsilon_s^2-1)^2\right| \le C\,R_n\,(1+\epsilon_s^4).
\end{equation*}

Finally, we have 
\begin{equation*}
|\mathfrak{T}_{1,n}| \le C\,R_n\times\frac1n\sum_{s=1}^n(1+\epsilon_s^4).
\end{equation*}
By Assumption \ref{errors}(iii), the law of large numbers,  and $R_n=o_{\mathbb{P}}(1)$ (Lemma \ref{cor:residual-approx}), we conclude
\begin{equation*}
\mathfrak{T}_{1,n}\xrightarrow{\mathbb{P}} 0.
\end{equation*}
 
Combining the two terms gives ${\omega_n}\xrightarrow{\mathbb{P}}\omega$, and Slutsky's Theorem completes the proof.

\subsection*{Proof of Theorem \ref{thm:Hill-residual}}

We have
\[
\sqrt{n(1-\alpha_n)}\bigl(\widehat\gamma-\gamma\bigr)
=
\sqrt{n(1-\alpha_n)}\bigl(\widehat\gamma-\widecheck\gamma\bigr)
+
\sqrt{n(1-\alpha_n)}\bigl(\widecheck\gamma-\gamma\bigr)
=: T_{1,n}+T_{2,n}.
\]
Then, it remains to show that $T_{1,n}=o_{\mathbb P}(1)$ and that $T_{2,n}$ converges in
distribution to the normal law in~\eqref{eq:Hill-residual-CLT}.

\medskip\noindent
\emph{Step 1: distribution of oracle Hill estimator.}
Under Assumption~\ref{cgamma} and the conditions on $k_n$ and $A(\cdot)$,  the classical extreme-value theory applies
(e.g.\ \cite{deHaanFerreira2006}, Theorem~3.2.5). In particular,
\begin{equation}\label{eq:Hill-oracle-CLT}
\sqrt{n(1-\alpha_n)}\bigl(\widecheck\gamma-\gamma\bigr)
\xrightarrow{\mathcal D}
\mathcal N\!\left(\frac{\lambda}{1-\rho},\,\gamma^2\right).
\end{equation}
Thus $T_{2,n}$ already has the desired limit. It remains to prove that
$T_{1,n}=\sqrt{n(1-\alpha_n)}(\widehat\gamma-\widecheck\gamma)=o_{\mathbb P}(1)$.


\noindent \emph{Step 2: control of $T_{1,n}$.}
Write
\[
\widehat\gamma-\widecheck\gamma
=
\frac{1}{k_n}
\sum_{j=1}^{k_n}
\Bigl\{
\log\widehat\varepsilon_{n-j+1:n}
-
\log\varepsilon_{n-j+1:n}
\Bigr\}
-
\frac{1}{k_n}
\sum_{j=1}^{k_n}
\Bigl\{
\log\widehat\varepsilon_{n-k_n:n}
-
\log\varepsilon_{n-k_n:n}
\Bigr\}
=:A_{1,n}-A_{2,n}.
\]
Observe that, using Lemma \ref{cor:residual-approx} and Lemma A.3 in \cite{Girardannals2021} one has
\[
\max_{1\le j\le k_n}
\biggl|
\log\left(\dfrac{\widehat\varepsilon_{n-j+1:n}}{\varepsilon_{n-j+1:n}}\right)
\biggr|
=\mathcal{O}_{\mathbb P}\left(h^{\min(b_1,b_2)} + \sqrt{\dfrac{\log n}{n\phi(h)}}\right) = o_{\mathbb{P}}(1).
\]
Consequently, using conditions \eqref{condh}, one obtains
\begin{eqnarray*}
|A_{1,n}| &=& \sqrt{n(1-\alpha_n)}|\widehat\gamma-\widecheck\gamma|
\le
\sqrt{n(1-\alpha_n)} \max_{1\le j\le k_n}
\biggl|
\log\left(\dfrac{\widehat\varepsilon_{n-j+1:n}}{\varepsilon_{n-j+1:n}}\right)
\biggr|\\
&=& \mathcal{O}_{\mathbb{P}}\left(\sqrt{n(1-\alpha_n)}h^{\min(b_1,b_2)} + \sqrt{\dfrac{n(1-\alpha_n) \log n}{n\phi(h)}} \right) = o_{\mathbb P}(1).
\end{eqnarray*}
Similarly, using the same arguments, one can show that 
$
|A_{2,n}|
=
o_{\mathbb P}(1).
$
Therefore
\[
\sqrt{n(1-\alpha_n)}|\widehat\gamma-\widecheck\gamma|
\le
\sqrt{n(1-\alpha_n)}(|A_{1,n}|+|A_{2,n}|)
=o_{\mathbb P}(1),
\]
that is,
\begin{equation}\label{eq:T1n}
T_{1,n}=\sqrt{n(1-\alpha_n)}(\widehat\gamma-\widecheck\gamma)=o_{\mathbb P}(1).
\end{equation}

\medskip\noindent
Finally, combining \eqref{eq:Hill-oracle-CLT} and \eqref{eq:T1n} with Slutsky’s
theorem yields
\[
\sqrt{n(1-\alpha_n)}\bigl(\widehat\gamma-\gamma\bigr)
=
\underbrace{\sqrt{n(1-\alpha_n)}\bigl(\widehat\gamma-\widecheck\gamma\bigr)}_{T_{1,n}}
+
\underbrace{\sqrt{n(1-\alpha_n)}\bigl(\widecheck\gamma-\gamma\bigr)}_{T_{2,n}}
\xrightarrow{d}
\mathcal N\!\left(\frac{\lambda}{1-\rho},\,\gamma^2\right),
\]
which completes the proof of the weak convergence of Hill estimator given in \eqref{eq:Hill-residual-CLT}.

Regarding the proof of \eqref{CLTinter}, note first that because we are considering $k_n =  \left\lfloor n\left(1-\alpha_n\right)\right\rfloor$, then $\widehat{\mu}_{\alpha_n}(\varepsilon) = \widehat{\varepsilon}_{n - k_n:n}$. Therefore, the result in Lemma  \ref{cor:residual-approx} as well as assumptions of Theorem \ref{thm:Hill-residual} allow to conclude the proof of the weak consistency of the intermediate quantile of $\varepsilon$ given in \eqref{CLTinter} by making use of Corollary 2.1 in \cite{Girardannals2021}.
 

\subsection*{Proof of Corollary \ref{corES}}
In view of \eqref{eq:VARESrelation}, for $0 < \gamma < 1$ and
$\alpha_n \uparrow 1$, one has
\begin{eqnarray}\label{decomESeps}
\dfrac{\widehat{\text{ES}}_{\alpha_n}(\varepsilon)}
      {\text{ES}_{\alpha_n}(\varepsilon)} - 1
&=& \left(\dfrac{1/(1-\widehat{\gamma})}{1/(1-\gamma)} -1 \right)
  + \left(\dfrac{\widehat{\mu}_{\alpha_n}(\varepsilon)}
                {\mu_{\alpha_n}(\varepsilon)} -1 \right)\nonumber\\
& & \quad + \left(\dfrac{1/(1-\widehat{\gamma})}{1/(1-\gamma)} -1 \right)
      \times \left(\dfrac{\widehat{\mu}_{\alpha_n}(\varepsilon)}
                         {\mu_{\alpha_n}(\varepsilon)} -1 \right)
      \nonumber\\
&=& \dfrac{\widehat{\gamma} - \gamma}{1-\widehat{\gamma}}
  + \left(\dfrac{\widehat{\mu}_{\alpha_n}(\varepsilon)}
                {\mu_{\alpha_n}(\varepsilon)} -1 \right)
  + \dfrac{\widehat{\gamma} - \gamma}{1-\widehat{\gamma}}
    \times \left(\dfrac{\widehat{\mu}_{\alpha_n}(\varepsilon)}
                       {\mu_{\alpha_n}(\varepsilon)} -1 \right),
\end{eqnarray}
where the second equality uses
$\frac{1-\gamma}{1-\widehat\gamma}-1
=\frac{\widehat\gamma-\gamma}{1-\widehat\gamma}$.

Result \eqref{CLTinter} in Theorem~\ref{thm:Hill-residual} gives
\begin{equation}\label{1stterm}
\dfrac{\widehat{\mu}_{\alpha_n}(\varepsilon)}
      {\mu_{\alpha_n}(\varepsilon)} -1
= \mathcal{O}_{\mathbb{P}}\left(\dfrac{1}{\sqrt{n(1-\alpha_n)}}\right).
\end{equation}
It remains to control $(\widehat{\gamma}-\gamma)/(1-\widehat{\gamma})$,
for which we must ensure that $1-\widehat\gamma$ stays bounded away from
zero in probability. To this end we show that $\widehat\gamma$ lies, with
probability tending to one, in the interval
$I:=\left[\tfrac{\gamma}{2},\tfrac{\gamma+1}{2}\right]$.

If $\widehat{\gamma}\notin I$ then either $\widehat{\gamma}<\gamma/2$, in
which case $|\widehat\gamma-\gamma|>\gamma/2$, or
$\widehat\gamma>(\gamma+1)/2$, in which case
$|\widehat\gamma-\gamma|>(1-\gamma)/2$. Hence
$\{\widehat\gamma\notin I\}\subset\{|\widehat\gamma-\gamma|>
\min(\gamma/2,(1-\gamma)/2)\}$, and the convergence in probability of
$\widehat\gamma$ to $\gamma$ established in
Theorem~\ref{thm:Hill-residual} yields
\[
\mathbb{P}\left(\widehat\gamma\notin I\right)
\leq \mathbb{P}\left(|\widehat\gamma-\gamma|
>\min(\gamma/2,(1-\gamma)/2)\right)\longrightarrow 0,
\]
so that $\mathbb{P}(\widehat\gamma\in I)\to1$ as $n\to\infty$. On this
event $1-\widehat\gamma\geq(1-\gamma)/2$, whence
\[
\left|\dfrac{\widehat\gamma-\gamma}{1-\widehat\gamma}\right|
\leq \dfrac{2}{1-\gamma}\,|\widehat\gamma-\gamma| .
\]

Finally, \eqref{eq:Hill-residual-CLT} in
Theorem~\ref{thm:Hill-residual} gives
\begin{equation}\label{termgammahat}
\dfrac{\widehat\gamma-\gamma}{1-\widehat\gamma}
= \mathcal{O}_{\mathbb{P}}\left(\dfrac{1}{\sqrt{n(1-\alpha_n)}}\right).
\end{equation}
Since the third term of \eqref{decomESeps} is the product of two terms
each of order $\mathcal{O}_{\mathbb{P}}(\{n(1-\alpha_n)\}^{-1/2})$, it is
of order $\mathcal{O}_{\mathbb{P}}(\{n(1-\alpha_n)\}^{-1})$ and hence
negligible. Combining \eqref{decomESeps}, \eqref{1stterm} and
\eqref{termgammahat} completes the proof.

\subsection*{Proof of Theorem \ref{thm:extremeIQeps}}

Let us denote $v_n = \dfrac{\sqrt{n(1-\alpha_n)}}{\log\left((1-\alpha_n)/(1-\tau_n) \right)}.$\\

\noindent{\textit{Proof of} \eqref{CLTexteps} }\\
First observe that, in view of \eqref{CLTinter}, we can write, as $n\to\infty$, 
$$
v_n\left(\dfrac{\widehat{\mu}_{\tau_n}(\varepsilon)}{\mu_{\tau_n}(\varepsilon)} - 1\right) = v_n \left(\log \widehat{\mu}_{\tau_n}(\varepsilon) -  \log \mu_{\tau_n}(\varepsilon)\right)(1+o_{\mathbb{P}}(1)).
$$
Then, the proof of statement \eqref{CLTexteps} will be concluded if we show that  
$$v_n \left(\log \widehat{\mu}_{\tau_n}(\varepsilon) -  \log \mu_{\tau_n}(\varepsilon)\right) \xrightarrow{d}
\mathcal N\!\left(\frac{\lambda}{1-\rho},\,\gamma^2\right)$$

Note that since $k_n =  \left\lfloor n\left(1-\alpha_n\right)\right\rfloor$, the Weissman estimator of $\mu_{\tau_n}(\varepsilon)$ computed on the unobserved errors $\varepsilon_1, \dots, \varepsilon_n$ is defined as follows
$$
\widecheck{\mu}_{\tau_n}(\varepsilon) = \left(\dfrac{1-\alpha_n}{1-\tau_n} \right)^{\widecheck\gamma_n}\,\varepsilon_{n - k_n:n}, 
$$
where $\widecheck\gamma$ is defined in \eqref{HillMAR}.

Consider now the following decomposition
\begin{eqnarray}\label{decomplog}
v_n \left(\log \widehat{\mu}_{\tau_n}(\varepsilon) -  \log \mu_{\tau_n}(\varepsilon)\right) &=& v_n \left(\log \widehat{\mu}_{\tau_n}(\varepsilon) -  \log \widecheck\mu_{\tau_n}(\varepsilon)\right) + v_n \left(\log \widecheck{\mu}_{\tau_n}(\varepsilon) -  \log \mu_{\tau_n}(\varepsilon)\right) \nonumber\\
&=: & I_{1,n} + I_{2,n}.
\end{eqnarray}
Now let us discuss the behavior of each of the above terms.

One can see that $|I_{1,n}| \leq v_n \left|\log \dfrac{\widehat\varepsilon_{n - k_n:n}}{\varepsilon_{n - k_n:n}} \right| + v_n |\widehat\gamma - \widecheck\gamma| \,\left|\log\left(\dfrac{1-\alpha_n}{1-\tau_n} \right) \right| =: I_{1,1,n} + I_{1,2,n}.$
Making use of Lemma  \ref{cor:residual-approx} and Corollary 2.1 in \cite{Girardannals2021}, together with conditions \eqref{condh} and \eqref{condalpha}, we conclude that 
\begin{equation}\label{I11n}
I_{1,1,n} = o_{\mathbb{P}}(1).
\end{equation}
The second term $I_{1,2,n}$ can be written as $I_{1,2,n} = |T_{1,n}| (1 + o_{\mathbb{P}}(1))$ where $T_{1,n}$ is defined in the proof of Theorem \ref{thm:Hill-residual}.  Then, in view of \eqref{eq:T1n}, one concludes that $I_{1,2,n} = o_{\mathbb{P}}(1).$ This allows to deduce that
\begin{equation}\label{I1n}
I_{1,n} = o_{\mathbb{P}}(1).
\end{equation}

Following similar steps as for $I_{1,n}$, we can show that the term $I_{2,n}$ in \eqref{decomplog} inherits the asymptotic properties of Hill estimator $\widecheck\gamma$. Therefore, Theorem 4.3.8 in \cite{deHaanFerreira2006} allows to say that 
\begin{equation}\label{I2n}
I_{2,n} \xrightarrow{d}
\mathcal N\!\left(\frac{\lambda}{1-\rho},\,\gamma^2\right).
 \end{equation}
 By combining \eqref{I1n} and \eqref{I2n} we conclude the proof of result in \eqref{CLTexteps}.\\

 \noindent{\textit{Proof of} \eqref{CLTESeps} }\\

 In view of Corollary \ref{corES} the following relation holds
 $$
 \dfrac{\widehat{\text{ES}}_{\tau_n}(\varepsilon)}{\text{ES}_{\tau_n}(\varepsilon)} - 1 = [\log \widehat{\text{ES}}_{\tau_n}(\varepsilon) - \log \text{ES}_{\tau_n}(\varepsilon) ] \times (1+o_{\mathbb{P}}(1)).
 $$
Now making use of \eqref{eq:VARESrelation} and for $\tau_n\uparrow 1$, one obtains
\begin{eqnarray*}
   v_n \left[\log \widehat{\text{ES}}_{\tau_n}(\varepsilon) - \log \text{ES}_{\tau_n}(\varepsilon)\right] &=& v_n [\log \widehat{\mu}_{\tau_n}(\varepsilon) - \log \mu_{\tau_n}(\varepsilon)] + v_n \left[\log\left(\widehat\gamma/(1-\widehat\gamma) \right) - \log\left(\gamma/(1-\gamma) \right)\right]\\
    &=:& J_{1,n} + J_{2,n}.
\end{eqnarray*}

Note that $J_{1,n} \xrightarrow{d}
\mathcal N\!\left(\lambda/(1-\rho),\,\gamma^2\right)$ as shown above in the proof of \eqref{CLTexteps}. Then, it remains to control the term $J_{2,n}.$  

Since $\gamma \in (0,1)$ and $\widehat\gamma \to \gamma$ in probability, then, with probability tending to one, $\widehat\gamma$ belongs to a compact subinterval of $(0,1).$ On such a set, the function $g(x) := \log(1/(1-x))$ is continuously differentiable.

By the mean value theorem, there exists a random value $\widetilde\gamma$ lying between $\widehat\gamma$ and $\gamma$ such that $g(\widehat\gamma) - g(\gamma) = g^\prime(\widetilde\gamma)(\widehat\gamma - \gamma).$ Moreover, since $\widehat\gamma \xrightarrow{\mathbb P} \gamma$, we have $\mathbb{P}(\gamma/2 \leq \widehat\gamma \leq (1+\gamma)/2) \to 1.$

On this high-probability event, both $\gamma$ and $\widehat\gamma$ belong to $[\gamma/2, (1+\gamma)/2]$, and therefore $\widetilde\gamma$ also lies in this interval. Consequently,

$$|g^\prime(\widetilde\gamma)| \leq \dfrac{1}{(\gamma/2)(1-(1+\gamma)/2)} = \dfrac{4}{\gamma(1-\gamma)} =: C_\gamma.$$

Under condition \eqref{condalpha}, it follows that, 
$$
|J_{2,n}| \leq v_n C_\gamma |\widehat\gamma - \gamma| +o_{\mathbb{P}}(1).
$$

Finally, by Theorem \ref{thm:Hill-residual} we have $J_{n,2} \xrightarrow{d}
\mathcal N\!\left(\lambda/(1-\rho),\,\gamma^2\right)$ which completes the proof of \eqref{CLTESeps}.

\subsection*{Proof of Theorem \ref{thm:extremeQ}}

Let us denote $v_n = \dfrac{\sqrt{n(1-\alpha_n)}}{\log\left((1-\alpha_n)/(1-\tau_n) \right)}.$\\

\noindent{\textit{Proof of} \eqref{CLTextcondQ} }\\
Observe that, using definitions in \eqref{eq:extreme_CVaR} and \eqref{estimatorcvar}, one can obtain the following decomposition
\begin{eqnarray*}
v_n \left(\dfrac{\widehat{\mu}_{\tau_n}(x)}{\mu_{\tau_n}(x)} - 1\right) &=& \dfrac{v_n (m_n(x) - m(x))}{m(x) + \sigma(x) \mu_{\tau_n}(\varepsilon)} + \left(\dfrac{m(x)}{m(x) + \sigma(x) \mu_{\tau_n}(\varepsilon)} + 1 \right)^{-1}\times \dfrac{v_n(\sigma_n(x) - \sigma(x))}{\sigma(x)}\\
&+& v_n \left(\dfrac{\widehat\mu_{\tau_n}(\varepsilon)}{\mu_{\tau_n}(\varepsilon)}-1 \right)\left(\dfrac{m(x)}{m(x) + \sigma(x) \mu_{\tau_n}(\varepsilon)} + 1 \right)^{-1} \left(\dfrac{\sigma_n(x) - \sigma(x)}{\sigma(x)} \right).
\end{eqnarray*}
Under the conditions $\sqrt{n(1-\alpha_n)} h^{\min{(b_1,b_2)}}\to 0$, $(1-\alpha_n)/(1-\tau_n) \to \infty$, $n(1-\alpha_n)\log n/(n\phi(h)) \to 0$, $\log n/(n\phi(h))\to 0$ as $n\to \infty$ (i.e. conditions \eqref{cond1}, \eqref{condh}, and \eqref{condalpha}) and the almost sure convergence of $m_n(x)$ (see Theorems \ref{T1}), one has, 
   \begin{equation}\label{vnmn}
   v_n (m_n(x) - m(x)) = o_{\mathbb{P}}(1).
   \end{equation}
   Under the same conditions stated above and using the almost sure convergence of $\sigma_n(x)$ (see Theorem \ref{T2}) we have
   \begin{equation}\label{vnsigman} 
   v_n(\sigma_n(x) - \sigma(x)) = o_{\mathbb{P}}(1).
   \end{equation}
   In view of $\mu_{\tau_n}(\varepsilon) \to \infty$ as $n\to\infty$, \eqref{vnmn}, \eqref{vnsigman}, and  Theorem \ref{T2}, and $\sigma(x) <\infty$, one finally obtain
   $$
  v_n \left(\dfrac{\widehat{\mu}_{\tau_n}(x)}{\mu_{\tau_n}(x)} - 1\right) = v_n \left(\dfrac{\widehat\mu_{\tau_n}(\varepsilon)}{\mu_{\tau_n}(\varepsilon)}-1 \right) \times (1 + o_{\mathbb{P}}(1)) + o_{\mathbb{P}}(1).
   $$
   Finally, making use of \eqref{CLTexteps} in Theorem \ref{thm:extremeIQeps}, one concludes the proof of the result in \eqref{CLTextcondQ}.\\

   \noindent{\textit{Proof of} \eqref{CLTcondES} }\\

   Note that the following decomposition holds
   \begin{eqnarray*}
v_n \left(\dfrac{\widehat{\text{CES}}_{\tau_n}(x)}{\text{CES}_{\tau_n}(x)} - 1\right) &=& \dfrac{v_n (m_n(x) - m(x))}{m(x) + \sigma(x) \text{ES}_{\tau_n}(\varepsilon)} + \left(\dfrac{m(x)}{m(x) + \sigma(x) \text{ES}_{\tau_n}(\varepsilon)} + 1 \right)^{-1}\times \dfrac{v_n(\sigma_n(x) - \sigma(x))}{\sigma(x)}\\
&+& v_n \left(\dfrac{\widehat{\text{ES}}_{\tau_n}(\varepsilon)}{\text{ES}_{\tau_n}(\varepsilon)}-1 \right)\left(\dfrac{m(x)}{m(x) + \sigma(x) \text{ES}_{\tau_n}(\varepsilon)} + 1 \right)^{-1} \left(\dfrac{\sigma_n(x) - \sigma(x)}{\sigma(x)} \right).
\end{eqnarray*}

   The proof of \eqref{CLTcondES} can be concluded by combining \eqref{vnmn}, \eqref{vnsigman}, \eqref{CLTESeps}, conditions \eqref{condalpha} and the fact that $\text{ES}_{\tau_n}(\varepsilon) \to\infty$ as $\tau_n \to 1.$

\end{document}